\documentclass[12pt]{article}
\usepackage{amsmath}
\usepackage{amscd}
\usepackage{amssymb}
\usepackage{array}
\usepackage{makeidx}
\usepackage{ifpdf}
\ifx\pdfoutput\undefined
   \pdffalse
 \else
   \pdfoutput=1
   \pdftrue
  \usepackage[pdftex]{hyperref}
\fi

\newtheorem{remark}{Remark}[section]
\newtheorem{definition}[remark]{Definition}
\newtheorem{theorem}[remark]{Theorem}
\newtheorem{example}[remark]{Example}

\newtheorem{lemma}[remark]{Lemma}
 \DeclareMathOperator{\sign}{sign}

\newcommand{\ft}[2]{{\textstyle\frac{#1}{#2}}}
\def\rmi{{\rm i}}
\def\rmd{{\rm d}}
\def\rme{{\rm e}}
\newcommand{\DLC}{D}
\def\ib{{\bar \imath}}
\def\jb{{\bar \jmath}}
\def\kb{{\bar k}}

\makeindex

\begin{document}

\newcommand{\id}{\relax{\rm 1\kern-.28em 1}}

\newcommand{\R}{\mathbb{R}}
\newcommand{\C}{\mathbb{C}}
\newcommand{\Z}{\mathbb{Z}}
\newcommand{\Hb}{\mathbb{H}}

\newcommand{\rspan}{\mathrm{span}}

\newcommand{\rGL}{\mathrm{GL}}
\newcommand{\rU}{\mathrm{U}}
\newcommand{\rSU}{\mathrm{SU}}
\newcommand{\rO}{\mathrm{O}}
\newcommand{\rSO}{\mathrm{SO}}
\newcommand{\rSL}{\mathrm{SL}}
\newcommand{\rSp}{\mathrm{Sp}}
\newcommand{\rId}{\mathrm{Id}}

\newcommand{\rhor}{\mathrm{hor}}

\newcommand{\rank}{\mathrm{rank}}

\newcommand{\cM}{\mathcal{M}}
\newcommand{\cV}{\mathcal{V}}
\newcommand{\cF}{\mathcal{F}}
\newcommand{\cK}{\mathcal{K}}
\newcommand{\cH}{\mathcal{H}}

\newcommand{\cX}{\mathcal{X}}
\newcommand{\cC}{\mathcal{C}}
\newcommand{\cA}{\mathcal{A}}
\newcommand{\cL}{\mathcal{L}}
\newcommand{\cU}{\mathcal{U}}
\newcommand{\cD}{\mathcal{D}}

\newcommand{\fgl}{\mathfrak{gl}}
\newcommand{\fsl}{\mathfrak{sl}}
\newcommand{\fso}{\mathfrak{so}}
\newcommand{\fsp}{\mathfrak{sp}}
\newcommand{\fg}{\mathfrak{g}}
\newcommand{\fs}{\mathfrak{s}}
\newcommand{\fa}{\mathfrak{a}}
\newcommand{\fh}{\mathfrak{h}}
\newcommand{\fb}{\mathfrak{b}}
\newcommand{\fR}{\mathfrak{R}}

\newcommand{\be}{\mathbf{E}}

\newcommand{\CP}{\mathbb{C}\mathrm{P}}

\begin{titlepage}
\begin{flushright}
KUL-TF-06/34\\
IFIC/06-29\\
FTUV-06/1216\\
hep-th/0612210
\end{flushright}
\vspace{.5cm}

  \centerline{\LARGE \bf Special geometry for arbitrary signatures }

  \vskip 0.7 cm

  \centerline{Mar{\'\i}a A. Lled{\'o}$^\sharp$, {\'O}scar Maci{\'a}$^\sharp$, Antoine Van Proeyen$^\flat$ and Veeravalli S. Varadarajan$^\natural$
  }
 \vskip 0.5 cm

\begin{center}
  {\it $^\sharp$Departament de F{\'\i}sica Te{\`o}rica,
Universitat de Val{\`e}ncia and IFIC,\\
C/ Dr.
Moliner, 50, E-46100 Burjassot (Val{\`e}ncia), Spain.\\[3mm]
 $^\flat$Instituut voor Theoretische Fysica, Katholieke
Universiteit Leuven.\\

 Celestijnenlaan 200D B-3001 Leuven, Belgium.\\[3mm]
 $^\natural$Department of Mathematics, UCLA.
 Los Angeles, CA 90095-1555, USA}\\[4mm]

 {\footnotesize E-mail: Maria.Lledo@ific.uv.es,
Oscar.Macia@ific.uv.es,\\
Antoine.VanProeyen@fys.kuleuven.be,vsv@math.ucla.edu }

\end{center}

%%%%%%%%%%%%%%%%%%%%%%%%%%%%%%%%%%%%%%%%%%%%%%%%%%%%%%%%%%%%%%%%%%%%%%
%%%%%%%%%%%%%%%%%%%% abstract %%%%%%%%%%%%%%%%%%%%%%%%%%%%%%%%%%%%%%%%
%%%%%%%%%%%%%%%%%%%%%%%%%%%%%%%%%%%%%%%%%%%%%%%%%%%%%%%%%%%%%%%%%%%%%%

\vskip 0.5cm

\begin{abstract}
In this paper we generalize special geometry to arbitrary signatures in
target space. We formulate the definitions in a precise mathematical
setting and give a translation to the coordinate formalism used in
physics. For the projective case, we first discuss in detail projective
K{\"a}hler manifolds, appearing in $N=1$ supergravity. We develop a new point
of view based on the intrinsic construction of the line bundle. The
topological properties are then derived and the Levi-Civita connection in
the projective manifold is obtained as a particular projection of a
Levi-Civita connection in a `mother' manifold with one extra complex
dimension. The origin of this approach is in the superconformal formalism
of physics, which is also explained in detail.
 Finally, we specialize these results to projective special K{\"a}hler
 manifolds and provide explicit examples with different choices of
 signature.
\end{abstract}
\vspace{2mm} \vfill \hrule width 3.cm
 \vspace{1mm}

\noindent Contribution to the handbook on pseudo-Riemannian geometry and
supersymmetry, ed. V. Cort{\'e}s, published by the European Mathematical
Society in the series ``IRMA Lectures in Mathematics and Theoretical
Physics".
\end{titlepage}
\addtocounter{page}{1}
 \tableofcontents{}
\newpage
%%%%%%%%%%%%%%%%
\section{Introduction}

{\it Special K{\"a}hler geometry}\index{special K{\"a}hler geometry} is the
geometry of the manifold spanned by the scalars of vector multiplets of
$D=4$, $N=2$ supersymmetry. The {\it rigid} version, that occurs in
supersymmetry without gravity, appeared first in Refs.
\cite{Sierra:1983cc,Gates:1984py}. The construction for $N=2$
supergravity appeared in Ref. \cite{deWit:1984pk}, and it is called {\it
projective special K{\"a}hler geometry}\index{projective special K{\"a}hler
geometry}. It played an important role in several developments of string
theory.

These first formulations offered a local point of view. For the rigid
case, the condition for a K{\"a}hler geometry to be `special' is the
existence of a preferred set of holomorphic coordinates $z^i$, called
{\it special coordinates}\index{special coordinates}  in which the K{\"a}hler
potential $\cK$ can be expressed in terms of a holomorphic function, the
{\it prepotential}\index{prepotential} $\cF$,
\begin{equation}g_{i\jb }=\frac{\partial \cK}{\partial z^i\partial \bar z^\jb }, \qquad \cK=2\Im(\frac {\partial \cF} {\partial z^k}\bar
z^{\bar k}).\label{prep}
\end{equation}
It is seen then as a further restriction on the metric, compatible with
the complex structure. For the projective case, the original construction
was based on superconformal tensor calculus and involves a
projectivization of the manifold due to the extra vector field, the
graviphoton, which does not have associated a scalar. In simple words,
one has a rigid special manifold with a dilation symmetry and a non
physical scalar, which is projected out by fixing the symmetry.

The property of being a special K{\"a}hler manifold is then a purely
geometrical one, and can be formulated independently of supersymmetry. It
is given though in terms of a preferred set of coordinates. Although this
local formulation is not incomplete (there has to exists an {\it open
cover} of the manifold by special coordinates), it remains the intriguing
question if there is a way of defining what is a special K{\"a}hler manifold
with global statements,
 independent of coordinates. The first attempts were in Refs.
 \cite{Strominger:1990pd,Castellani:1990tp,Castellani:1990zd}.
A set of equivalent definitions was found in Ref.
 \cite{Craps:1997gp}, and later on, a mathematical formulation
 appeared in Ref. \cite{Freed:1997dp}.

One fundamental ingredient in the global approach is the
  existence of a certain flat {\it symplectic bundle}\index{symplectic bundle}.
  Peculiar to Freed's formulation \cite{Freed:1997dp} is that the
  symplectic bundle is recognised as the tangent bundle, so the
  construction is {\it intrinsic}. In fact, the rigid case (see
  Definition \ref{rsdef})
   comes
  out very elegantly, and for this part we will follow closely Freed's work
  (with the  exception of the pseudo-Riemannian case, which we will mention later). The
  projective case is much more involved. We define a projective
  special manifold in terms of a rigid special manifold with a
  {\it homothetic Killing vector} (see Definition \ref{psdef}). In this way, the definition
   is not only intrinsic
  but directly related to the way in which it is obtained in supergravity
  \cite{deWit:1984pk}. The point of contact of this definition
  with Freed's work is in his Proposition 4.6.

  \bigskip

So far as for Riemannian, special K{\"a}hler manifolds. Pseudo-Riemannian
special K{\"a}hler manifolds\index{pseudo-Riemannian special K{\"a}hler
manifold}\footnote{Note that all discussions on the signature in this
work concern the signature of the K{\"a}hler manifold, i.e. the target
manifold of the supergravity theory. This is unrelated to the signature
of spacetime, which we keep Minkowskian to have the standard special
geometries. Discussions on generalizations to Euclidean spacetime
signature are in Refs.
\cite{Cortes:2003zd,Cortes:2005uq,Mohaupt:2006ph}.} are very relevant in
supergravity. A physically sensible supergravity theory must have a
positive definite
  target-space metric. From the conformal calculus approach it is known that in
  order to get such positive definite metric   the rigid K{\"a}hler
  manifold before projection has to signature $(2, 2n)$.
  But pseudo-Riemannian special K{\"a}hler manifolds present an additional
complication.
  Special coordinates \index{special coordinates} are complex coordinates constructed  from a set of
  flat Darboux coordinates \index{Darboux coordinates} $(q^i, p_i)$ by taking the holomorphic extension of the
  $q^i$'s (or, alternatively, of the $p_i's$). They have then the prepotential property (\ref{prep}).
   When the signature of
  the metric is indefinite, this holomorphic extension does not always result in a set of
  $n$ independent holomorphic coordinates. There is a subclass of
  Darboux systems that have this property. It is important thought
  that one can always make a constant symplectic rotation to coordinates
  $({q'}^i,p'_i)$ such that the ${q'}^i$'s extend to special
  coordinates, so there is still a covering of the manifold by special coordinates.
  But not all flat Darboux systems are
  suitable to obtain special coordinates. As a consequence, the structure group of the bundle is reduced
  to a subgroup of the symplectic group. This was first observed in \cite{Ceresole:1995jg}.
Nevertheless, flat Darboux coordinates which do not lead to special
coordinates nor prepotential are  very relevant. They were used
  to prove that one can break $N=2$ supersymmetry  partially to $N=1$ \cite{Ferrara:1995gu} and not
  necessarily to $N=0$, as it was thought before. This is an
  extremely important property for phenomenological applications.
It is then one of the main motives of  this work (which was missing in
Ref. \cite{Freed:1997dp}) to generalize the construction of special
geometry to arbitrary signatures.

  \bigskip

In another context, it has been recently shown
\cite{LopesCardoso:2006bg,Ferrara:2006at,Ferrara:2006js} that relating
flat
  Darboux coordinates with the real central charges and attractor equations
  would have a simplifying role in the description of the attractor mechanism of black holes in
   $N=2$ supergravity  (see Refs.
   \cite{Ferrara:1995ih,Strominger:1996kf,Ferrara:1996dd} for the attractor
   mechanism).
 \bigskip

Pseudo-Riemannian, projective special K{\"a}hler manifolds appear also as
dimensional reductions of supergravity theories in eleven dimensions and
exotic signatures, obtained by duality transformations from the standard
Minkowskian signature. These are the theories $M^*$ and $M'$, in
signatures (9,2) and (6,5) proposed in Ref. \cite{Hull:1998ym}. The
pseudo-Riemannian special manifolds arising in $D=4$ are discussed in
Ref. \cite{Ferrara:2001sx}.
\bigskip

 An important part of the work in dealing with projective special K{\"a}hler
  manifolds concerns in fact a more general class of K{\"a}hler manifolds,
the so-called K{\"a}hler-Hodge manifolds\index{K{\"a}hler-Hodge
manifold}.\footnote{A K{\"a}hler-Hodge manifold is a K{\"a}hler manifold with
integer  K{\"a}hler cohomology class.} It was found in Ref.
\cite{Witten:1982hu} that the K{\"a}hler geometries of $N=1$ supergravity
should be K{\"a}hler-Hodge. We
  propose an intrinsic definition of projective K{\"a}hler  manifolds (see Definition \ref{psdef}),
inspired in the conformal calculus approach used in physics. Then we show
that they have integer K{\"a}hler cohomology class, so they are K{\"a}hler-Hodge.

  \bigskip

  The paper is as much self contained as possible, so we have
  included vast review material. On the other hand, having in mind
  the connection to physics, we have tried to work everything out in
  coordinates, as to have the sometimes difficult translation between
  two languages, the physicist's  and the mathematician's one,
   each of them with its own advantages. We have also taken
   time in explaining some examples, which may clarify the
   abstract definitions.

  The paper is organized as follows.

  In Section \ref{ss:Kahler} we review basic material
  on complex and K{\"a}hler manifolds and Hermitian bundles.
  It is used extensively in the paper, so with it we set the basic
  notation. The reader  can also skip it and come back to it punctually
  when some concept is called for.

  Section \ref{ss:rigid} is devoted to the rigid case. We start
  with some geometric preliminaries not included in Section
  \ref{ss:Kahler} and then we take up the definition of rigid
  special K{\"a}hler manifolds. We follow the lines of Ref.
  \cite{Freed:1997dp}, explaining carefully how the formulas in
  coordinates are obtained from the abstract definition. Then we
  treat the pseudo-Riemannian case, giving some clarifying simple
  examples. We come back to Freed's treatment for the holomorphic
  cubic form, which will be used later on.

  Section \ref{ss:projective} is dedicated to projective K{\"a}hler
  manifolds as a previous step towards projective special
  geometry. We introduce some concepts on affine transformations
  and homothetic Killing vectors and derive some of its
  properties. This material is not new, but perhaps not so widely
  known, so it is fundamental to have it at hand. Then we define  projective
  K{\"a}hler manifolds starting from a  K{\"a}hler manifold
  $\tilde \cM$ (with arbitrary signature)
   which has an action of $\C^\times$ (as well as other properties,
   see Definition \ref{psdef}). We then consider $\cM=\tilde \cM/\C^\times$ and
   construct the symplectic and the line
  bundles over it. The line bundle has a Hermitian metric induced from the K{\"a}hler metric
  on $\tilde \cM$ (here the importance of the intrinsic approach), whose
  Ricci  form turns out to be closed and non degenerate, defining then a
  symplectic structure (actually, a K{\"a}hler one). Since it is the
  first Chern class of a line bundle, the manifold is
  K{\"a}hler-Hodge.

  We then propose an alternative and beautiful way of understanding the
  Levi-Civita connection in $\cM$, directly induced from the one in
  $\tilde \cM$. It is a bit
  involved, but it really gives precious insight into the geometry of $\cM$.

  Section \ref{ss:confcal} is a brief excursion on the origin of
  projective K{\"a}hler geometry as it is seen from a model in
  physics. It is the simplest one to consider, and it does not
  include supersymmetry. Indeed, the ideas of conformal calculus
  are more general than their applications to supergravity.

In Section \ref{ss:psk} we impose on $\tilde \cM$ the condition to be
rigid special K{\"a}hler, then  $\cM$ will be a projective special K{\"a}hler
manifold. The precise definition is Definition \ref{defpsk}, and the
consequences are analysed in the sequel. In particular, we obtain the
holomorphic cubic form and then the formula for the curvature. We
conclude with some examples, in particular the pseudo-Riemannian space
$$\frac{\rSU(1,2)}{\rSU(1,1)\times \rU(1)}.$$

\section{K{\"a}hler manifolds}
 \label{ss:Kahler}
This first section recapitulates the basic definitions on complex
manifolds and K{\"a}hler manifolds in particular. It is essentially a summary
of part of Chapter IX in Ref. \cite{Kobayashi1996}. It can be skipped by
readers familiar with K{\"a}hler manifolds or used just to set the
conventions.

\subsection{Generalities on quasicomplex and complex manifolds}
Let $\cM$ be a quasicomplex (or almost complex) manifold\index{almost
complex manifold}\index{quasicomplex manifold} of dimension $2n$, with
$J:T\cM\mapsto T\cM$ the quasicomplex structure, $J^2=-1$.

\begin{remark}\label{complexmanifold}\end{remark}

Suppose that $\cM$ is a complex manifold\index{complex manifold} and that
$(z^1, \dots z^n)$ are complex coordinates on an open set $U\subset \cM$,
$z^j=x^j+\rmi y^j$. Then, $(x^1,\dots x^n, y^1,\dots y^n)$ is a
coordinate system in $U$ and we have that
$$J\left(\frac{\partial}{\partial x^j}\right)=\frac{\partial}{\partial y^j},
\quad J\left(\frac{\partial}{\partial
y^j}\right)=-\frac{\partial}{\partial x^j}, \qquad j=1,\dots
n.$$\hfill$\square$

\bigskip

Let $T_m^c\cM$ denote the complexification of the tangent space at $m\in
\cM$. We denote by $T_m^{1,0}\cM$ and $T_m^{0,1}\cM$ the eigenspaces of
$J$ at $m$ with eigenvalues $\rmi$ and $-\rmi $ respectively. Then
$$Z=X-\rmi JX\in T_m^{1,0}\cM, \qquad \bar Z=X+\rmi JX\in T_m^{0,1}\cM$$
for any real vector $X\in T_m\cM$. The operation $Z\mapsto \bar Z$ is a
real linear endomorphism called {\it complex conjugation\index{complex
conjugation}}. From now on we will denote the (complexified) tangent
space simply as $T_m=T_m\cM$.

Let ${T_m^*}^c$ denote the complexification of the cotangent space at $m$
and ${T^*}^c$ the complexified cotangent bundle of $\cM$. Let $\omega\in
T^*$. The pull back, at each point $m$, of $\omega$ through $J$,
$$J^*\omega_m(X)=\omega_m(JX), \qquad \forall X\in T_m$$
defines  an endomorphism
 $$J^*:T^*\rightarrow T^*,$$
with $(J^*)^2=-1$, which extends in the obvious way to the complexified
cotangent space. The eigenspaces of eigenvalues $\rmi$ and $-\rmi $ of
$J^*$ at $m$ are denoted as
$$\Omega_m^{1,0}=\Lambda^{1,0}{T_m^*}^c\quad \mathrm{and}\quad
\Omega_m^{0,1}=\Lambda^{0,1}{T_m^*}^c$$ respectively. One has that
\begin{eqnarray*}\Omega_m^{0,1}=\{\omega_m\in {T_m^*}^c\;/\;
\omega_m(Z)=0 \quad \forall Z\in
T_m^{1,0}\}\\
\Omega_m^{1,0}=\{\omega_m\in{T_m^*}^c\;/\; \omega_m(Z)=0 \quad
\forall Z\in T_m^{0,1}\}
\end{eqnarray*}

Since the exterior product space,
${\Omega_m}=\sum_{r=0}^{2n}\Lambda^r{T_m^*}^c$, is generated by
$\Omega_m^{0,0}, \Omega_m^{1,0}$ and $\Omega_m^{0,1}$, $\Omega_m$ has a
bigrading
$$\Omega_m=\sum_{p,q=0}^n\Omega_m^{p,q},$$ and so has the space of
complex forms
$$\Omega=\sum_{p,q=0}^n\Omega^{p,q}.$$

\begin{remark}\label{complexmanifold2}\end{remark}If $\cM$ is a complex manifold, as in
Remark \ref{complexmanifold}, then
\begin{eqnarray*}T^{1,0}_m=\mathrm{span}_\C\left\{\frac{\partial}{\partial
z^j}\Big|_m=\frac{1}{2}\left(\frac{\partial}{\partial x^j}-\rmi
\frac{\partial}{\partial
y^j}\right)\Big|_m\right\}_{j=1}^n, \\
T^{0,1}_m=\mathrm{span}_\C\left\{\frac{\partial}{\partial \bar z^\jb
}\Big|_m=\frac{1}{2}\left(\frac{\partial}{\partial x^j}+\rmi
\frac{\partial}{\partial y^j}\right)\Big|_m\right\}_{j=1}^n.
\end{eqnarray*}
For the complex forms we have
\begin{eqnarray*}
\Omega_m^{1,0}=\mathrm{span}_\C\left\{\rmd z^j|_m=(\rmd x^j+\rmi\rmd
y^j)|_m\right\}_{j=1}^n,
\\
\Omega_m^{0,1}=\mathrm{span}_\C\left\{\rmd\bar z^\jb |_m=(\rmd
x^j-\rmi\rmd y^j)|_m\right\}_{j=1}^n.
\end{eqnarray*}
The set of forms
\begin{eqnarray*}
\{\rmd z^{j_1}\wedge \rmd z^{j_2}\wedge\dots \rmd z^{j_p}\wedge \rmd\bar
z^{\kb_1}\wedge \rmd\bar z^{\kb_2}\wedge\dots \rmd\bar z^{\kb_q}\}, \\
1\leq j_1\leq j_2\leq \dots j_p\leq n,\quad 1\leq \kb_1\leq \kb_2\leq
\dots \kb_q\leq n,
\end{eqnarray*}
is a local basis of $\Omega^{p,q}$.

\bigskip

For a complex manifold one can prove \cite{Kobayashi1996} that the
differential
$$\rmd\Omega^{p,q}\subset \Omega^{p+1,q}+\Omega^{p,q+1}. $$ Then we
can define $\partial:\Omega^{p,q}\rightarrow \Omega^{p+1,q}$ and
$\bar\partial:\Omega^{p,q}\rightarrow \Omega^{p,q+1}$ as
$$\rmd=\partial+\bar \partial,$$
and since $\rmd^2=0$ we have
$$\partial^2=0, \qquad \bar\partial^2=0, \qquad \partial
\circ\bar\partial+\bar\partial\circ \partial=0.$$

A form $\omega\in \Omega^{p,0}$ is said to be
holomorphic\index{holomorphic form}
 if $\bar\partial \omega=0$. A form $\omega\in \Omega^{0,p}$ is said to be
 antiholomorphic\index{antiholomorphic form}
 if $\partial \omega=0$. A function is holomorphic if
 $$\frac {\partial}{ \partial\bar z^\jb }f=0, \qquad j=1,\dots n$$
 (respectively, antiholomorphic). A holomorphic vector field $Z$ is a
 complex vector field of type $(p,0)$ such that $Zf$ is holomorphic
 for every holomorphic $f$. Locally,
 $$Z=\sum_{j=1}^nf^j\frac {\partial}{\partial z^j}$$ with all the
 $f^j$ holomorphic.
 \hfill$\square$

 \subsection{Hermitian metrics and K{\"a}hler metrics}
A {\it Hermitian metric\index{Hermitian metric}} on a quasicomplex
manifold $\cM$ with quasicomplex structure $J$ is a Riemannian metric $g$
such that
 $$g(JX,JY)=g(X,Y), \qquad \forall X,Y\in T\cM.$$
Every paracompact quasicomplex manifold admits a Hermitian metric. This
is because for a given Riemannian metric $h$ and a quasicomplex structure
$J$ we can obtain a Hermitian metric by setting
 $$g(X,Y)=h(X,Y)+h(JX,JY) \qquad \forall X,Y\in T\cM,$$
$g$ is  extended to $T^c$ by linearity. It is easy to check that
\label{propertiesg}

 \bigskip

\noindent 1. $g(Z,W)=0$ for $ Z, W$ of type $(1,0)$,

\noindent 2. $g(Z,\bar Z)>0$,

\noindent 3. $g(\bar Z, \bar W)=\overline{g(Z,W)}$.

\bigskip

The fundamental 2-form of a Hermitian metric is
$$\Phi(X,Y)=g(X,JY)\qquad \forall X,Y\in T\cM.$$
It is non degenerate at each point of the manifold.

\begin{remark} Almost complex linear connections\index{almost complex linear connections}\end{remark} The
{\it torsion} of a quasicomplex structure\index{quasicomplex
structure!torsion} $J$ is the tensor field (1-covariant, 2-contravariant)
$$N(X,Y)=2\{[JX,JY]-[X,Y]-J[X,JY]-J[JX,Y]\}.$$

A quasicomplex structure is said to be integrable if it has no torsion.
This is equivalent to saying that the commutator of two vector fields of
type $(1,0)$ (alternatively $(0,1)$) is a  vector field of type $(1,0)$
(alternatively $(0,1)$). To see this, let $Z, W$ be such that $JZ=\rmi Z$
and $JW=\rmi W$, then if $N(Z,W)=0$ it is immediate that $J[Z,W]=\rmi
[Z,W]$. In the other direction, a real vector field can be always written
as the sum $X=Z+\bar Z$, where $Z$ is $(1,0)$ and $\bar Z$ is $(0,1)$.
Let also $Y=W+\bar W$. Then it is immediate to prove that $N(X,Y)=0$.

\medskip

A quasicomplex structure is a complex structure if and only if  it is
integrable. This is the Newlander-Nirenberg
theorem\index{Newlander-Nirenberg theorem} \cite{Newlander1957}.

\medskip

We say that a linear connection is quasicomplex if the covariant
derivative of the quasicomplex structure is zero (which is equivalent to
being a connection in the bundle of complex linear frames). Every
quasicomplex manifold admits a quasicomplex affine connection whose
torsion $T$ is proportional to the torsion $N$ of the quasicomplex
structure. \hfill$\square$
\bigskip

In general, the Riemannian connection associated to a Hermitian metric
is not quasicomplex. If it is so, then  the quasicomplex structure has
no torsion and the fundamental form is closed. The converse is also
true: for a complex manifold, the Riemannian connection of a Hermitian
metric is quasicomplex if and only if the fundamental 2-form $\Phi$ is
closed. (The proof of these statements can be read in Ref.
\cite{Kobayashi1996}, Chapter IX.)

A quasicomplex manifold, with a Hermitian metric is a {\it quasi-K{\"a}hler}
(or almost K{\"a}hler) manifold\index{quasi-K{\"a}hler manifold}\index{almost
K{\"a}hler manifold} if the fundamental form is closed.

Let  $\cM$ be a differential manifold.  A {\it symplectic
structure\index{symplectic structure}} on $\cM$ is a  2-form $\Phi$ such
that

\medskip

\noindent {\bf i.} It is closed,  $\rmd \Phi = 0$,

\noindent {\bf ii.} It is non degenerate: for every  $X\in T\cM$, there
exists $Y\in T\cM$ such that $\Phi(X,Y)\neq 0$.

\medskip

The couple $(\cM,\Phi)$ is  a {\it symplectic manifold\index{symplectic
manifold}}, and $\cM$ has always even dimension. In any symplectic
manifold, we have local {\it Darboux coordinates\index{Darboux
coordinates}} defined by the following theorem:

\begin{theorem}(Darboux) If $\cM$ is a symplectic manifold, $\dim\cM=2n$, for each $m\in \cM$
there is a chart $(U, \varphi:U\rightarrow \R)$ such that $\varphi(m)=0$
and for $u\in U$,\\
 $\varphi(u)=\bigl(x^1(u),\dots x^n(u),y_1(u),\dots
y_n(u)\bigr)$ and $\Phi$ on the open set  $U$ is
$$\Phi\bigr|_U=\sum_{i=1}^n\rmd x^i\wedge \rmd y_i.$$\hfill$\square$
\end{theorem}

Notice that a {\it quasi-K{\"a}hler} manifold is a {\it symplectic
manifold}, since the fundamental 2-form is non degenerate. If, in
addition, the manifold is complex then it is a {\it K{\"a}hler
manifold\index{K{\"a}hler manifold}}. Moreover, let $\DLC$ be a Riemannian
connection, so $\DLC_X g=0$  for every vector field $X$ on $\cM$. We
have that
\begin{equation}
  \DLC_X\Phi (Y,Y')=\DLC_Xg(Y, JY')+g(Y, (\DLC_XJ)Y')=0,
 \label{DLCPhi0}
\end{equation}
which means that the Riemannian connection is trivially a {\it symplectic
connection}.

The holonomy of a K{\"a}hler manifold\index{K{\"a}hler manifold!holonomy} of
complex dimension $n$  is a subgroup of $\rU(n)\simeq\rO(2n)\cap
\rGL(n,\C)$, since the Riemannian connection is quasicomplex. Here
$\rGL(n,\C)$ is taken in its real representation
$$A+\rmi B\longrightarrow
\begin{pmatrix} A&B\\-B&A\end{pmatrix}.$$ One can prove that if
the manifold is Ricci flat then the {\it restricted holonomy
group\index{restricted holonomy group}} (that is, considering only
parallel displacements along paths that are homotopic to a point) is
contained in $\rSU(n)$.

\paragraph{K{\"a}hler manifolds in coordinates.}

On a quasicomplex manifold, we can consider the principal bundle of
unitary frames\index{principal bundle}, that is the bundle of complex
frames that are orthonormal with respect to the Hermitian metric. Its
structural group is $\rU(n)$. We will denote this bundle by $\rU(\cM)$.
We want to give the metric, connection and curvature of a K{\"a}hler manifold
in coordinates.

\bigskip

Let $\cM$ be a complex manifold, with Hermitian metric $g$ and
complex structure $J$. We use the notation of Remarks
\ref{complexmanifold}, and \ref{complexmanifold2} and denote
$$Z_i=\frac{\partial}{\partial z^i}, \qquad \bar Z_i=Z_{\ib }=\frac{\partial}{\partial
\bar z^{i}}.$$ It is easy to see that
\begin{equation}
g_{ij}=g(Z_i,Z_j)=0, \qquad g_{\ib \jb }=g(\bar Z_i,\bar Z_j)=0,
\label{metricomp}
\end{equation}
and the only non zero components are of the form $g_{i\jb }=g( Z_i,\bar
Z_j)=g_{\jb  i}$, so\footnote{Note the conventions with factors for
symmetric products and for forms. A symmetric product of forms
$\alpha\beta $ is $\ft12(\alpha \otimes \beta +\beta \otimes \alpha )$.
Similarly a wedge product is taken as $\alpha \wedge \beta =\ft12(\alpha
\otimes \beta -\beta \otimes \alpha )$.}
$$g=g_{i\jb }(\rmd z^i\otimes \rmd\bar z^\jb+\rmd\bar z^\jb\otimes \rmd z^i)=2g_{i\jb }\rmd z^i \rmd\bar z^\jb .$$
Since $J(Z_i)=\rmi Z_i$ and $J(\bar Z_{\ib })=-\rmi \bar Z_{\ib }$, the
fundamental 2-form is
\begin{equation}
\Phi=-2\rmi g_{i\jb }\rmd z^i\wedge \rmd\bar z^\jb .
\label{fundamentalform}
\end{equation}
 If $\cM$ is a K{\"a}hler manifold, the fundamental form  is closed, so
$$\rmd\Phi=-2\rmi \bigl(\frac{g_{i\jb }}{\partial z^k}\rmd z^k\wedge \rmd z^i\wedge \rmd\bar z^\jb + \frac{g_{i\jb }}{\partial \bar z^{\bar k} }\rmd\bar z^{\bar k} \wedge \rmd z^i\wedge
\rmd\bar z^\jb \bigr)=0,$$ which implies
\begin{equation}\frac{g_{i\jb }}{\partial z^k}-\frac{g_{k\jb }}{\partial z^i}=0, \qquad \frac{g_{i\jb }}{\partial \bar z^{\bar k} }-\frac{g_{i\bar k}}{\partial \bar z^\jb
}=0.
\label{kahler}
\end{equation}
These equations are the integrability
condition for the existence of a real valued function
 $\cK$ such that
$$g_{i\jb }=\frac{\partial \cK}{\partial z^i\bar z^\jb }.$$
For any real function $\cK$, the tensor $g_{i\jb }$ satisfies $ (g_{i\jb
})^*=g_{j\ib }$, and property (\ref{kahler}). If it is positive
definite, then it is a K{\"a}hler metric on $\cM$. So any K{\"a}hler metric can
be written locally in this way. Notice also that $\cK$ is defined modulo
a holomorphic function $f$,
$$\cK\rightarrow \cK+f(z)+f(\bar z).$$
$\cK$ is the {\it K{\"a}hler potential\index{K{\"a}hler potential}}

\medskip

We will denote by $I$ an arbitrary index in $\{1,2,\dots 2n\}$ and by
$\{x^I\}$ arbitrary coordinates in $\cM$. Let
$Y=Y^I\frac{\partial}{\partial x^I}=Y^I\partial_I $ be  a vector field on
$\cM$. The covariant derivative of $Y$ with respect to a linear
connection can be written as
$$(\DLC_JY)^I=\frac{\partial Y^I}{\partial
x^J}+\Gamma_{JK}^IY^K,$$ where the {\it Christoffel symbols}
$\Gamma$ are
$$\Gamma_{JK}^I=(\DLC_J\partial_K)^I.$$
The Levi-Civita connection\index{Levi-Civita connection} is the only
torsionfree connection satisfying $\DLC g=0$. The Christoffel
symbols\index{Christoffel symbols} are
\begin{equation}\Gamma_{JI}^K=\frac 12g^{KL} \left( \frac{\partial
g_{LI}}{\partial x^J}+\frac{\partial g_{JL}}{\partial x^I}-
\frac{\partial g_{JI}}{\partial
x^L}\right).\label{christoffel}\end{equation} For a complex manifold, we
can extend the covariant differentiation to complex vector fields by
linearity. We can then consider $I=i $ for $I=1,\dots n$ and $I=n+\ib $
for $I=n+1, \dots 2n$. The Christoffel symbols become complex, and it is
easy to see that
$$\bar \Gamma^I_{JK}=\Gamma^{\bar I}_{\bar J\bar K},$$
where we have denoted $\bar I= I+ n$ for $I=i$ and $\bar I= I- n$ for
$I=n+i$.

A linear connection is  quasicomplex if the complex structure is
parallel. For $J=\rmi \rmd z^j\otimes \partial_j-\rmi \rmd\bar z^\jb
\otimes
\partial_{\jb }$ this means
$$(\DLC_AJ)^B_C=\Gamma_{AD}^BJ_C^D-\Gamma_{AC}^DJ_D^B=0\;\Rightarrow\;\Gamma_{A\bar
c}^d= \Gamma_{Ac}^{\bar d}=0.$$ If the connection is  torsionfree we
have,
$$T(X,Y)=\DLC_X
Y-\DLC_YX-[X,Y]=0\;\Rightarrow\;\Gamma^A_{BC}=\Gamma^A_{CB},$$ so the
only non zero Christoffel symbols of the quasicomplex connection are
\begin{equation}
  \Gamma^i_{jk}=\Gamma^i_{kj}, \qquad \Gamma^{\ib }_{\jb \bar
k}=\Gamma^{\ib }_{\bar k\jb }.
 \label{Gammanon0}
\end{equation}

If the Riemannian connection is quasicomplex then the manifold is
a K{\"a}hler manifold, and  we have from (\ref{christoffel}) and
(\ref{kahler})
\begin{equation}\Gamma^i_{jk}=g^{i\bar \ell }\partial_jg_{\bar \ell  k},
\qquad \Gamma^{\ib }_{\jb \bar k}=g^{\ib  \ell }\partial_{\jb }g_{\ell
\bar k}. \label{criskal}
\end{equation}

The  curvature tensor associated to a linear connection is a
3-contravariant 1-covariant tensor given by
$$R(X,Y)Z=[\DLC_X,\DLC_Y]Z-\DLC_{[X,Y]}Z, \qquad X,Y,Z\in
T\cM,$$ and in components
$$R^I{}_{JKL}=\left(\frac {\partial \Gamma^I_{LJ}}{\partial x^K}-
\frac {\partial \Gamma^I_{KJ}}{\partial
x^L}\right)+\sum_M\bigl(\Gamma^M_{LJ}\Gamma^I_{KM}-\Gamma^M_{KJ}\Gamma^I_{LM}\bigr).$$
It satisfies $R^I{}_{JKL}=-R^I{}_{JLK}$.

If the connection is torsionfree, the curvature tensor satisfies the {\it
Bianchi identities\index{Bianchi identities}}
\begin{eqnarray}
 &&R(X,Y)Z +R(Z,X)Y+R(Y,Z)X=0,\label{BI1}\\
 &&\DLC_X R(Y,Z)+\DLC_Z R(X,Y)+\DLC_Y R(Z,X)=0.\label{BI2}
\end{eqnarray}
(If the torsion is not zero, then the Bianchi identities are modified by
terms containing the torsion, see \cite{Kobayashi1996}, volume I page
135.)

% The curvature tensor of a  quasicomplex, torsionfree connection has
% the following independent components:
%$$R^{ i}{}_{jk \ell  }, \quad R^{ i}{}_{jk \bar \ell  }, \quad  R^{ i}{}_{ \jb k \ell  },
%\quad R^{ i}{}_{\jb \bar k \bar \ell  },\quad R^{ i}{}_{j \bar k \bar
%\ell  },\quad R^{ i}{}_{\jb   k \bar \ell  }.$$ The rest can be
%obtained by complex conjugation or using  the symmetry property
%$R^I{}_{JKL}=-R^I{}_{JLK}.$ (One could also use the first Bianchi
%identity to reduce the number of independent components, but we prefer
%not to do it, for clarity).

 It is immediate to
see that a for a quasicomplex connection
$$R^{ i}{}_{ \jb  k \ell}=0, \quad R^{ i}{}_{\jb \bar k \bar \ell}=0,\quad R^{ i}{}_{j \bar k \bar
\ell}=0,\quad R^{ i}{}_{\jb   k \bar \ell}=0.$$
  From (\ref{criskal}) one finds that for a K{\"a}hler metric
$$R^{ i}{}_{jk \ell}=0,$$
and the only components that can be different from zero are
\begin{eqnarray}R^i{}_{jk\bar \ell}=-\partial_{\bar
\ell}\Gamma^i_{kj},\qquad R^i{}_{j\bar k \ell}=\partial_{\bar
k}\Gamma^i_{j\ell}, \nonumber\\
R^{\ib }{}_{\jb  \bar k \ell}=-\partial_{\ell}\Gamma^{\ib }_{\bar k\jb
},\qquad R^{\ib }{}_{\jb  k \bar \ell}=\partial_{k}\Gamma^{\ib }_{\jb
\bar \ell},
 \label{curvkm}
\end{eqnarray}
and those obtained using  the symmetry property
$R^I{}_{JKL}=-R^I{}_{JLK}$. The upper and lower line are related by
complex conjugation.

The Ricci tensor\index{Ricci tensor} is the contraction
$R_{AB}=R^C{}_{ACB}$. We have that
$$R_{ij}=R_{\ib  \jb }=0, \qquad R_{i\jb }=\bar R_{\ib
j}=-\partial_{\jb }\Gamma^k_{ik}=-\partial_{\jb }(g^{k\bar
\ell}\partial_ig_{k\bar \ell}).$$ Let $G=\det g_{i\jb }$, then
\begin{equation}\partial_i G=Gg^{k\bar \ell}\partial_ig_{k\bar \ell}\;\Rightarrow\;
R_{i\jb }=- \partial_i\partial_{\jb }\log |G|.\label{ricci}\end{equation}

\hfill$\blacksquare$

\begin{example} The complex projective space\index{complex projective space} $\CP^1$
\end{example} We consider the complex projective space of
1-dimensional subspaces in $\C^2$. Let $z^1, z^2$ be the natural
coordinate system in $\C^2$, $z^i:\C^2\rightarrow \C$. They are complex
linear maps. Let $U_1$ be the set of subspaces $S$ such that $z^1|_S\neq
0$. Then $z^1|_S$ spans the dual space to $S$, so we may write
 \begin{equation}z^{2}|_S=t^1z^1|_S, \qquad t^1\in
 \C.\label{p1}\end{equation}
 Each equation as (\ref{p1}) defines a subspace in $U_1$, so
 $t^1$ is a   complex coordinate in $U_1$. In the same way we can
 define $U_2$ as the set of subspaces $S$ such that $z^2|_S\neq
 0$. Then we have that
 $$z^{1}|_S=t^2z^2|_S, \qquad t^2\in
 \C$$
 and $t^2$ is a complex coordinate in $U_2$. $\{(U_1, t^1), (U_2, t^2)\}$
  is a complex atlas of $\CP^1$. In the intersection
 $U_1\cap U_2$ the gluing condition is
 $$t^2=\frac 1{t^1}.$$

We want to define a K{\"a}hler metric on $\CP^1$. On $U_1$ and $U_2$ we
consider, respectively, the following real-valued functions:
$$f_1=(t^1\bar t^1 +1), \qquad f_2=(t^2\bar t^2 +1).$$ It is
easy to see that the two 2-forms defined by $$\Phi_1=4\rmi \partial \bar
\partial \ln f_1=-4\rmi \frac 1{f_1^2} \rmd t^1\wedge \rmd\bar t^1, \qquad
\Phi_2=4\rmi\partial \bar
\partial \ln f_2=-4\rmi\frac 1{f_2^2} \rmd t^2\wedge \rmd\bar t^2 $$
coincide in the intersection, so they define globally a closed
2-form $\Phi$. The K{\"a}hler metric is then
$$g(X,Y)=\Phi(JX,Y).$$ One can see that it is
positive definite by computing it in an open set:
$$\rmd s^2=\frac 4{(1+t^1\bar t^1)^2}\rmd t^1\rmd\bar t^1.$$\hfill$\square$

\subsection{Hermitian line bundles and fiber metrics}
 \label{ss:hermLineBundles}

Hermitian fiber metrics\index{Hermitian fiber metric} are introduced here
and will be used later, in Section~\ref{ss:projective}. The definitions
and statements in this section can be found in Refs.
\cite{Kobayashi1996,Kobayashi1987}.

 Let $E\rightarrow \cM$ be a rank $k$ complex
vector bundle over the complex manifold $\cM$. Then the fiber at $m\in
\cM$, $E_m$, is a complex vector space of complex dimension $k$.

Let us assume that the total space $E$ has a complex structure, that the
projection $\pi:E\rightarrow \cM$ is a {\it holomorphic
map\index{holomorphic map}}\footnote{That is, a map preserving the
complex structures.} between complex manifolds and that there is a local
trivialization $\{\cU_A\}_{A\in I}$ such that the maps
$$\pi^{-1}(\cU_A)\rightarrow \cU_A\times \C^k$$
and their inverses are holomorphic with $\pi^{-1}(m)=E_m\approx\C^k$.
Then we say that $E$ is a {\it holomorphic vector
bundle\index{holomorphic vector bundle}} over $\cM$.

\begin{example} The tangent bundle of a complex manifold\index{complex manifold!tangent bundle}.
\end{example}  Let $\cM$ be a complex manifold.  Let $(x^j,
y^j)$, $j=1,\dots n$ be coordinates on a neighbourhood $\cU$ of $m\in
\cM$ such that the complex structure on $\cM$ is given by
\begin{equation}J\left(\frac{\partial}{\partial
x^j}\right)=\frac{\partial}{\partial y^j}, \qquad
J\left(\frac{\partial}{\partial y^j}\right)=-\frac{\partial}{\partial
x^j}, \qquad j=1,\dots n.\label{holtanbundle}
\end{equation}
A vector on $m$ is of the form
$$V_m=X^j_m\frac{\partial}{\partial x^j}+Y^j_m\frac{\partial}{\partial
y^j}.$$ The components $X^j_m, Y^j_m$ are coordinates on $T_m(\cM)$. On
$\pi^{-1}(\cU)$ we have coordinates $(x^j,y^j,X^j,Y^j)$, and a
quasicomplex structure on $T\cM$ is given by (\ref{holtanbundle}) and
$$J\left(\frac{\partial}{\partial X^j}\right)=\frac{\partial}{\partial Y^j},
\quad J\left(\frac{\partial}{\partial
Y^j}\right)=-\frac{\partial}{\partial X^j}, \qquad j=1,\dots n.$$
 The quasicomplex structure is integrable and we have complex coordinates
on $\pi^{-1}(\cU)$:
$$(z^j=x^j+\rmi y^j,\;\; Z^j=X^j+\rmi Y^j).$$
 $T\cM$ is a complex manifold and a holomorphic vector
bundle over $\cM$.\hfill$\square$

\bigskip

 A {\it  fiber metric\index{fiber metric}} on a
vector bundle $E\rightarrow
 \cM$ is a smooth assignment of an inner product on each fiber $$h_m:E_m\times
 E_m\longrightarrow \R.$$
 If the fiber has a complex structure $J_m$ we require that the inner product is Hermitian,
 $$ h(JX,JY)=h(X,Y)\quad \forall X,Y\in E_m,$$ and we
 say that $E$ is a {\it Hermitian vector bundle}\index{Hermitian vector bundle}.

 A connection $\nabla$ on $E$ is {\it metric\index{metric connection}} if
 $\nabla h=0$ (the connection is extended to $E^*\otimes
 E^*$). If the bundle is holomorphic, we can ask the covariant
 derivative of a holomorphic section to be holomorphic,
 \begin{equation}
 \nabla s\in \Omega^{1,0}(E)\quad \hbox{for } s\hbox{
holomorphic}.\label{complexconn}
\end{equation}
There is a unique metric connection satisfying (\ref{complexconn}); it is
the {\it Hermitian connection} of the Hermitian vector bundle (see for
example Ref. \cite{Kobayashi1996}).

Let $\{s_a\}_{a=1,\dots k}$ be a holomorphic frame of the bundle $E$ on
a neighbourhood $\cU$ of $m\in \cM$ (that is, $k$ independent local
sections) and $\{\alpha^a\}$ the dual coframe. The connection 1-form on
$\cU$, for a connection satisfying (\ref{complexconn}), is
$$(\nabla s_b)=\Gamma^a{}_bs_a=\Gamma_i{}^a{}_ b\rmd z^i\otimes  s_a, \qquad i=1,\dots n,\quad a,b=1,\dots
k,$$
 so the covariant derivative of a holomorphic section $s=a^as_a$ is
$$\nabla s=(\partial_i a^a +\Gamma_i{}^a{}_b a^b) \rmd
z^i\otimes s_a.$$ The hermiticity of the fiber metric means
$$h=2h_{a\bar b} \alpha ^a\bar \alpha ^b,$$ and the condition for the
connection to be metric is
\begin{equation}
\partial_i h_{a\bar b}-\Gamma_i{}^c{}_a h_{c\bar b}=0\quad
\Rightarrow\quad \Gamma_i{}^a{}_ b= h^{a\bar c}\partial_i h_{\bar c b},
\label{metricconnection}
\end{equation}
where $h^{a\bar c}$ is the inverse matrix of $h_{a\bar c}$. The curvature
is then
$$R^a{}_{bi\jb  }=-h^{a\bar c}\partial_i\partial _{\jb }h_{b\bar c}+h^{a\bar c}h^{d\bar e}\partial_{i}h_{b\bar e}\partial_{\jb }h_{d\bar c}.$$ We can define the {\it Ricci form} of the Hermitian
bundle as the trace of the curvature tensor,
$$\rho=-2\rmi R^a{}_{ai\jb  }\rmd z^i\wedge\rmd \bar z^{j}.$$

If $E$ is a Hermitian {\it line bundle}, that is, it has rank 1, then the
metric is just
$$h= \theta(z,\bar z)\alpha\bar\alpha,$$ and the Ricci form becomes
\begin{equation}
\rho=-2\rmi \bar \partial\partial \log |\theta|=-2\rmi
\partial_i\partial_{\jb }\log|\theta| \rmd z^i\wedge \rmd \bar z^\jb . \label{ricciform}
\end{equation}

\section{Rigid special K{\"a}hler manifolds\index{rigid special K{\"a}hler manifold|(} \label{ss:rigid}}

In this section we will deal with rigid special K{\"a}hler geometry, or
simply special K{\"a}hler geometry, as opposed to projective special K{\"a}hler
geometry, which will be the subject of Section \ref{ss:psk}.

\subsection{Some geometric preliminaries}\label{geometricpreliminaries}

This part is inspired in the second chapter of Ref. \cite{Donaldson1990}.

Let $E$ be a vector bundle over $\cM$ with a connection $\nabla$. For
every vector field $X$ of $\cM$ (section of $T\cM$), $\nabla$ sends
sections of $E$ to sections of $E$,
$$\nabla_X:\Gamma(E)\longrightarrow \Gamma(E).$$
Let $\Omega_\cM^p(E)=\Lambda^p(\cM)\otimes \Gamma(E)$ be the space of
\textit{$E$-valued $p$-forms on $\cM$}, with $\Omega_\cM^0(E)=\Gamma(E)$.
We are going to define the {\it covariant differential\index{covariant
differential}} $\rmd_\nabla: \Omega_\cM^p(E)\rightarrow
\Omega_\cM^{p+1}(E)$.
 For 0-forms we define
$$\begin{CD}\Omega_\cM^0(E)@>\rmd_\nabla>> \Omega_\cM^1(E)\\
F@>>>\rmd_\nabla F\end{CD} \qquad \hbox{such that}\; \rmd_\nabla
F(X)=\nabla_X F, \quad X\in T\cM.$$ This definition can be extended to
$\Omega_\cM^p(E)$,
$$\begin{CD}\Omega_\cM^p(E)@>\rmd_\nabla>> \Omega_\cM^{p+1}(E)\\
F@>>>\rmd_\nabla F\end{CD}$$ assuming the condition
$$\rmd_\nabla(\alpha\wedge F)= \rmd\alpha\wedge F +
(-1)^p\alpha\wedge \rmd_\nabla F,$$ for $\alpha \in \Lambda^p(\cM)$ and
$F\in \Omega_\cM^q(E)$. For example, if $F\in \Omega_\cM^1(E)$, locally
$F=\rmd x^i\otimes \alpha_i=\rmd x^i\wedge \alpha_i$ with $\alpha_i\in
\Omega_\cM^0(E)$, so
$$\rmd_\nabla F (X,Y)=-\rmd x^i\wedge
\rmd_\nabla\alpha_i(X,Y)=-\left(
X^i\nabla_Y\alpha_i-Y^i\nabla_X\alpha_i\right) .$$

 If $A^I_J=A^I_{\mu J} \rmd x^\mu$ is the 1-form connection matrix in an open set
$U\subset \cM$, then
$$(\rmd_\nabla F)^I=\rmd F^I+A^I_{J}\wedge F^J,$$ from which it is
easy to deduce the standard transformation rule under a local fiber
(gauge) transformation $${F'}= UF, \quad (\rmd_\nabla F)'=U\rmd_\nabla
F,\quad \Rightarrow\quad A'=-\rmd UU^{-1}+UAU^{-1},$$
 where $(\rmd_\nabla F)'\equiv \rmd F'+A'F'$.
 Differently than
for the ordinary differential, $\rmd_\nabla^2$ is not zero in general. In
fact,
$$(\rmd^2_\nabla F)^I= (\rmd A^I_K +A^I_J\wedge A^J_K)\wedge F^K=R^I_K\wedge
F^K,$$
where $R$ is the (Lie algebra valued) curvature 2-form associated
to the connection. A flat connection then defines a complex. The de Rham
complex is associated to the trivial connection on the trivial bundle
$E=\cM\times V$.

It is easy to check the Bianchi identity $\rmd_\nabla R=0$. In the
associated bundle with typical fiber  the Lie algebra, the group acts
with the adjoint representation. The covariant differential in such
bundle is then
$$\rmd_\nabla R^I_J=\rmd R^I_J+A^I_K\wedge R^K_J -A^K_J\wedge R^I_K=0.$$

\subsection{Definition of rigid special K{\"a}hler manifolds\index{rigid special K{\"a}hler manifold!definition}}\label{definition}

Here we follow the first section of Ref. \cite{Freed:1997dp}.
\begin{definition} \label{rsdef}Let $\cM$ be a K{\"a}hler manifold with K{\"a}hler
form $\Phi$ and complex structure $J$. A special K{\"a}hler structure on
$\cM$ is a real, flat, torsionfree, symplectic connection $\nabla$
satisfying
\begin{equation}
\rmd_\nabla J=0.
\label{dnablaj}
\end{equation} \hfill$\square$\end{definition}

$J$ is seen here as a 1-form with values in the tangent bundle $T\cM$,
and the covariant differential must be interpreted in the sense described
in Section \ref{geometricpreliminaries}.

As we have seen, a K{\"a}hler manifold is always symplectic, being the K{\"a}hler
form $\Phi$ its symplectic form. On a symplectic manifold, a linear
connection $\nabla$ is said to be symplectic\index{symplectic connection}
if
\begin{equation}
  \nabla \Phi =0.
 \label{NablaPhi0}
\end{equation}

We want to see what is the meaning of the ingredients in this definition.
We first examine the implications of the existence of a flat, torsionfree
connection.

Let $U$ be  an open set, with coordinates $\{x^I\}_{I=1}^{2n}$ and a
(matrix-valued) connection 1-form $A^K_L=A^K_{ML}\rmd x^M$.

  Due to
the flatness condition, $\rmd_\nabla^2=0$, so  a $\rmd_\nabla$-closed
form is locally $\rmd_\nabla$-exact.

Let $\id$ be the identity endomorphism in $T\cM$. It can be seen as a
$T\cM$ valued 1-form, $\id=\partial_I\otimes dx^I$. The torsionfree
condition can be expressed as
\begin{equation}  \rmd_\nabla
\id= (\rmd_\nabla \id)^K\partial_K=A^K_{LM}\rmd x^L\wedge \rmd x^M\otimes
\partial_K=0\;\Rightarrow\; A^K_{LM}=A^K_{ML}.\label{torsion}\end{equation}

A local frame on $T\cM$ on $U\subset_{\mathrm{open}} \cM$ is a set
$\{\xi_\alpha=\xi_\alpha^J\partial_J\}_{\alpha=1}^{2n}$ of $2n$ local
sections of $TU\subset T\cM$ that are linearly independent for each point
$x\in U$. Since the connection is flat (the curvature tensor is zero),
there exists a {\it flat frame\index{flat frame}}, that is,
\begin{equation}\nabla_I\xi_\alpha=0 \quad \mathrm{for}\; I=1, \dots 2n,
\quad \hbox{or equivalently,}\quad
\rmd_\nabla\xi_\alpha=0.\label{flatframe}\end{equation} This is because
the integrability condition of (\ref{flatframe}) is
$$\rmd^2_\nabla\xi_\alpha=R_K^I\wedge\xi^K_\alpha\partial_I=0$$
for the $2n$ independent sections $\xi_\alpha$, which implies necessarily
that $R^I_K=0$.

 Let
$\theta^\alpha=\theta^\alpha_I\rmd x^I$ be the dual coframe, that is
$\theta^\alpha(\xi_\beta)=\delta^\alpha_\beta$. We have that
$$\theta^\alpha_I\xi^I_\beta=\delta_\alpha^\beta,
\quad\Leftrightarrow\quad
\xi^J_\alpha\theta_I^\alpha=\delta^J_I.$$
 Then we can express
$$\id=\theta^\alpha\otimes\xi_\alpha= \theta^\alpha\wedge\xi_\alpha,$$
and $$\rmd_\nabla\id=0\quad\Rightarrow\quad
\rmd_\nabla\theta^\alpha\wedge\xi_\alpha-\theta^\alpha\wedge
\rmd_\nabla\xi_\alpha=0 \quad\Rightarrow\quad
\rmd_\nabla\theta^\alpha=\rmd\theta^\alpha=0.$$

This means that $\theta^\alpha=\rmd t^\alpha$ for some functions
$t^\alpha$. Then $\xi_\alpha=\partial/\partial t_\alpha$ and $t^\alpha$
are local coordinates on $U$ called {\it flat coordinates\index{flat
coordinates}}.

Up to here, we used the fact that the connection is real, flat and
torsionfree. We introduce now the additional condition that the
connection is symplectic, that is, $\nabla_I\Phi=0$. We denote the
symplectic matrix as
\begin{equation}P=\begin{pmatrix}0&\id_{n\times n}\\-\id_{n\times
n}&0\end{pmatrix}.\label{symplectic}\end{equation} The coordinates
$t^\alpha$ are {\it Darboux} coordinates if
$$\Phi(\xi_\alpha,\xi_\beta)=P_{\alpha\beta},\quad
\hbox{so}\quad \Phi= \frac 12P_{\alpha\beta}\rmd t^\alpha\wedge \rmd
t^\beta.$$ It is possible to choose the  flat coordinates $t^\alpha$ in
such a way that they are Darboux. This is because
$$\partial_I(\Phi(\xi_\alpha,\xi_\beta))=\nabla_I(\Phi)(\xi_\alpha,\xi_\beta)-
\Phi(\nabla_I\xi_\alpha,\xi_\beta)-\Phi(\xi_\alpha,\nabla_I\xi_\beta)=0,$$
so $\Phi(\xi_\alpha,\xi_\beta)$ is a constant (antisymmetric, non
degenerate)  matrix which can always be brought to the form
(\ref{symplectic}) by a linear change of coordinates.

We see that the existence of a flat, torsionfree, symplectic connection
on $\cM$ is  equivalent to having a covering by {\it flat Darboux
coordinates} (it is also said that $\cM$ has a flat symplectic
structure). If $\{q^\alpha\}_{\alpha=1}^{2n}$ are also flat Darboux
coordinates, we have that the transition functions satisfy
$$\rmd q^\alpha=\frac{\partial q^\alpha}{\partial t^\beta}\rmd t^\beta,
\quad \nabla_{t^\gamma}\rmd q^\alpha=\frac{\partial q^\alpha}{\partial
t^\beta\partial t^\gamma}\rmd t^\beta=0$$ since
\[\nabla_{t^\gamma}\rmd t^\beta = \nabla_{t^\gamma}\theta ^\beta =0.\]
This implies $$ \frac{\partial q^\alpha}{\partial t^\beta\partial
t^\gamma}=0\Rightarrow q^\alpha= A_\beta^\alpha t^\beta +c^\alpha,
$$ with $A_\beta^\alpha $ and $c^\alpha$  constant. It follows that
$A\in \rSp(2n)$.

\bigskip

Let us now consider the condition (\ref{dnablaj}). In arbitrary
coordinates $\{x^I\}_{I=1}^{2n}$ it becomes
\begin{equation}
  J=J^I\partial_I=J^I_K\partial_I\otimes \rmd x^K,
\qquad (\rmd_\nabla J)^I= \rmd J^I +A_L^I\wedge J^L=0,
 \label{dnablaJ}
\end{equation}
which in components reads
$$\frac 1 2 (\partial_MJ_N^I-\partial_NJ_M^I)+\frac
12(A_{ML}^IJ_N^L-A_{NL}^IJ_M^L)=0.$$ (The factor $1/2$ appears when $M$
and $N$ are arbitrary, so each {\it strict component} is counted twice).
This implies, assuming that the connection is torsionfree, that
\begin{equation}A_{i\jb }^I=A_{\ib
j}^I=0.\label{complexcondition}\end{equation}

\bigskip

The connection $\nabla$ is a linear connection (a connection on the frame
bundle of $\cM$), so one can compute
\begin{equation}
 (\nabla_IJ)^K_L=\partial_IJ^K_L+A_{I M}^KJ_L^M-A_{IL}^MJ_M^K.
 \label{NablaJ}
\end{equation}
The condition $\nabla_IJ=0$ together with the torsionfreeness
implies, in addition to (\ref{complexcondition}) that
$$A^i_{\bar m\bar n}=A^{\ib }_{mn}=0.$$
  If the connection is torsionfree and  $\nabla_IJ=0$ then we have that
$\rmd_\nabla J=0$, but the converse is not necessarily true. Then the
flat symplectic connection is not necessarily complex.

\medskip

The complex structure can be written locally in terms of the
complex coordinates $\{z^j\}_{j=0}^n$ as
$$J=\rmi (\partial_{z^j}\otimes \rmd z^j-\partial_{\bar z^\jb }\otimes \rmd\bar z^\jb )=\rmi (\pi^{(1,0)}-\pi^{(0,1)}),$$ where
\begin{equation}
\pi^{(1,0)}=\partial_{z^j}\otimes \rmd z^j\qquad \hbox{and} \qquad
\pi^{(0,1)}=\partial_{\bar z^\jb }\otimes \rmd\bar z^\jb
 \label{defpi10}
\end{equation}
are the projectors onto the $T\cM ^{(1,0)}$ and $T\cM ^{(0,1)}$ spaces
respectively. The condition $\rmd_\nabla J=0$, together with the
torsionfreeness, is equivalent to $$\rmd_\nabla \pi^{(1,0)}=0.$$ Indeed,
one can also write
\begin{equation}
  \id=\pi^{(1,0)}+\pi^{(0,1)},
 \label{idinpi}
\end{equation}
and the torsionfree condition was expressed as $\rmd_\nabla \id=0$.

Using the Poincar{\'e} lemma, $\rmd_\nabla \pi^{(1,0)}=0$ implies that
locally there exists a complex vector field $\chi$ such that
$$\nabla\chi=\rmd_\nabla\chi=\pi^{(1,0)},$$
which is unique up to a flat complex vector field.

Let $\{x^j,y_j\}_{j=1}^n$ be a flat Darboux coordinate system, that is,
\begin{equation}
  \Phi=\rmd x^j\wedge \rmd y_j,\quad \mathrm{and}\qquad \rmd_\nabla\left(\frac{\partial}{\partial x^j}\right)=0,
\quad \rmd_\nabla\left(\frac{\partial}{\partial y_j}\right)=0.
 \label{PhiDarboux}
\end{equation}
 In this coordinate system we denote
\begin{equation}
\chi=\frac 12\left(\eta^i\frac{\partial}{\partial
x^i}-\lambda_j\frac{\partial}{\partial y_j}\right), \label{sympsec}
\end{equation}
where $\eta^j, \lambda_j$ are complex functions ($\chi$ is a complex
vector field). Taking the covariant differential and using
(\ref{PhiDarboux}) we obtain
\begin{equation}
 \pi^{(1,0)}=\rmd_\nabla\chi=\frac 12\left(\rmd\eta^j\otimes\frac{\partial}{\partial x^j}
 -\rmd\lambda_j\otimes \frac{\partial}{\partial
y_j}\right).
 \label{pi10Darboux}
\end{equation}
$\pi^{(1,0)}$ is a (1,0) tensor, so it follows that $\eta^j$ and
$\lambda_j$ are holomorphic functions. Taking the real part of this
equation we have, using (\ref{idinpi}),
\begin{equation}
 \Re(\pi^{(1,0)})=\frac{1}{2}\id =\frac{1}{2}\left(
 \rmd x^j\otimes\frac{\partial}{\partial x^j}+\rmd y_j\otimes \frac{\partial}{\partial
y_j}\right),
 \label{Repi10}
\end{equation}
so we can identify
\begin{equation}\Re (\rmd\eta^j)=\rmd x^j, \qquad \Re
(\rmd\lambda_j)=-\rmd y_j.\label{realpart}
\end{equation}
Together with
the condition that $\eta^j$ and $\lambda_j$ are holomorphic, we have that
\begin{equation}
\rmd\eta^j=\rmd x^j-\rmi J^*\rmd x^j,\qquad \rmd\lambda_j=-\rmd y_j+\rmi
J^*\rmd y_j.\label{differentials}
\end{equation}

We want to see under what conditions the sets $\{\eta^j\}$ and
$\{\lambda_j\}$ are sets of complex coordinates. Let $\{z^1,...,z^n\}$ be
complex coordinates with
\begin{eqnarray}&&z^l=\xi^l+\rmi \omega^l, \qquad
 \rmd z^l=\rmd\xi^l+\rmi\rmd\omega^l,\qquad
 \frac{\partial}{\partial z^l}=\frac{1}{2}\left(\frac{\partial}{\partial
\xi^l}-\rmi\frac{\partial}{\partial \omega^l}\right),\nonumber\\&&
\frac{\partial}{\partial \xi^l}=\frac{\partial}{\partial
z^l}+\frac{\partial}{\partial \bar{z}^l},\qquad \frac{\partial}{\partial
\omega^l}=\rmi\left(\frac{\partial}{\partial
z^l}-\frac{\partial}{\partial
\bar{z}^l}\right).\label{jacobian}\end{eqnarray} We have
\begin{equation}
\rmd\eta^j=\alpha^j_{\phantom{j}l}(z)\rmd z^l,\qquad
\rmd\lambda_j=\beta_{jl}(z)\rmd z^l.\label{alphabeta}
\end{equation}
$\pi^{(1,0)}$ is the projector on the space of holomorphic vectors. This
means that it kills all the antiholomorphic vectors and its image is the
set of all the holomorphic vectors. Using (\ref{pi10Darboux}),
(\ref{jacobian}) and (\ref{alphabeta}) we have
\begin{eqnarray}
\pi^{(1,0)}\left(\frac{\partial}{\partial \xi^j}\right)=\frac12\left(
\alpha^l_{\phantom{j}j}\frac{\partial}{\partial
x^l}-\beta_{lj}\frac{\partial}{\partial y_l}\right),\label{basis1}\\
\pi^{(1,0)}\left(\frac{\partial}{\partial
\omega^j}\right)=\frac{\rmi}{2}\left(
\alpha^l_{\phantom{j}j}\frac{\partial}{\partial
x^l}-\beta_{lj}\frac{\partial}{\partial y_l}\right). \label{basis2}
\end{eqnarray}
The set  $\{\partial/\partial\xi^l,
\partial/\partial\omega^l\}$ forms a basis of the tangent space, so the vectors in (\ref{basis1})
and (\ref{basis2}) are linearly independent. This implies that the
$2n\times n$ matrix
\begin{equation}
  (\alpha^T,-\beta^T)
 \label{2nnmatrix}
\end{equation}
has rank $n$ (the superindex $T$ indicates the usual transpose). From the
set of $2n$ functions $(\eta^j, \lambda_j)$ we can always select $n$
independent holomorphic functions that form a set of holomorphic
coordinates.

{}From the fact that the symplectic form $\Phi=\rmd x^i\wedge \rmd y_j$
is of type $(1,1)$ and using (\ref{realpart}) and (\ref{alphabeta}) in
(\ref{PhiDarboux}), and compare with (\ref{fundamentalform}), we obtain
as conditions for $\alpha$ and $\beta$
\begin{eqnarray}
 &&-\alpha^T\beta +\beta^T\alpha=0,\nonumber\\
 &&\beta^T\bar\alpha - \alpha^T\bar \beta=8\rmi g,
 \label{alphabetaT}
\end{eqnarray}
where $g$ is the $n\times n$ matrix $g_{i\jb }$.

These are the equations that we can obtain in general where we have not
used any information on the signature of the metric.

\subsection{The signature of the metric}
\label{ss:signatureSpecial}

Let us first assume that the metric is positive definite (Riemannian
metric). We want to show that $\alpha$ itself has rank $n$. Suppose that
 $\mathrm{rank}(\alpha)<n$. Then, there exists a holomorphic  vector $c$ such that
$$\alpha^j_{\phantom{j}l}(z)c^l(z)=0.$$ But then
$\beta_{jl}c^l(z)\neq 0$, since otherwise the total rank of the matrix
(\ref{2nnmatrix}) would be lower than $n$. This means that there exists a
non zero, holomorphic linear combination of the vectors
$\{{\partial}/{\partial y^k}\}$, namely
$$
\gamma=\sum \tilde{c}_k(z)\frac{\partial}{\partial y_k}\neq 0, \qquad
\tilde{c}_j(z)=\beta_{j k}c^k.$$ Then, as $\gamma $ has only
$y$-components,
$$
0=\Phi(\gamma,\bar\gamma)=g(\gamma,J\bar\gamma)=g(\gamma,-\rmi
\bar\gamma)=-\rmi g(\gamma,\bar\gamma).
$$
For a K{\"a}hler manifold (with positive definite metric)
\begin{equation}
g(\gamma,\bar\gamma)=0\;\Leftrightarrow\;\gamma=0, \label{property}
\end{equation}
so we have a contradiction and $\alpha$ must have rank $n$. Looking now
to (\ref{alphabeta}) we can conclude that $\{\eta_j\}$ is a set of
holomorphic coordinates.

In the same way we can prove that $\beta$ has rank $n$ so $\{\lambda_j\}$
is also a set of holomorphic coordinates.

\begin{remark}Symplectic transformations (Riemannian case).\label{rem:sympltr}\end{remark}

One can see independently that a real symplectic transformation cannot
change the rank of $\alpha$ and $\beta$, provided they satisfy the
following conditions:

\smallskip

\noindent 1. $\,\alpha$ and $\beta$ have rank $n$.

\noindent 2. $\,-\alpha^T\beta +\beta^T\alpha=0$,

\noindent 3. $\,\beta^T\bar\alpha - \alpha^T\bar \beta=8\rmi g$, where
$g$ is the $n\times n$ matrix $g_{i\jb }$. \vspace{2mm}

In our case, conditions 2. and 3. were obtained in (\ref{alphabetaT}).

Let us first introduce the vielbein for the metric
$$\begin{pmatrix}0&g\\g^T&0\end{pmatrix}=E\begin{pmatrix}0&\id\\\id&0\end{pmatrix}E^T,\quad E=\begin{pmatrix}e&0\\ 0&\bar e\end{pmatrix},\quad
g=e\bar e^T.$$
We can define
$$\alpha=2\alpha'e^T, \qquad \beta=2\beta'e^T,$$
so we can express 2. and 3. as

\smallskip

\noindent 2. $\,-{\alpha'}^T\beta' +{\beta'}^T\alpha'=0$,

\noindent 3. $\,{\beta'}^T\bar\alpha' - {\alpha'}^T\bar
\beta'=2\rmi\id$.\vspace{2mm}

Let us denote $\alpha'=\alpha'_0+\rmi\alpha'_1$,
$\beta'=\beta'_0+\rmi\beta'_1$ with $\alpha_i'$ and $\beta_i'$ real. We
define the matrix
$$S_0=\begin{pmatrix}\alpha'_0&\alpha'_1\\\beta'_0&\beta'_1\end{pmatrix}.$$
Then properties 2. and 3. are equivalent to
\begin{equation}S_0^T\begin{pmatrix}0&\id\\-\id & 0\end{pmatrix} S_0=\begin{pmatrix}0&\id\\-\id & 0\end{pmatrix}.\label{symplecticmatrix}\end{equation}
(\ref{symplecticmatrix}) means that $S_0$ is a symplectic matrix, $S_0\in
\rSp(2n,\R)$.

We are ready now to prove the statement above. We have that
 $$S_0\begin{pmatrix}\id\\ \rmi\id\end{pmatrix}
 =\begin{pmatrix}\alpha'\\\beta'\end{pmatrix}.$$
We assume that $\rank(\alpha')=\rank(\beta')=n$ and that
(\ref{symplecticmatrix}) holds. We want to prove that
 $$\hbox{for}\; \hat S=\begin{pmatrix}\hat A&\hat B\\\hat C&\hat D
\end{pmatrix}\in \rSp(2n, \R), \; \hbox{the matrix}\;\begin{pmatrix}\hat \alpha\\\hat\beta
\end{pmatrix}= \hat S\begin{pmatrix}\alpha'\\\beta'
\end{pmatrix}$$
is such that $\rank(\hat\alpha)=\rank(\hat\beta)=n$. Let us write
$$\hat S\begin{pmatrix}\alpha'\\\beta'
\end{pmatrix}=\hat S\,S_0\, S_0^{-1}\begin{pmatrix}\alpha'\\\beta'
\end{pmatrix}=S\begin{pmatrix}\id\\\rmi\id
\end{pmatrix}, \quad \hbox{with}\quad S=\hat S\,S_0.$$
$S$ is an arbitrary matrix in $\rSp(2n,\R)$, so all we have to prove is
that
$$S\begin{pmatrix}\id\\ \rmi\id
\end{pmatrix}=\begin{pmatrix}A&B\\ C&D
\end{pmatrix}\begin{pmatrix}\id\\ \rmi\id
\end{pmatrix}=\begin{pmatrix}A+\rmi B\\C+\rmi D
\end{pmatrix}$$
is such that $\rank(A+\rmi B)=\rank(C+\rmi D)=n$. We consider the
matrices
$$M=\rmi(A+\rmi B),\qquad N=(C+\rmi D).$$
We have that $$\cA\equiv M^\dag N=\id -\rmi \cA_H,\qquad \cA_H=
A^TC+B^TD,$$ since $S$ is a symplectic matrix,
 $$A^TC=C^TA,\quad B^TD=D^TB,\quad A^TD-C^TB=\id.$$
The matrix $\cA_H$ is therefore also symmetric, and can be diagonalized
such that $\cA$ is diagonalized with eigenvalues of the form $(1+\rmi
a)\neq 0$. The determinant of $\cA$ is the product of its eigenvalues, so
it is different from zero. This implies that $\det M\neq 0$, $\det N\neq
0$, so our statement is proven.
 \hfill$\square$

\bigskip

\begin{remark} Symplectic transformations (pseudo-Riemannian case).\label{rem:sympltrps}\end{remark}
If $g$ has pseudo-Riemannian signature, there are symplectic
transformation changing the rank of $\alpha$ and $\beta$ satisfying 1 to
3 in Remark \ref{rem:sympltr}. It is enough to give one of such
symplectic matrices. First we realize that, as before,  conditions 2 and
3 can be put as
\smallskip

\noindent 2. $\,-{\alpha'}^T\beta' +{\beta'}^T\alpha'=0$,

\noindent 3. $\,{\beta'}^T\bar\alpha' - {\alpha'}^T\bar
\beta'=2\rmi\eta$,\vspace{2mm}

\noindent where $\eta$ is the flat pseudo-Riemannian metric. For
definiteness, let us assume that the signature of $\eta$ is $(n-1, 1)$
(the other cases can be obtained in the same way). We take $\eta$ in the
standard form
$$\eta=\begin{pmatrix}
1&0&0&0&\cdots\\
0&1&0&0&\cdots\\
\cdots\\
0&0&\cdots&0&1\\
0&0&\cdots&1&0\end{pmatrix}
$$
 $\eta^2=\id$ and the vielbein is defined
accordingly. We have that the matrix
$$S_0=\begin{pmatrix}\alpha'_0&\alpha'_1\eta\\\beta'_0&\beta'_1\eta\end{pmatrix}$$
is a symplectic matrix, condition that is equivalent to 2 and 3. Also, we
have that
$$S_0\begin{pmatrix}\id\cr\rmi\eta\end{pmatrix}=\begin{pmatrix}\alpha'\cr\beta'\end{pmatrix},$$
so we can bring $\begin{pmatrix}\alpha'\cr\beta'\end{pmatrix}$ to the
standard form $\begin{pmatrix}\id\cr\rmi\eta\end{pmatrix}$ with the
symplectic transformation $S_0^{-1}$. It is enough to consider the $n=2$
case. The symplectic matrix
$$S=\begin{pmatrix}1&0&0&0\\0&0&0&-1\\0&0&1&0\\0&1&0&0\end{pmatrix}$$
has the property
$$S\begin{pmatrix}\id\cr\rmi\eta\end{pmatrix}=S\begin{pmatrix}1&0\\0&1\\0&\rmi\\\rmi&0\end{pmatrix}=
\begin{pmatrix}1&0\\-\rmi&0\\0&\rmi\\0&1\end{pmatrix}=\begin{pmatrix}\alpha''\\\beta''\end{pmatrix},$$
with the property that $\det\alpha'=\det\beta'=0$, as we wanted to show.
The proof in the remark \ref{rem:sympltr} is not valid here because the
real part of $\cA$, determined by $(A+\rmi B\eta )$ and $(C+\rmi D\eta
)$, would be zero.

 \hfill$\square$

\bigskip

We consider now a pseudo-Riemannian metric (pseudo-K{\"a}hler manifold).
Notice that in this case (\ref{property}) is not true since we can have
null vectors. In fact, assume that we have a holomorphic vector field
$\gamma$ such that $\Phi(\gamma,\bar\gamma)=0$ and let us consider the
vectors
$$N_\pm=\gamma\pm \bar\gamma.$$ Since $g$ is of type (1,1) and $\Phi(\gamma,\bar\gamma)=
-\rmi g(\gamma,\bar\gamma)=0$, each one of the terms below is separately
0,
 $$g(N_\pm,
N_\pm)=g(\gamma,\gamma)+g(\bar\gamma,\bar\gamma)\pm
2g(\gamma,\bar\gamma)=0,
 $$
and then $N_+$ and $\rmi N_-$ are null, real  vectors. On the other hand,
if $N$ is a null, real vector, its holomorphic and antiholomorphic
extensions $\gamma=N-\rmi JN$ satisfy $g(\gamma ,\bar \gamma )=0$ and
thus $\Phi(\gamma ,\bar \gamma)=0$. We will treat the case of special
pseudo-K{\"a}hler manifolds in section \ref{sspseudok}.
   \hfill$\blacksquare$

We have thus proven in this section that for a positive definite metric,
the matrices $\alpha $ and $\beta $ are each of rank $n$. When the metric
is not positive definite, this proof breaks down due to null vectors that
may be zero modes of these matrices. However, these matrices might even
then still be invertible (see the example in Section \ref{ss:aspecial}).
In fact, in  Ref. \cite{Craps:1997gp} it is proven that with a symplectic
rotation we can always bring $\alpha$ to be non degenerate. A sketch of
the proof is given in Appendix \ref{ss:sometec}.

\subsection{\label{prepotential}The prepotential\index{prepotential|(}}
We come back to the positive definite metric, or, at least that $\alpha$
and $\beta $ are invertible. Then $\{\eta^j\}_{j=1}^n$ and
$\{\lambda_j\}_{j=1}^n$ are called {\it conjugate} coordinate systems.
Eqs. (\ref{defpi10}) and (\ref{alphabeta}) then imply
$$\pi^{(1,0)}=\rmd\eta^j\otimes \frac\partial{\partial \eta^j},$$
from which, comparing with (\ref{pi10Darboux}),
\begin{equation}
  \frac\partial{\partial \eta^j}=\frac 12\left(\frac{\partial}{\partial x^j}-\tau_{jk}\frac{\partial}{\partial
y_k}\right), \qquad \tau_{jk}=\frac{\partial\lambda_k}{\partial \eta^j}.
 \label{detatau}
\end{equation}

The K{\"a}hler form is
$$\Phi=\rmd x^j\wedge \rmd y_j=-\frac 14(\rmd\eta^j+\rmd\bar \eta^\jb)\wedge(\tau_{jk} \rmd\eta^k +\tau_{\jb \bar k}\rmd\bar \eta^{ k}).$$
Since it is of type (1,1), it follows that
\begin{equation}
\tau_{ij}=\tau_{ji},
\label{symmetry}
\end{equation} so
$$\Phi=\frac 14 \rmd\eta^i\wedge \rmd\bar \eta^\jb(\tau_{ij}-\bar \tau_{\ib \jb }).$$ Comparing to (\ref{fundamentalform}), we see the metric and the
K{\"a}hler form become
\begin{equation}
g_{i\jb }=-\frac 14\Im(\tau_{ij}), \qquad \Phi=\frac 12\rmi
\Im(\tau_{ij})\rmd\eta^i\wedge \rmd\bar \eta^\jb. \label{rigidmetric}
\end{equation}
Because of (\ref{symmetry}), there exists a local holomorphic function,
determined up to a constant, such that
\begin{equation}
\lambda_j=-8\frac {\partial \cF}{\partial\eta^j}, \qquad
\tau_{ij}=-8\frac {\partial^2 \cF}{\partial \eta^i\partial\eta^j}.
 \label{lambdataurigid}
\end{equation}
 $\cF$ is called the {\it
holomorphic prepotential\index{holomorphic prepotential|see
{prepotential}}}. In terms of it, the K{\"a}hler potential becomes
\begin{equation}
\cK=-\frac 14\Im(\lambda_k\bar \eta^{ k})=2\Im\left(\frac {\partial \cF}
{\partial\eta^k}\bar \eta^{ k}\right). \label{prepotential2}
\end{equation}
The coordinate system $\{\eta^j\}_{j=1}^n$ is a {\it special coordinate
system\index{special coordinates}}.

In the particular case in which $\tau_{jk}=\rmi\delta_{jk}$, then
$\eta^j=x^j+\rmi y_j$, and
$$\Phi=\frac{\rmi}2\rmd\eta^i\wedge \rmd\bar \eta^\ib,$$
so the manifold is locally isometric to $\C^n$.

\paragraph{Recovering the flat connection.} A structure of special
geometry can be given, in an open set, by a holomorphic function
$\cF(\eta)$ such that $\Im (\tau_{ij})$ with $\tau_{ij}$ as in
(\ref{lambdataurigid}), is a non singular, negative definite matrix. The
holomorphic coordinates are declared to be special coordinates. From the
knowledge of $\cF$ we can recover the flat symplectic
coordinates\index{flat symplectic coordinates}
$$x^i=\Re(\eta^i), \quad y_j=-\Re(\lambda_j)=
8\Re\left(\frac{\partial\cF}{\partial\eta^j}\right)$$ (up to a constant)
and also reconstruct the symplectic section
$$\chi=\frac 12\left(\eta^i\frac{\partial}{\partial x^i}-\lambda_j\frac{\partial}{\partial
y_j}\right).$$

In the flat coordinates the coefficients of the flat connection are zero
(that is, the covariant derivatives are usual derivatives). If we want to
use the holomorphic coordinates, these coefficients are not zero anymore.
We want to compute them in the coordinates $(\eta^j)$. In order to do
this, we perform the coordinate change so the connection transforms as
$$A'=A+\Lambda ^{-1}\rmd\Lambda $$
with $A=0$ and
$$\begin{pmatrix}\rmd x\\\rmd y\end{pmatrix}=
\Lambda
\begin{pmatrix}\rmd\eta\\\rmd\bar \eta\end{pmatrix}=
\frac 12\begin{pmatrix}\id&\id\\-\tau&-\bar\tau\end{pmatrix}
\begin{pmatrix}\rmd\eta\\\rmd\bar \eta\end{pmatrix}.$$
$A$ and $A'$ are considered here as matrices $A^I{}_J$, $\bar \tau $ is
$\tau _{\ib \jb }$ and $\Lambda _{IJ}$ is written in terms of blocks of
size $n\times n$. We have
$$\Lambda^{-1}=2\begin{pmatrix}\beta\bar\tau&\beta\\-\beta\tau&-\beta\end{pmatrix},
\qquad \beta=(\bar\tau-\tau)^{-1}=-\frac{1}{8}\rmi g^{-1}.$$ Then
\begin{equation}
A'=\Lambda^{-1}\rmd\Lambda=\begin{pmatrix}
-\beta&0\\\beta&0\end{pmatrix}\rmd\tau+
\begin{pmatrix}0& -\beta\\0&\beta\end{pmatrix}\rmd\bar\tau,\qquad
\rmd\tau_{ij}=\frac{\partial
\tau_{ij}}{\partial\eta^k}\rmd\eta^k,
\label{connection}
\end{equation}
$$\hbox{with}\qquad  A'=
\begin{pmatrix}
A'^j_k&A'^j_{\bar k}\\[1mm] A'^{\jb }_k&A'^{\jb }_{\bar k}
\end{pmatrix}.$$
{}From this expression one can check that conditions
(\ref{complexcondition}) are satisfied.

Let us compute the covariant differential of a vector with only
holomorphic components, $H=H^i\partial/\partial\eta^i$. Notice that,
acting on such vector, only the first term in (\ref{connection})
contributes, so
\begin{eqnarray}\nabla H&=&(\nabla_i
H^j)\rmd\eta^i\otimes\frac {\partial}{\partial\eta^j}+ (\nabla_{\ib } H^j)\rmd\bar \eta^{\ib}\otimes\frac {\partial}{\partial\eta^j}\nonumber\\
&=&\partial_iH^j\rmd \eta^i\otimes \frac{\partial}{\partial \eta^j}
+A'^j_kH^k\otimes \frac{\partial}{\partial \eta^j}+A'^{\jb }_kH^k\otimes
\frac{\partial}{\partial \bar \eta^\jb}+
\partial_{\ib }H^j\rmd \bar \eta^{\ib}\otimes \frac{\partial}{\partial
\eta^j}\nonumber\\
&=&\partial_iH^j\rmd \eta^i\otimes \frac{\partial}{\partial \eta^j}+
\partial_{\ib }H^j\rmd \bar \eta^{\ib}\otimes \frac{\partial}{\partial
\eta^j} -\frac 12 \frac {\partial \tau_{jl}}{\partial\eta^k}H^j\rmd
\eta^k\otimes\frac{\partial}{\partial
y_l},\label{flatderivative}\end{eqnarray}

$$\hbox{since}\qquad \frac{\partial}{\partial
y_k}=2(\beta^{jk}\frac{\partial}{\partial \eta^{
j}}-\beta^{jk}\frac{\partial}{\partial \bar \eta^\jb}).$$

\hfill$\blacksquare$

\begin{example}
The flat metric on $\C^n$\end{example} Let $z^1, \dots ,z^n$ be the
standard coordinates in $\C^n$. To have a rigid special K{\"a}hler structure
it is enough to give a holomorphic function $\cF(z^1, \dots ,z^n)$ such
that the matrix
$$-\Im\left(\frac{\partial^2\cF}{\partial z^i\partial
z^j}\right)$$
 is positive definite and non degenerate. If we take
$$\cF=\frac{1}{4}\rmi\bigl((z^1)^2+\cdots +(z^n)^2\bigr),$$
we obtain the flat metric on $\C^n$
$$g_{i\jb }=\delta_{ij}.$$\hfill$\square$
\begin{example}
The upper half plane\index{upper half plane}\end{example} In one complex
dimension, we consider the holomorphic prepotential
$$\cF=-\frac 1{24} \eta^3, $$ giving the metric
$$\rmd s^2=\Im(\eta) \rmd\eta \rmd\bar\eta,$$
which is positive definite and non degenerate on $$\{\eta\in
\C/\Im(\eta)>0\}.$$ From $\cF$ we can recover the symplectic
coordinates
$$x=\Re(\eta), \qquad y=8\Re\left(\frac{\partial\cF}{\partial \eta}\right)=-\Re(\eta^2).$$
denoting $\eta=x+\rmi p$, we have that $y=p^2-x^2$ and
$$\Im(\eta)=p=+\sqrt{y+x^2}, \qquad y>-x^2,$$ from which the metric
reads
\begin{equation}
\rmd s^2=\frac {4(y+2x^2)\rmd x^2+\rmd y^2+4x\rmd
x\rmd y}{4\sqrt{y+x^2}}.\label{rigidisc}
\end{equation}
This metric is not the Poincar{\'e} metric on the upper half plane. From
(\ref{ricci}) we can see that it has non constant curvature
$$R=\frac1{4(\Im(\eta))^3}.$$\hfill$\square$
\index{prepotential|)}

\subsection{The pseudo-K{\"a}hler case\label{sspseudok}}

As we have seen at the end of section \ref{ss:signatureSpecial} that in
the pseudo-K{\"a}hler case we cannot conclude the independence of the
$\eta^i$, so they may not form a complex coordinate system. Nevertheless,
the $2n\times n$ matrix $(\alpha^T,-\beta^T)$ has still rank $n$, so at
each point we can always perform a linear transformation $A$
 $$\begin{pmatrix}\rmd\eta'\\\rmd\lambda'\end{pmatrix}=A\begin{pmatrix}\rmd\eta\\\rmd\lambda\end{pmatrix}=
 \begin{pmatrix}\alpha'\\\beta'\end{pmatrix} (\rmd z),$$
such that the matrix
 $$\begin{pmatrix}\alpha'\\\beta'\end{pmatrix}=A\begin{pmatrix}\alpha\\\beta\end{pmatrix}
 $$
has $\alpha'$ of rank $n$. Moreover, the linear transformation $A$ can be
chosen as a transformation of the symplectic group. A proof of this fact
is given in  Lemma A1 in Ref. \cite{Craps:1997gp}. We reproduce a sketch
of the proof and some further comments in Appendix \ref{ss:sometec},
Lemma \ref{lemmatoine} and Remark \ref{remarklemmatoine}.

 The conclusion is that there exists a locally finite covering by
 flat Darboux coordinates such that in each open set the matrix $\alpha$
 has rank $n$ and then the functions $\eta_i$ are a system of
 complex coordinates. These will be also called {\it special
coordinates\index{special coordinates}}. The calculation of the
prepotential in these coordinates follows as in  section
 \ref{prepotential}.

The lesson to learn here is that, unlike the K{\"a}hler case, in the
pseudo-K{\"a}hler case not all the Darboux coordinates are suitable to
construct special complex coordinates, but one can equally cover the
manifold with special coordinates. These systems of {\it special Darboux
coordinates} transform in the intersections between charts with matrices
belonging to a subgroup of $\rSp(2n,\R)$, the subgroup that preserves the
maximal rank of the block $\alpha$ in the $2n\times n$ matrix
$$V=\begin{pmatrix}\alpha\\ \beta\end{pmatrix}.$$
It is easy to determine this subgroup. First, we notice that the matrices
$$M_0=\begin{pmatrix}A&0\\C&(A^T)^{-1}\end{pmatrix}\in
\rSp(n,\R)$$ form a subgroup, and this subgroup is maximal (we relegate
the proof to the Appendix, Lemma \ref{lemmaraja}). For matrices of this
form, we have that $\det A\neq 0$, so the rank of $\alpha$ is preserved.
On the other hand, as we proved in Remark \ref{rem:sympltrps}, there
exists always a symplectic transformation that does not preserve the
rank of $\alpha$. The conclusion is that the flat symplectic structure
of the tangent bundle is reduced to the subgroup  of matrices
$$\left\{\begin{pmatrix}A&0\\C&(A^T)^{-1}\end{pmatrix}\right\}\subset
\rSp(n,\R).$$

\subsection{A special pseudo-K{\"a}hler manifold \label{ss:aspecial}}
Let $(z^1,z^2)$ be holomorphic coordinates on
 $\C^2$ and consider the prepotential
 $$\cF=-\frac{1}{8}\rmi z^1 z^2, $$ then
\begin{eqnarray*}
\tau_{ij}&=&-8\frac{\partial^2 \cF}{\partial z^i \partial z^j}=
\begin{pmatrix}
0&\rmi\\
\rmi&0\end{pmatrix}=\rmi\begin{pmatrix}
0&1\\
1&0\end{pmatrix},\nonumber\\
g_{i\jb }&=&-\frac 1 4\Im(\tau_{ij})=-\frac 1
4\begin{pmatrix}0&1\\1&0\end{pmatrix},\nonumber\\
\rmd s^2&=&2g_{i\jb }\rmd z^i \rmd \bar z^{ j}= -\frac{1}{2}
\Im(\tau_{ij})\rmd z^i\rmd\bar z^{ j}=-\frac{1}{2}\left(\rmd z^1\rmd\bar
z^2+\rmd z^2\rmd\bar z^1\right),
\end{eqnarray*}
which clearly has signature $(2,2)$ (null vectors always come in pairs,
one holomorphic and one antiholomorphic). The K{\"a}hler form is
$$\Phi=-2\rmi g_{i\jb }\rmd z^i\wedge \rmd\bar z^{ j}=\frac{\rmi}{2} \Im(\tau_{ij})\rmd z^i \wedge \rmd\bar z^{ j}
=\frac \rmi 2 (\rmd z^1\wedge \rmd\bar z^2+\rmd z^2\wedge \rmd\bar
z^1).$$

Let us denote
\begin{equation}
z^1=x^1+\rmi y_2, \qquad z^2=x^2+\rmi y_1,
\end{equation}
the real and imaginary parts of the complex holomorphic coordinates.
These are the Darboux coordinates of (\ref{PhiDarboux}). Then the  K{\"a}hler
form takes the standard form
$$\Phi=\rmd x^1\wedge \rmd y_1+\rmd x^2\wedge \rmd y_2,$$ so $(x^i, y_i)$ are
symplectic coordinates. For these symplectic coordinates, there is
associated a special holomorphic system of coordinates, just as in the
Riemannian case.

We want to show now that not all the symplectic coordinate systems have
this property when the metric is pseudo-Riemannian. Let us make the
following symplectic change of coordinates,
\begin{eqnarray}&&
x'^1=x^1, \qquad x'^2=y_2, \qquad y'_1=y_1, \qquad y'_2=-x^2, \nonumber\\
\hbox{with}\qquad&& z^1=x'^1+\rmi x'^2, \qquad z^2=-y'_2+\rmi y'_1.
\label{newvariables}
\end{eqnarray}

We have
\begin{eqnarray}\pi^{(1,0)}&=&\rmd z^i\otimes \frac{\partial}{\partial
z^i}=\rmd z^1\otimes\frac{1}{2}\left(\frac{\partial}{\partial
x^1}-\rmi\frac{\partial} {\partial y^2}\right)+\rmd
z^2\otimes\frac{1}{2}\left(\frac{\partial}{\partial
x^2}-\rmi \frac{\partial}{\partial y^1}\right)\nonumber \\
&=&\frac{1}{2}\left( \rmd z^1\otimes \frac{\partial}{\partial x^1}+\rmd
z^2\otimes \frac{\partial}{\partial x^2}-\rmi\rmd z^2\otimes
\frac{\partial}{\partial y^1}-\rmi \rmd z^1\frac{\partial}{\partial
y^2}\right) .
\end{eqnarray}

Comparing this equation with (\ref{pi10Darboux})
$$\rmd\eta^1=\rmd z^1, \qquad \rmd\lambda _1=\rmi \rmd z^2, $$
$$\rmd\eta^2=\rmd z^2, \qquad \rmd\lambda _2=\rmi \rmd z^1,$$
and thus, following (\ref{alphabeta})
$$\alpha^T=\left(\begin{array}{cc}1&0\\0&1\end{array}\right),\qquad \qquad
\beta^T=\rmi \left(\begin{array}{cc}0&1\\1&0\end{array}\right).$$

We can use the new variables $(x',y')$ defined in (\ref{newvariables}) to
calculate $(\eta',\lambda')$
\begin{eqnarray}
\pi^{(1,0)}=\rmd z^1\otimes\frac{1}{2}\left(\frac{\partial}{\partial
x'^1}-\rmi \frac{\partial}{\partial x'^2} \right)+\rmd
z^2\otimes\frac{1}{2}\left(-\frac{\partial}{\partial y'_2}-\rmi
\frac{\partial}{\partial y'_1}\right).
\end{eqnarray}
Comparing this equation with (\ref{pi10Darboux})
$$\rmd\eta'^1=\rmd z^1, \qquad \rmd\lambda'^1=\rmi \rmd z^2,$$
$$\rmd\eta'^2=-\rmi\rmd z^1, \qquad \rmd\lambda'^2=\rmd z^2,$$
then
$$\alpha'^T=\left(\begin{array}{cc}1&-\rmi \\0&0\end{array}\right),\qquad \qquad
\beta'^T=\left(\begin{array}{cc}0&0\\ \rmi&1\end{array}\right).$$

We compute now the null vector, following the general case explained at
the beginning of Section \ref{ss:signatureSpecial}. Let $c=(c_1,c_2)$ be
such that
$$c\alpha'^T=(c_1,c_2)\left(\begin{array}{cc}1&-\rmi \\0&0\end{array}\right)=(0,0)\quad \Rightarrow\quad c_1=0.$$
For any $c=(c_1, c_2)$ we have that
$$c\beta'^T=(c_1,c_2)\left(\begin{array}{cc}0&0\\ \rmi&1\end{array}\right)=c_2 (\rmi,1),$$ so $(0, c_2)$ is not a null
vector of $\beta'^T$. The vector $\gamma$ and its complex conjugate
$\bar \gamma$ are then
$$\gamma=\rmi \frac{\partial}{\partial z^1}+\frac{\partial}{\partial z^2},$$
$$\bar\gamma=-\rmi \frac{\partial}{\partial \bar z^1}
+\frac{\partial}{\partial \bar z^2}.$$
 Then
\begin{equation}
\Phi(\gamma,\bar\gamma)=\frac12 \rmi (\rmd z^1\wedge \rmd\bar z^2 + \rmd
z^2\wedge \rmd\bar z^1)(\gamma\otimes \bar\gamma)=0.
\end{equation}

\subsection{The holomorphic cubic form\index{holomorphic cubic form|(}}\label{cubic}

Let $\cM$ be a rigid special K{\"a}hler manifold. We want to compute the
difference between the Levi-Civita connection $\DLC$ and the flat
connection $\nabla$. Using the same notation as in Ref.
\cite{Freed:1997dp}, we define the tensor $B_\R$ as
$$B_\R\equiv
 \nabla-D,\qquad B_\R\in \Omega^1(\cM, \mathrm{End}_\R T\cM ).$$
Since both connections are symplectic, $\DLC\Phi =0$ and $\nabla \Phi =0$
(see (\ref{DLCPhi0}) and (\ref{NablaPhi0})), we have that
\begin{eqnarray}
 \partial_u(\Phi(v,w))&=&\nabla_u(\Phi(v, w))=\Phi(\nabla_u(v), w)+\Phi(v, \nabla_u(w)), \nonumber\\
 \partial_u(\Phi(v,w))&=&\DLC_u(\Phi(v, w))=\Phi(\DLC_u(v), w)+\Phi(v, \DLC_u(w)), \label{SymplBR}\\
 0&=&\Phi((B_\R)_u(v), w)+\Phi(v, (B_\R)_u(w)).\nonumber
\end{eqnarray}
This says that the endomorphism $(B_\R)_u$, for arbitrary $u$, is  in the
Lie algebra $\fsp(2n,\R)$ defined by $\Phi$. In components, using
(\ref{Gammanon0}) and (\ref{complexcondition}), we get
\begin{eqnarray*}
B_\R=\nabla-D&=&(A_{ij}^k-\Gamma_{ij}^k)\rmd\eta^i\otimes
\rmd\eta^j\otimes
\partial_k + A_{ij}^{\bar k}\rmd\eta^i\otimes \rmd\eta^j\otimes
\partial_{\bar k} \\&+& (A_{\ib \jb }^{\bar k}
-\Gamma_{\ib \jb }^{\bar k})\rmd\bar \eta^\ib\otimes \rmd\bar
\eta^\jb\otimes
\partial_{\bar k} + A_{\ib \jb }^{ k}\rmd\bar \eta^\ib\otimes \rmd\bar \eta^\jb\otimes
\partial_{ k}.
\end{eqnarray*}
Let $u, v, w$ vectors of type (1,0). Then $(B_\R)_u(\bar w)=0$, and the
last line of  (\ref{SymplBR}), with $w$ replaced by $\bar w$, implies
that $\Phi((B_\R)_u(v), \bar w)=0$. In components, this means
$$(A_{ij}^k-\Gamma_{ij}^k)=0,\qquad ( A_{\ib \jb }^{\bar k}-\Gamma_{\ib \jb }^{\bar
k})=0,
$$
where the second one follows by complex conjugation. One can define $B\in
\Omega^{1,0} (\mathrm{Hom}(T\cM ,\overline{T\cM }))$ such that
$$B= A_{ij}^{\bar k}\rmd\eta^i\otimes \rmd\eta^j\otimes
\partial_{\bar k}               \qquad \mathrm{so} \qquad B_\R=B+\bar B.$$
Lowering the antiholomorphic index with the metric, we can define locally
a holomorphic 3-tensor,
\begin{equation}
\Xi_{ijk}=-2\rmi g_{i\bar \ell }A^{\bar \ell }_{jk},
\quad\Leftrightarrow\quad A^{\ib }_{jk}=\frac 1{2}\rmi g^{\ib
\ell}\Xi_{\ell jk}. \label{xib}
\end{equation}
Using (\ref{connection}) and the fact that $g^{\jb  \ell}=8\rmi
\beta^{\jb  \ell}$ we get
\begin{equation}
\Xi_{ijk}=-\frac 14\frac{\partial\tau_{ij}}{\partial \eta^k}=2
\frac{\partial^3\cF}{\partial \eta^i\partial \eta^j\partial \eta^k},
 \label{xicom}
\end{equation}
from which it follows that $\Xi$ is holomorphic and symmetric.

\bigskip

In Ref. \cite{Freed:1997dp} the following global definition is given for
this tensor:
\begin{equation}
\Xi(X,Y,Z)=\Phi(\pi^{(1,0)}X, (\nabla_Y\pi^{(1,0)})Z).
\label{holcubic}
\end{equation}
In fact, since $\DLC\pi^{(1,0)}=0$, we can substitute $\nabla $ by $B_\R$
in (\ref{holcubic}) so
\begin{eqnarray*}
\Xi(X,Y,Z)&=&\Phi(\pi^{(1,0)}X,[(B_\R)_Y, \pi^{(1,0)}]Z)\nonumber\\
&=&\Phi(\pi^{(1,0)}X,[(B+\bar B)_Y, \pi^{(1,0)}]Z)\\
&=&\Phi(\pi^{(1,0)}X,B_Y\pi^{(1,0)}Z),
\end{eqnarray*}
which in components, using (\ref{fundamentalform}), gives (\ref{xib}).

It is then clear that given the flat connection $\nabla$ we can determine
the cubic form $\Xi$. Conversely, assume that we are given a holomorphic
symmetric cubic form $\Xi$ on a K{\"a}hler manifold. We can determine a
tensor $B_\R=B+\bar B$ from (\ref{xib}). Then, a new connection is
defined by $\nabla= \DLC +B_\R$. The symmetry of $\Xi$ guarantees that
$\nabla$ is torsionfree, symplectic and satisfies (\ref{dnablaj}), as it
follows straightforwardly  from (\ref{xib}). The flatness condition
imposes some restrictions on $\Xi$. We have to impose $\rmd_\nabla^2=0$,
with
$$\rmd_\nabla F=\rmd_\DLC F +B\wedge F+\bar B\wedge F,$$ for $F\in
\Omega^p_\cM(T\cM )$. Then, if $R$ is the curvature of the Levi-Civita
connection, $\frac{1}{2}R^I{}_{JKL}\rmd x^K\wedge \rmd x^L$, then
$$\rmd_\nabla^2=0  \quad\Leftrightarrow\quad R+\rmd_DB+\rmd_D\bar
B+B\wedge \bar B+\bar B\wedge B=0.$$ Analysing  the holomorphic and
antiholomorphic components in this equation, we obtain that the following
expressions have to cancel separately,
\begin{eqnarray}&R+B\wedge\bar B+\bar B\wedge B=0\label{curvature}\\&
\rmd_\DLC B=0\label{integrability}\\&\rmd_\DLC \bar B=0.\label{integrability2}
\end{eqnarray}
Equations (\ref{integrability}) and (\ref{integrability2}) are the
complex conjugate of each other. (\ref{curvature}) imposes a constraint
on the curvature of the K{\"a}hler manifold\index{rigid special K{\"a}hler
manifold!curvature}. It should be expressed solely in terms of the
holomorphic cubic form. In coordinates this means
\begin{equation}
R^{\ell  }{}_{ij \bar k}=A^{\bar p}_{ji}A^{\ell  }_{\bar p\bar k} =\frac
14g^{\ell  \bar \ell  '} g^{p'\bar p}\Xi_{p'ji}\bar\Xi_{\bar p\bar \ell
'\bar k}. \label{curvaturecom}
\end{equation}
(\ref{integrability}) imposes a constraint on the metric and the cubic
tensor. In components we have
$$B^{\bar k}_j=A_{ij}^{\bar k}\,\rmd\eta^i=-\frac 1{2\rmi}g^{\bar k
\ell}\Xi_{\ell ij}\,\rmd\eta^i,$$ so
$$-2\rmi \rmd_DB^{\bar k}_j=\rmd_D g^{\bar k
\ell}\wedge \Xi_{\ell ij}\,\rmd\eta^i+g^{\bar k \ell } \rmd_D(\Xi_{\ell
ij}\,\rmd\eta^i)=0.$$ Here, $g^{\bar k \ell }$ is understood as the
components of  an element of $\Omega^0(\cM,T\cM\otimes\overline{T\cM}
)$, so its covariant differential is the covariant derivative and it is
0. Only the last term survives, so we have
\begin{equation}\rmd_D(\Xi_{\ell ij}\,\rmd\eta^i)=(\partial_k\Xi_{lij}-\Gamma_{k\ell }^{\ell '} \Xi_{\ell 'ij}
-\Gamma_{kj}^{j'} \Xi_{\ell
ij'})\,\rmd\eta^k\wedge\rmd\eta^i=0.\label{integrabilitycom}
\end{equation}
It is easy to see that for a metric as in (\ref{rigidmetric}) and a cubic
form as in (\ref{holcubic}) this equation is satisfied identically. What
this argument proves is that (\ref{curvature}) and (\ref{integrability}),
or the equivalent statements (\ref{curvaturecom}) and
(\ref{integrabilitycom}), are sufficient conditions to have a flat
symplectic connection satisfying the requirements of a rigid special
K{\"a}hler structure.

Indeed, given a Hermitian metric and and arbitrary holomorphic cubic form
with components $\Xi _{ijk}=\Xi _{(ijk)}$, one can construct a
torsionfree, symplectic connection as $\nabla=D+B_\R$, where the
connection coefficients for $B_\R$ are determined by (\ref{xib}). This
connection, by construction, satisfies $d_\nabla J=0$. Then
(\ref{curvature}), (\ref{integrability}) are equivalent to the statement
that $\nabla$ is flat, ($\rmd^2_\nabla=0$). So given a Hermitian metric
and a holomorphic cubic form, they will in this case define a special
K{\"a}hler structure.\index{holomorphic cubic form|)} \index{rigid special
K{\"a}hler manifold|)}

\section{Projective K{\"a}hler (K{\"a}hler-Hodge) manifolds\label{ss:projective}}
\index{projective K{\"a}hler manifold|(} \index{K{\"a}hler-Hodge manifold}

\subsection{Affine transformations, isometries and homothetic Killing vectors}
  \label{ss:AffineHomoth}
For the results in this section, see ref. \cite{Kobayashi1996}, Chapter
VI.

\bigskip

An {\it affine transformation\index{affine transformation}} of a manifold
$\cM $ with linear connection $\nabla$ is a diffeomorphism $f:\cM
\rightarrow \cM $ whose tangent map $T\!f:T\cM \rightarrow T\cM $ maps
any parallel vector field along a curve $\gamma$ into a parallel vector
field along the curve $f(\gamma)$. The push-forward by $f$ of a vector
field $X$ on $\cM$ is
$$f_*(X)=Tf\circ X\circ f^{-1},$$ or, in components,
\begin{equation}
  f_* X (x)= \frac{\partial f^I }{\partial x^J}X^J(f^{-1}(x)) \partial _I.
 \label{compf*X}
\end{equation}

If $Y$ and $Z$ are two vector fields on $\cM$ and $f$ is an affine
transformation,  then
\begin{equation}
(f_*\nabla_YZ)= \nabla_{f_*Y}(f_*Z).
 \label{affine}
\end{equation}

Let $K$ be a vector field on $\cM $ and let $\varphi_t:U\rightarrow \cM $
be the flow of $K$ on a neighbourhood $U$ of $x\in \cM $,  $\, t\,\in\,
]-\epsilon,\epsilon[\, $. $\varphi_t$ is a local uniparametric group of
transformations, and for each $x\in \cM $, $\varphi_t(x)$  is an integral
curve of $K$:
$$
\frac{\rmd\varphi_t(x)}{\rmd t}=K(\varphi_t(x)).
$$
We say that $K$ is an {\it infinitesimal affine
transformation\index{affine transformation!infinitesimal}} of $\cM $ if
$\varphi_t$ is an affine transformation of $U$ (the connection being the
restriction of $\nabla$ to $U$). Specifying $f=\varphi_t$ in
(\ref{affine}) and taking a derivative with respect to $t$ and putting
$t=0$ one obtains\footnote{One uses here $\left.\frac{\rmd}{\rmd t}
\varphi _{t*} X\right|_{t=0}=-[K,X]=-{\cal L}_K X$, where $\left.\varphi
_t(x)\right|_{t=0}=x$. The first can be derived from (\ref{compf*X}) with
$f=\varphi _t$.}
\begin{equation}
\cL_K\circ \nabla_Y-\nabla_Y\circ \cL_K=\nabla_{[K,Y]}, \quad \hbox{for
every vector field $Y$ on } \cM,
 \label{infiaffine}
\end{equation}
which characterizes $K$ as an infinitesimal affine transformation. (Here
$\cL_K$ stands for the Lie derivative with respect to $K$). In
components, this condition reads
\begin{equation}
  K^J \partial _J \nabla _I   Z^L -\nabla _\sigma  \left( K^J \partial
  _J Z^L \right) + Z^J \nabla _I \partial _J K^L +(\partial
  _I  K^J )\nabla _J Z^L =0.
 \label{compaffine}
\end{equation}
The infinitesimal affine transformations form a subalgebra of the Lie
algebra of vector fields on $\cM$.

For torsionfree connections, (\ref{infiaffine}) reduces to
\begin{equation}
  R(K,Y)Z+\nabla _Y\nabla _Z K -\nabla _{\nabla _YZ}K=0,
 \label{affinetorsionfree}
\end{equation}
or, in components (as $Z$ is arbitrary)
\begin{equation}
  K^J R _{ J IK }{}^L+ \nabla _I \nabla _K K^L  =0,
 \label{compaffinetorsionfree}
\end{equation}
which was used in Ref. \cite{Bergshoeff:2002qk} as the definition of
symmetry of the physical sigma model, independently of the action (in
fact, such action may not exist).

%We have the following
%
%\begin{theorem} If $\cM $ is a connected manifold of dimension $n$, with an affine
%connection $\nabla$,  then the Lie algebra of infinitesimal
%transformations of $\cM $ is, at most, of dimension $n^2+n$. If the
%dimension is exactly $n^2+n$ then the torsion and curvature of $\cM $
%vanish.\end{theorem}
%
%{\it Proof.} See Ref. \cite{Kobayashi1996}, Chapter VI, Theorem
%2.3.\hfill$\square$
%
%\bigskip

A vector field $X$ on $\cM $ is  {\it complete\index{complete vector
field}} if each integral curve $\varphi_t(x)$ extends to $t\in
]-\infty,+\infty[$. This means that the local uniparametric group extends
to a global uniparametric group
$$\begin{CD}\R\times \cM @>>> \cM \\(t,x)@>>>\varphi_t(x).\end{CD}$$

We say that $\nabla$ is a {\it complete linear connection\index{complete
linear connection}} if every geodesic extends to $t\in]-\infty,+\infty[$.

The Lie algebra of the Lie group\footnote{It is necessary to assume that
$\cM $ has a finite number of connected components.} of affine
transformations consists of all infinitesimal affine transformations
that are complete. Moreover,
 if $\nabla$ is  {\it complete} then all the infinitesimal affine transformations are complete.

\bigskip

Let $\cM $ be a manifold with Riemannian metric $g$ and Riemannian
connection $\nabla$. An {\it isometry\index{isometry}} of $\cM $ is a
transformation that leaves the metric invariant. An isometry is
necessarily an affine transformation of $\cM $ with respect to the
Riemannian connection.

A vector field $X$ is an {\it infinitesimal isometry} (or {\it Killing
vector\index{Killing vector}}) if the uniparametric group of
transformations generated by $X$ in a neighbourhood of $x\in \cM $ (for
arbitrary $x$) consists of local isometries. An infinitesimal isometry
$X$ is characterized by
$$\cL_X g=0,$$
which implies that the set of infinitesimal isometries is a Lie algebra.
%We have the following
%
%\begin{theorem} The Lie algebra of infinitesimal isometries of a connected
%Riemannian manifold $\cM $ of dimension $n$ has at most dimension $\frac 12
%n(n+1)$. If this dimension is maximal, then it is a space of constant
%curvature.
%\end{theorem}
%
%{\it Proof.} See Ref. \cite{Kobayashi1996}, Chapter VI, theorem 3.3
%\hfill$\square$

As in the case of affine transformations, the Lie algebra of the Lie
group of isometries is the Lie algebra of all complete infinitesimal
isometries, and if $\cM $ is complete all the infinitesimal isometries
are complete.

%\bigskip
%
%A Riemannian manifold is said to be {\it irreducible} if the holonomy
%group (with base point at $x\in \cM $) of the Riemannian connection acts
%irreducibly on $T_x\cM $. Otherwise the manifold is {\it reducible}. The
%following theorem is known as the {\it decomposition theorem of de Rham}.
%
%\begin{theorem} A connected, simply connected and complete Riemannian manifold $\cM $
%is isometric to the direct product $\cM _0\times \cM _1\times \cdots \times
%\cM _k$ where $\cM _0$ is a Euclidean space (possibly of dimension 0) and
%$\cM _1,\dots ,\cM _k$ are simply connected, complete, irreducible Riemannian
%manifolds. Such decomposition is unique up to an order.\end{theorem}
%
%{\it Proof.} See Ref. \cite{Kobayashi1996}, Chapter IV, theorem
%6.2.\hfill$\square$
%
%\bigskip
%
%If the Riemannian  manifold is not simply connected we can always work in
%the universal covering and apply the decomposition theorem of de Rham. It
%can be proven that, in such case the identity component of the groups of
%affine transformations and isometries can be computed as the product of
%similar groups of the pieces $\cM _i$ in the de Rham decomposition. It is
%then important to single out such groups for irreducible Riemannian
%manifolds.

We say that a transformation $\phi$ of a (pseudo) Riemannian manifold
$\cM $ is {\it homothetic\index{homothetic!transformation}} if there is a
positive constant $a^2$ (which depends on $\phi$) such that
\begin{eqnarray}&
(\phi^*g)_x(X,Y)=g_{\phi(x)}(T\phi(X),T\phi(Y))=a^2g_x(X,Y), \nonumber\\
&\forall X,Y\in T\cM\hbox{ and } x\in \cM. \label{homotrans}
\end{eqnarray}
Notice that the Christoffel symbols for the metrics $g$ and $a^2g$ are
the same,
 so the covariant derivatives are the same.
It is then easy to see that a homothetic transformation is an affine
transformation of the Levi-Civita connection.

An infinitesimal transformation $K$ of a (pseudo) Riemannian manifold is
homothetic if its flow is a homothetic transformation in a neighbourhood
of each point $x\in \cM $. Infinitesimal homothetic transformations are
also called {\it homothetic Killing vectors\index{homothetic!Killing
vectors}} and can be characterized as
\begin{equation}
  {\cal L}_K g = c g,
 \label{homotheticK}
\end{equation}
for a constant $c$. This can be seen by substituting $\phi=\varphi_t$,
the flow of $X$, in (\ref{homotrans}) and taking the derivative with
respect to $t$ at $t=0$. We obtain also
$$c=\frac {\rmd a^2}{\rmd t}\big|_{t=0}.$$
If $\DLC$ is the Levi-Civita connection, then (\ref{homotheticK}) is
equivalent to the statement that
\begin{equation}
 g(X,\DLC_Y K)+g(Y,\DLC_X K)= c g(X,Y) \qquad  \forall X,Y\in T\cM ,
 \label{homothetic2}
\end{equation}
since $\DLC_KX-\cL_KX=\DLC_XK$. In components we have
$$\DLC_IK_J+\DLC_JK_I=cg_{IJ}.$$

Let us consider the 1-form $g_K(X)=g(K,X)$. If $\nabla$  is a torsionfree
connection we have that
$$\rmd g_K(X,Y)=\nabla_X (g_K)(Y)-\nabla_Y(g_K)(X).$$
This is true for any 1-form. In our case,
$$\nabla_X(g_K)(Y)=\nabla_X(g)(K,Y)+g(\nabla_XK, Y).$$
If  the connection is compatible with the metric, $\nabla_X(g)=0$, we
have
$$\rmd g_K(X,Y)=g(\nabla_XK, Y)-g(\nabla_YK, X),$$ so
\begin{equation}\rmd g_K=0\quad \Leftrightarrow\quad g(\nabla_XK, Y)-g(
\nabla_YK,X)=0\qquad \forall X,Y\in T\cM
,\label{closedhomothetic1}
\end{equation}
in components
$$\nabla_IK_J-\nabla_JK_I=0.$$
We say that $K$ is a {\it closed homothetic Killing vector} if it is a
homothetic Killing vector such that $g_K$ is a closed 1-form.

If $K$ is a closed, homothetic Killing vector\index{homothetic!closed
homothetic Killing vector} and $\DLC$ is the Levi-Civita connection, then
equations (\ref{homothetic2}) and (\ref{closedhomothetic1}) imply that
\begin{equation}
  \DLC_Y K=\ft12 c Y, \qquad \forall Y\in T\cM.
 \label{closedhomothetic}
\end{equation}
This condition is also sufficient. In components we have that
$$\DLC_I K^J=\frac 12 c \delta _I{}^J.$$
Observe that the statement (\ref{closedhomothetic}) involves only the
connection, so we can use it to generalize the concept of closed
homothetic Killing vector to any linear connection. For a torsionfree
connection $\nabla$, we will say that a vector field is a closed
homothetic Killing vector if
\begin{equation}
\nabla_Y K=\ft12 c Y, \qquad
\forall Y\in T\cM.
\label{closedhomolinear}
\end{equation}
We would like to see if such a vector is in fact an infinitesimal affine
transformation for the linear connection. For an arbitrary torsionfree
connection (\ref{infiaffine}) is reduced to (\ref{affinetorsionfree}).
Using (\ref{closedhomolinear}) the last two terms of
(\ref{affinetorsionfree}) vanish, so the condition for a closed
homothetic Killing vector to be an infinitesimal affine transformation is
\begin{equation}
  R(K,Y)Z=0.
 \label{condAffineClosedHK}
\end{equation}
On the other hand, for an arbitrary connection, the integrability
condition of (\ref{closedhomolinear}) is $R(Y,Z)K=0$ for all $Y$ and $Z$,
which implies
$$R(K,Y)Z-R(K,Z)Y=0$$ by using the Bianchi identity (\ref{BI1}). The symmetric
combination in $Y$ and $Z$ is not zero in general.

We conclude that in general for a torsionfree connection, a closed
homothetic Killing vector is not necessarily an infinitesimal affine
transformation. For a flat, torsionfree connection,
(\ref{condAffineClosedHK}) is trivial and thus in this case the closed
homothetic Killing vector is an infinitesimal affine transformation.

For the Levi-Civita connection (not necessarily flat), we have seen that
any homothetic transformation is an affine transformation. In fact,
because of the symmetries of the curvature tensor
$$R(K,Y)Z=0\;\Leftrightarrow\; R(Y,Z)K=0,$$
so $K$ is an infinitesimal affine transformation,
 even if the connection is not flat.
\bigskip

Let $\cM$ be a K{\"a}hler manifold with complex structure $J$ and let $g$ be
the Hermitian metric. Let $H$ be a holomorphic vector field. The
equations above can be extended to the complexified tangent space. We
assume that $H$ is a homothetic Killing vector. In components this reads
\begin{equation}\cL_H g_{\alpha\bar\beta}=g_{\gamma\bar \beta}
\DLC_{\alpha}H^{\gamma}=cg_{\alpha\bar\beta}\quad\Leftrightarrow\quad
D_YH=cY,\quad Y \hbox{ holomorphic}.\label{holohomoki}
\end{equation}
As before, the last expression does not involve explicitly the metric and
can be used as a generalization of holomorphic homothetic Killing
vector\index{homothetic!holomorphic homothetic Killing vector} for any
linear connection. Since the metric is Hermitian, it is easy to see that
$\bar H$ is homothetic with the same constant $c$ (real). It follows that
$K=\frac 12(H+\bar H)$ is also a homothetic Killing vector with constant
$c$ while $\hat K=-JK=\frac1{2 \rmi}(H-\bar H)$ is a Killing vector,
\begin{eqnarray}\cL_K g=\frac 12 (\cL_Hg+\cL_{\bar H}g)=c g,\\ \cL_{\hat
K} g= \frac 1{2\rmi} (\cL_Hg-\cL_{\bar
H}g)=0.\label{holohomo}\end{eqnarray} Notice that (\ref{holohomo}) can be
written in components as
$$g_{\gamma\bar \beta}
\DLC_{\alpha}H^{\gamma}-g_{\bar \gamma\alpha} \DLC_{\bar\beta}H^{\bar
\gamma}=0,$$
 which is just the requirement that  $K$
is closed, so if $H$ is a holomorphic, homothetic Killing vector then
$K=\frac 12(H+\bar H)$ is a  closed homothetic Killing vector.

The converse is also true: if the metric has a closed homothetic Killing
vector $K$,  then $JK$ is a Killing vector. It also implies the presence
of the holomorphic homothetic Killing vector $H=(1-\rmi J)K$, i.e.\
satisfying (\ref{holohomoki}).

\begin{example}Euclidean space.\end{example} We consider $\C^n$
with the metric
$$\rmd s^2=\rmd z^\alpha\rmd \bar z^\alpha.$$ We take
$H=z^\beta\partial/\partial z^\beta$. Then $\cL_Hg=g,$ so $H$ is an
holomorphic, homothetic Killing vector with $c=1$.
 We define $K=\frac 12(H+\bar H)=\frac 12(z^\beta\partial/\partial z^\beta+
\bar z^{ \beta}\partial/\partial \bar z^{\beta}).$ Then
 $$g_K=\frac 12(z^\alpha\rmd \bar z^{\alpha}+\bar z^\alpha\rmd z^\alpha)\quad\Rightarrow\quad  \rmd g_K=0.$$\hfill$\square$

\bigskip

\subsection{Definition of projective K{\"a}hler manifolds}\index{projective K{\"a}hler manifold!definition}

Let $\tilde\cM$ be a complex manifold and let $H$ be a holomorphic vector
field. Then $[H,\bar H]=0$, so $\{H,\bar H\}$ form an integrable
distribution on $T\tilde\cM$. By Frobenius theorem, there is a foliation
of $\tilde\cM$ whose leaves\footnote{The leaves of a
foliation\index{leaves of a foliation} are disjoint sets whose union is
the whole manifold.} are complex submanifolds of $\tilde\cM$ whose
tangent space is generated by  $H$ and $\bar H$. If $H\neq0$ at all
points of $\tilde\cM$ the foliation is {\it regular}; then all the leaves
have complex dimension~1. We can define an equivalence relation on
$\tilde\cM$ by declaring as equivalent two points if they belong to the
same leaf. Then, if the foliation is regular, the quotient of $\tilde\cM$
by this relation (the set of all equivalence classes) is a manifold.

Let $K$ and $\hat K$ be, as above, the real and imaginary parts of $H$,
respectively, so $H=K+\rmi \hat K$. Let $\varphi_\tau$ and $\hat
\varphi_\sigma$ the flows of $K$ and $\hat K$ respectively,
$$\frac {\rmd \varphi_\tau(x)}{\rmd \tau}=K(\varphi_\tau(x)),\qquad \frac {\rmd \hat\varphi_\sigma(x)}{\rmd \sigma}=\hat
K(\hat\varphi_\sigma(x)),\qquad \varphi_0(x)=\hat\varphi_0(x)=x.$$ Since
$[K,\hat K]=0$, $K$ is invariant under the flow of $\hat K$ and
viceversa. This in turn implies that
$$\varphi_\tau\circ\hat\varphi_\sigma=\hat\varphi_\sigma\circ
\varphi_\tau.$$ Let us define $\lambda =\tau -\rmi\sigma$ and
$\rho_\lambda=\varphi_\tau\circ\hat\varphi_\sigma$; then it is easy to
see that
\begin{equation}
\frac {\rmd \rho_\lambda(x)}{\rmd \lambda}=\frac {\rmd
\rho_\lambda(x)}{\rmd \tau}+\rmi \frac {\rmd \rho_\lambda(x)}{\rmd
\sigma}=H(\rho_\lambda(x)),\qquad
\rho_0(x)=x.
\label{integralcurve}
\end{equation}
$\rho_\lambda$ is a local, complex 1 dimensional group of
transformations,
$$\rho_\lambda\circ\rho_{\lambda'}=\rho_{\lambda+\lambda'}.$$
%If the vector fields $K, \hat K$ are complete, then we have a global
%group of transformations $G$.
$H$ is the {\it fundamental vector field\index{fundamental vector field}}
of the action of $G$.

We consider now on $\tilde\cM$ a holomorphic action of $\C^\times$.
\begin{equation}
  \begin{CD}\C^\times\times \cM@>>>\cM\\(b, x)@>>>\mathfrak{R}_b(x),\end{CD}
 \label{infinrholambda}
\end{equation}
with $b\in \C^\times$. Locally, $b=\exp \lambda$  and
$\mathfrak{R}_b(x)=\rho_\lambda(x)$ with $\rho_\lambda$ as in
(\ref{integralcurve}).
%\begin{equation}
%\begin{CD}\C^\times\times \cM@>>>\cM\\(\lambda, x)@>>>\rho_\lambda(x).\end{CD}
% \label{infinrholambda}
%\end{equation}
%$$\hbox{and for $x$ fixed denote}\quad
% \begin{CD}\C^\times@>x_\lambda>>\cM\\\lambda@>>>\lambda.\, x\end{CD}$$
%Let $x\in \cM$ be fixed and $\lambda_x:\C^\times\rightarrow \cM$.
%The fundamental vector field of this action is
%$$H(x)=Tx_\lambda\big|_\mathrm{id}.$$
Let $H$ be the (holomorphic) fundamental vector field of this action
(\ref{integralcurve}). The orbits of the action are the integral
submanifolds of the foliation defined by $H$. We assume also that the
action is free, so the orbits are diffeomorphic to $\C^\times$. Since the
group is abelian, the left action is also a right action, so $\tilde\cM$
is a principal $\C^\times$-bundle over the orbit space
$\tilde\cM/\C^\times$.

\begin{definition}(Projective K{\"a}hler manifold.)\label{psdef}
Let $\tilde\cM$ be a (pseudo) K{\"a}hler manifold with metric $\tilde g$. We
assume that on $\tilde\cM$ there is a free holomorphic action of
$\C^\times$ such that the fundamental vector field $H$ is a non null,
holomorphic homothetic Killing vector\index{homothetic!holomorphic
homothetic Killing vector} of the metric $\tilde g$ (or the Levi-Civita
connection $\tilde D$),
$$ \cL_H\tilde g=c\tilde g\quad\Leftrightarrow\quad\tilde D_YH=cY\quad\forall Y\in
T^{1,0}\tilde\cM,$$
such that $\tilde g(H,\bar H)\neq 0$.
 Then we say that $\cM=\tilde\cM/\C^\times$ is a (pseudo) projective K{\"a}hler manifold. \hfill$\square$
\end{definition}
We are going to show that $\cM$ is a K{\"a}hler manifold itself, of a
particular class. In order to do that, we need to construct a Hermitian
line bundle on $\cM$. It is in fact inherited from the tangent bundle on
$T\tilde\cM$.

\paragraph{The symplectic and line bundles and the fiber metric.}
$\tilde \cM$ has the structure of a principal $\C^\times$-bundle over
$\cM$, $\pi:\tilde \cM\rightarrow \cM$. As in ((\ref{infinrholambda})) we
denote the finite right action of $\C^\times$ on $\tilde\cM$
$$\begin{CD}\tilde\cM@>>>\tilde\cM\\
\tilde m@>>>\mathfrak{R}_b(\tilde m)\end{CD}\qquad b\in \C^\times,$$
 with $\mathfrak{R}_1\tilde m=\tilde m$.

Let $T\tilde \cM$ be the tangent bundle. The tangent of the action above
gives an action on  $T\tilde\cM$
$$\begin{CD}T\tilde\cM@>>>T\tilde\cM\\
(\tilde m, v_{\tilde m})@>>>(\mathfrak{R}_b(\tilde m), \rmd
\mathfrak{R}_b v_{\tilde m}).
\end{CD}$$ $T_{\tilde m}\tilde\cM$ is a complex vector space, so we
also have an action of $\C^\times$ on it. We will simply denote it by
multiplication,
$$\begin{CD}T_{\tilde m}\tilde\cM@>>>T_{\tilde m}\tilde\cM\\
(\tilde m, v_{\tilde m})@>>>(\tilde m, b v_{\tilde m}) .\end{CD}$$

We are going to define an associated bundle to $\tilde\cM$ using these
actions. We identify elements in $T\tilde \cM$ related by
\begin{equation}
(\tilde m, v_{\tilde m})\sim(\mathfrak{R}_b(\tilde m) , b^{-1}\rmd
\mathfrak{R}_b v_{\tilde m}). \label{assobundle}
\end{equation}
It is easy to see that this is an equivalence relation. The quotient
space is a complex vector bundle over $\cM$ of rank $n+1$, with
$\dim_\C\cM=n$. We will denote it by $\cH=T\tilde \cM/\sim$. It is a
bundle over $\cM$ associated to the $\C^\times$-principal bundle
$\tilde\cM\rightarrow \cM$ \cite{Kobayashi1996}, so there is an action of
$\C^\times$ on it. Also, the underlying real vector bundle of $\cH$ (and
its complexification) inherit the action of the symplectic group
$\rSp(2n+2,\R)$ from $T\tilde \cM$.

A  vector in the kernel of the projection $\ker(T\pi|_{\tilde m})\approx
\rspan_\C\{ H(\tilde m)\}$ is a vertical vector. We can consider the
subbundle of $T\tilde\cM$ consisting of vertical vectors. It is a trivial
line subbundle of $T\tilde\cM$, and it projects to a line bundle on
$\cM$. We will denote it by $L$. Two different trivializations $(\tilde
m, \lambda H(\tilde m))$ and $(\tilde m', \lambda' H(\tilde m'))$ (with
$\pi(\tilde m')=\pi(\tilde m)$) are related, according to
(\ref{assobundle}), by
$$(\tilde m, \lambda H(\tilde m))\sim(\tilde m', \lambda'
H(\tilde m'))\quad \Rightarrow\quad\tilde m'=\mathfrak{R}_b(\tilde m),\;
\lambda'=b^{-1}\lambda,
 $$
since (\ref{integralcurve}) implies for the finite transformation
$\rmd\mathfrak{R}_bH(\tilde m )=H(\mathfrak{R}_b(\tilde m) )$. The
transition functions of the bundle are of the form $b^{-1}$.

On $L$ we can define a fiber metric. Let $(\tilde m,\lambda H(\tilde m))$
be a representative of the equivalence class $(m, v_m)\in L$, and the
same for $(\tilde m,\sigma H(\tilde m))$ in $(m, u_m)\in L$. We set
$$h_{m}(v_m,\bar u_m)=\tilde g_{\tilde m}(\lambda H(\tilde m),\bar \sigma \bar H(\tilde
m))=(\lambda\bar \sigma) \tilde g_{\tilde m}( H(\tilde m), \bar H(\tilde
m)).$$
We remind that $H(\tilde m)$ is non null by assumption. We just
have to check that this definition is independent of the representatives
that we have used, so acting with $b\in\C^\times$ we have
\begin{eqnarray*}\lefteqn{\tilde g_{\tilde m b}(b^{-1}\lambda H(\mathfrak{R}_b(\tilde m) ),\bar b^{-1}\bar
\sigma \bar H(\mathfrak{R}_b(\tilde m )))}\nonumber\\
&=&(\lambda\bar
\sigma)(b\bar b)^{-1}\tilde g_{\mathfrak{R}_b(\tilde m) }(
H(\mathfrak{R}_b(\tilde m) ),\bar H(\mathfrak{R}_b(\tilde m
)))\\&=&(\lambda\bar \sigma)\tilde g_{\tilde m }( H(\tilde m ),\bar
H(\tilde m )),
\end{eqnarray*}
as we wanted to show. The last equality follows from (\ref{homotrans}),
taking $b\bar b=a^2$. \hfill$\blacksquare$

\bigskip

We can now define the K{\"a}hler structure on $\cM$.

\paragraph{The metric and the K{\"a}hler potential.}

Let  $\alpha$ be a local basis of $L^*$ (a coframe) dual to the frame
$\{H\}$ of $L$, so $\alpha(H)=1$. Using the formulae from the end of
section \ref{ss:hermLineBundles} we have (the index $a$ runs only over
one value and can be omitted)
\begin{equation}
  h= \theta\alpha\bar\alpha, \quad \theta= 2\tilde g_{\tilde m}(H(\tilde m),\bar H(\tilde
m)).
 \label{deftheta}
\end{equation}

We want to compute $\theta$ using convenient coordinates in $\tilde \cM$.
Let $z^i$, $i=1,\dots n$ be complex coordinates on an open set $U\subset
\cM$. Let $s:U\rightarrow \pi^{-1}(U)$ be a local section on $\cM$. Then
we can choose the local trivialization $\pi^{-1}(U)\thickapprox U\times
\C^\times$ given by
$$\tilde m=(m,\hat{y} s(m)),\qquad m=\pi(\tilde m), \; \hat{y}\in
\C^\times.$$ $(\hat{z}{}^i, \hat{y})$ are local coordinates on
$\pi^{-1}(U)\stackrel{\mathrm{open}}\subset \tilde\cM$. We define {\it
homogeneous coordinates} $\hat{\eta}^\alpha$ on $\pi^{-1}(U)$ as
\begin{equation}
\hat{\eta}^0=\hat{y},\qquad \hat{\eta}^\alpha=\hat{y}\hat{z}{}^i\; \hbox{
for }\alpha=i. \label{coorhom}
\end{equation}
The action of $\C^\times$ on $\tilde \cM$  defined in Eq.
(\ref{infinrholambda}), expressed

The action of  $\C^\times$, expressed in these coordinates is simply
$$\fR_b(\hat{z}{}^i, y)=(\hat{z}{}^i,b^p\hat{y})\qquad \fR_b(\hat{\eta}^\alpha)=b^p\hat{\eta}^\alpha,$$
for an arbitrary $p\in \mathbb{R}$. So the fundamental vector field is
\begin{equation}
  H=\hat{\eta}^\alpha\frac{\partial}{\partial\hat{\eta}^\alpha}=\hat{y}\frac{\partial}{\partial \hat{y}}.
 \label{Hfund}
\end{equation}
(One may choose a multiple of it, which by the definition
(\ref{holohomoki}) amounts to a rescaling of $c$.)  The homothety
condition is
$$\cL_H \tilde
g_{\alpha\bar\beta}=\hat{\eta}^\gamma\frac{\partial}{\partial\hat{\eta}^\gamma}
\tilde{g}_{\alpha\bar\beta}+\frac{\partial
\hat{\eta}^\gamma}{\partial\hat{\eta}^\alpha} \tilde{g}_{\gamma\bar
\beta} =c\tilde{g}_{\alpha\bar\beta}.$$
 We  make the change of variables
  $$\eta^\alpha =(\hat{\eta} ^\alpha)^c,\quad \Rightarrow\quad
  H=c\eta^\alpha\frac{\partial}{\partial\eta^\alpha}.$$
  In these coordinates the homothety condition is
$$\cL_H \tilde
g_{\alpha\bar\beta}=c\eta^\gamma\frac{\partial}{\partial\eta^\gamma}
\tilde{g}_{\alpha\bar\beta}+c\frac{\partial
\eta^\gamma}{\partial\eta^\alpha}
 \tilde{g}_{\gamma\bar \beta} =c\tilde{g}_{\alpha\bar\beta}.$$
 This condition becomes simply
\begin{equation}
\eta^\gamma\frac{\partial}{\partial\eta^\gamma}
\tilde{g}_{\alpha\bar\beta}=0. \label{homothety}
\end{equation}
Together with its complex conjugate, (\ref{homothety}) implies the
following property of the metric,
\begin{equation}
  \tilde g_{\alpha\bar\beta}(\lambda\eta,\bar\lambda\bar \eta)=\tilde g_{\alpha\bar\beta}(\eta,\bar
\eta).
 \label{tildegeta}
\end{equation}
 If we choose $p=1/c$, the action  $\fR_b$ in the coordinates $\eta^\alpha $ is
\begin{equation}
  \fR_b\eta ^\alpha =b\eta ^\alpha.
 \label{fRbeta}
\end{equation}
If we denote
\begin{equation}
  \eta ^0=y,\qquad \eta ^i=yz^i,
 \label{etayz}
\end{equation}
then $z^i$ are coordinates on $\mathcal{M}$. The coordinates on $\tilde
{\cal M}$ are also homogeneous coordinates, which we will further use,
and from now on $\partial _\alpha =\frac{\partial }{\partial \eta ^\alpha
}$. We have that
\begin{equation}
  H=c\eta ^\alpha \partial _\alpha = cy\frac{\partial }{\partial y}.
 \label{Hc}
\end{equation}

The metric in these coordinates can be written in terms of a
 K{\"a}hler potential
\begin{equation}
  \tilde g_{\alpha\bar\beta}=\partial_\alpha\partial_{\bar \beta}\cK.
 \label{metricalphabeta}
\end{equation}
The transformation (\ref{tildegeta}) leads to
$$\cK(\eta,\bar \eta)=\cK'(\eta,\bar \eta)+f(\eta)+f'(\bar \eta),$$ with
$$\cK'(\lambda\eta,\bar \lambda\bar
\eta)=(\lambda\bar\lambda)\cK(\eta,\bar\eta).$$
 Since $\cK$ is real, $f=\bar f'$ and with a K{\"a}hler transformation
$$\cK\longrightarrow \cK-f-\bar f$$ we can take $\cK'$ as the K{\"a}hler
potential. We will denote it as $\cK$ from now on, so we have
\begin{equation}
\cK(\lambda\eta,\bar\lambda\bar\eta)=(\lambda\bar\lambda)\cK(\eta,\bar\eta).
\label{kalpothom}
\end{equation}
In particular, this implies
\begin{equation}
  \eta^\gamma\partial_\gamma\cK=\cK, \quad \bar \eta ^\gamma\partial_{\bar \gamma}\cK=
\cK,\qquad \eta ^\gamma\bar \eta ^\delta\partial_\gamma \partial_{\bar
\delta}\cK=\cK.
 \label{HonKidentities}
\end{equation}
so the definition (\ref{deftheta}) gives
\begin{equation}
 \theta  =  \tilde g_{\alpha \bar \beta }H^\alpha \bar H^{\bar\beta}=
c^2\eta ^\alpha \bar \eta ^{\bar\beta} \partial _\alpha \bar
\partial _{\bar\beta}\cK=c^2\cK.
 \label{theta}
\end{equation}

Let us consider the exact (1,1)-form
\begin{equation}
\tilde \rho = 2\tilde \rho_{\alpha\bar\beta}\rmd
\eta^\alpha\wedge\rmd\bar\eta^\beta=2\left(-\frac 1{\cK^2}\frac{\partial
\cK}{\partial \eta^\alpha}\frac{\partial \cK}{\partial \bar
\eta^\beta}+\frac 1{\cK}\frac{\partial^2 \cK}{\partial
\eta^\alpha\partial \bar \eta^\beta}\right)\rmd\eta^\alpha\wedge\rmd
\bar\eta^\beta,\label{exact}
\end{equation}
and let us denote by $\rmi \rho$ its pull-back  by the section $s$,
$$
 s^*\tilde \rho  =  2\tilde \rho_{\alpha\bar\beta}\frac{\partial\eta^\alpha}{\partial z^i}
 \frac{\partial\bar\eta^\beta}{\partial \bar z^\jb }\rmd z^i\wedge\rmd\bar z^\jb =\rmi\rho.
 $$
Using (\ref{kalpothom}), we can see that the result is independent of the
section $s$ used.
 In fact, we have that $\tilde\rho=\rmi\pi^*\rho$,
 \begin{equation}\tilde\rho_{\alpha\bar\beta}=\rmi \rho_{i\bar
 j}\frac{\partial z^i}{\partial\eta^\alpha}\frac{\partial \bar z^\ib }{\partial\bar
 \eta^\beta },\label{projexact}
 \end{equation}
where $z^i(\eta^\alpha)$ is the expression of the projection map
$\pi:\tilde {\cal M}\rightarrow {\cal M}$ in coordinates.

The tensor $\tilde \rho_{\alpha\bar\beta}$ is degenerate. Indeed,
$H^\alpha=c\eta ^\alpha $ is a zero eigenvector due to the identities
(\ref{HonKidentities}). We want to show that there is no other zero mode,
under the assumption that $\tilde g_{\alpha \bar \beta }$ is
non-degenerate. Let us write it as
$$\tilde\rho_{\alpha\bar\beta}=\frac1\cK\tilde
g_{\alpha\bar\beta}-\mu_\alpha\bar\mu_{\beta}=\partial_\alpha\partial_{\bar\beta}\log|\cK|,\qquad
\mu_\alpha\equiv\frac 1\cK
\partial_\alpha\cK=\partial_\alpha\log|\cK|.$$ We assume now that there is a vector $v^\alpha$ such
that $v^\alpha\tilde \rho_{\alpha\bar \beta}=0$, then we find that
\[v^\alpha =\cK (v^\gamma\mu_\gamma) \tilde g^{\bar \beta \alpha }\bar\mu_\beta ,\]
where $\tilde g^{\bar \beta \alpha }$ is the inverse of $\tilde g_{\alpha
\bar \beta }$. Hence any zero eigenvector is proportional to $\tilde
g^{\bar \beta \alpha }\bar\mu_\beta $, and thus there is only one zero
mode. In particular, we also obtain
\[H^\alpha =\nu\bar\mu_\beta \tilde g^{\bar \beta \alpha }, \]
for some undetermined function $\nu(\eta ,\bar \eta )$.

The vectors $\partial _i=\eta^0{\partial}/{\partial\eta^i}$ are
transversal to $H$, thus the matrix
$$\rho_{i\jb }=-\rmi \tilde \rho_{\alpha\bar\beta}\frac
{\partial \eta^\alpha}{\partial z^i}\frac{\partial
\bar\eta^\beta}{\partial\bar z^\jb }
$$ is non degenerate. This matrix (or a matrix proportional to it) can
therefore be taken to be the metric on $\cM$.

We define therefore the metric on ${\cal M}$
\begin{equation}
  g_{i\jb }=\partial _i\partial _{\jb }\left[\pm \log |{\cal K}| \right] =\pm \rmi\rho_{i\jb }\,, \qquad \pm =\sign {\cal K}.
 \label{defgprojK}
\end{equation}
The reason for the $\pm $ convention will be explained below.

The Ricci form of the Hermitian bundle agrees according to
(\ref{ricciform}) with
\begin{eqnarray*}
 \rho & = & 2\rho _{i\jb }\rmd z^i\wedge\rmd \bar z^\jb =-2\rmi \frac{\partial^2 \log|\cK|}{\partial
z^i\partial\bar z^\jb }\rmd z^i\wedge\rmd \bar z^\jb \\
   & = & -2\rmi\left(-\frac
1{\cK^2}\frac{\partial \cK}{\partial z^i}\frac{\partial \cK}{\partial
\bar z^\jb }+\frac 1{\cK}\frac{\partial^2 \cK}{\partial z^i\partial \bar
z^\jb }\right)\rmd z^i\wedge\rmd \bar z^\jb .
\end{eqnarray*}

\bigskip

We can also compute the signature of the matrix $\tilde \rho _{\alpha
\bar \beta }$. A vector $V=V^\alpha\partial_\alpha$ is orthogonal to $H$
if $\mu_\alpha V^\alpha=0$ and the space of such vectors  has dimension
$n$. For two such vectors, $V$ and $V'$ we have
$$\tilde g(V, \bar V')= \cK\tilde\rho(V,\bar V'),$$
 so the signature of $\tilde \rho$ in the space orthogonal to $H$ the same
than the signature of $\tilde g$ in such space up to a sign. Furthermore,
the sign in the remaining direction of $\tilde g$ is the sign of ${\cal
K}$ as it follows from (\ref{HonKidentities}).

We can choose a section $s$ such that the vectors $\partial_i$ have a
lift $s_*\partial_i$ orthogonal to $H$. Then,
 $$\rho_{i\jb}=-\rmi s^*\tilde \rho(\partial_i,\partial_\jb),$$
 which is actually independent of the section. So the signature of
the metric $g$ in $\cM$,  (\ref{defgprojK}), is the same than the
signature of $\tilde g$ in the space orthogonal to $H$.

\bigskip

We conclude that $\rho$ defines a symplectic structure compatible with
the complex structure, so $\cM$ is a K{\"a}hler manifold with K{\"a}hler metric
as in (\ref{defgprojK}). The K{\"a}hler form is in the first Chern class of a
line bundle. This implies that the K{\"a}hler form is integer. Such manifolds
are called {\it K{\"a}hler-Hodge manifolds\index{K{\"a}hler-Hodge manifold}} in
the literature. When $\cM$ is compact, this condition implies that $\cM$
is a projective variety, so it is embedded in projective space. This is
the {\it Kodaira embedding theorem\index{Kodaira embedding theorem}}, see
for example Ref. \cite{Griffiths}, page 181.

What we have proven here is that a projective K{\"a}hler manifold is a
K{\"a}hler-Hodge manifold.

\subsection[The Levi-Civita connection on a K{\"a}hler-Hodge manifold]
{The Levi-Civita connection on a K{\"a}hler-Hodge \\ manifold}

The previous part leads to consider (\ref{defgprojK}) as the metric on
$\cM$. One can compute its Levi-Civita connection. However, there is a
natural way of inducing a connection on $\cM$ from the Levi-Civita
connection in $\tilde\cM$, which gives the same result. It clarifies the
geometrical meaning of the metric in the quotient manifold $\cM$. We will
perform two projections of the connection, first to the bundle
$\cH=T\tilde \cM/\sim$ and then to the tangent bundle $T\cM$.

\paragraph{Projecting down to the symplectic bundle.}
Let $X$ be a vector field on $\cM$ and $\xi$ a section of $\cH$. Let $\pi
:\tilde \cM\rightarrow \cM$ and $p:T\tilde \cM\rightarrow \cH$ the
natural projections. Let $\tilde\cD$ denote a linear connection on
$T\tilde\cM$.

The idea is to find adequate lifts $\tilde X$ of $X$ and $\tilde \xi$ of
$\xi$, both vector fields on $\tilde \cM$, in such way that the covariant
derivative $\tilde\cD_{\tilde X}\tilde\xi (\tilde m)$ projects through
$p$ to the same vector on $\cH$, independently of the point $\tilde m$ in
the fiber $\pi^{-1}(m)$ where it has been computed. This will define
immediately a covariant derivative on $\cH$ as
\begin{equation}\cD_X\xi(m)=p\left(\tilde\cD_{\tilde X}\tilde\xi (\tilde
m)\right),\qquad \pi(\tilde m)=m.\label{indconnection}\end{equation}

Let us first define the respective lifts. A local section $\xi$ of $\cH$
is specified by associating an equivalence class $[(\tilde m,v_{\tilde
m})]$ to any point $m$, with $\pi(\tilde m)=m$. We can choose an
arbitrary $\tilde m\in \pi^{-1}(m)$ and set $\tilde \xi(\tilde
m)=v_{\tilde m}$. Then $\tilde \xi$ is a vector field on $\tilde \cM$
satisfying (see eq.(\ref{assobundle}))
\begin{equation}
\tilde\xi(\mathfrak{R}_b(\tilde m))=b^{-1}\rmd \mathfrak{R}_b
\tilde\xi(\tilde m)\quad \Leftrightarrow\quad {\mathfrak{R}_b}_*\tilde
\xi=b\tilde\xi. \label{localsection}
\end{equation}
There is a one to one correspondence between the set of local sections of
$\cH$ and the set of local sections of $T\tilde\cM$ satisfying
(\ref{localsection}). So $\tilde \xi$ is a natural lift of~$\xi$.

Notice that (\ref{indconnection}) means just that $\tilde\cD_{\tilde
X}\tilde\xi$ is a vector field on $\tilde\cM$ satisfying
(\ref{localsection}), so it defines a section of $\cH$. For any affine
transformation $\mathfrak{R}_b$ of the connection $\tilde \cD$, we have
that (\ref{affine})
$${\mathfrak{R}_b}_*(\tilde\cD_{\tilde X}\tilde \xi)=\tilde\cD_{{\mathfrak{R}_b}_*\tilde X}\left({\mathfrak{R}_b}_*\tilde
\xi\right),$$
so all we need to complete the definition
(\ref{indconnection}) are lifts satisfying
\begin{eqnarray}
{\mathfrak{R}_b}_*\tilde\xi=b\tilde \xi,\label{lift1}\\
{\mathfrak{R}_b}_*\tilde X=\tilde X. \label{lift2}
\end{eqnarray}
(\ref{lift1}) is already guaranteed. There are many lifts of the vector
field $X$ to $T\tilde \cM$, but we have a connection on the principal
bundle $\tilde \cM$ (or on its associated bundle $L$), so it is natural
to consider the {\it horizontal lift\index{horizontal lift}}. Horizontal
lifts satisfy (\ref{lift2}), so this will show the existence of the
induced connection on $\cH$. Note that for the Levi-Civita connection or
for an arbitrary flat connection, $\mathfrak{R}_b$ are affine
transformations, so the result applies for these cases of special
interest.

To understand the horizontal lift we introduce  the definition of
connection on a principal bundle as a Lie algebra valued 1-form. The
relation with the standard covariant derivative in the associated vector
bundles can be found in many places, (see for example Ref.
\cite{Kobayashi1996}). For completeness we give a brief outline in the
Appendix~\ref{cpb}.

\begin{definition}\label{defcpb}Let $G$ be a Lie group and $\fg$ its Lie algebra. A connection
on a principal $G$-bundle $P\stackrel{\pi}\rightarrow \cM$ can be given
by a $\fg$-valued 1-form $\omega$ on $P$ such that:
\renewcommand{\theenumi}{(\roman{enumi})}
\begin{enumerate}
  \item  \label{condomega1} If $A$ is a fundamental vector field, generating the action of $G$ on the fibre,
associated to $\hat A\in \fg$ then $\omega(A)=\hat A$.
  \item \label{condomega2} ${\mathfrak{R}_b}^*\omega=\mathrm{Ad}_{b^{-1}}\omega=b^{-1}\omega b,\;
b\in G$. ($\mathrm{Ad}$ is the adjoint representation of~$G$).
\end{enumerate}

\hfill$\square$
\end{definition}
A horizontal vector $X_u$ is a vector in $T_uP$ satisfying
$\omega(X_u)=0$. In fact, $\omega$ defines a distribution of {\it
horizontal spaces\index{horizontal space}} on $T\!P$, denoted by $TP^h$.
At each point $u$ with $\pi(u)=m$, the horizontal space is mapped
isomorphically to $T_m\cM$. If $T\!P^v=\ker(T\!\pi|_u)$ is the set of
{\it vertical vectors} tangent to the fiber, then $T\!_uP=T\!_uP^h\oplus
T\!_uP^v$. Moreover, \ref{condomega2} implies that the distribution is
equivariant, that is
\begin{equation}T\!_{\mathfrak{R}_b(u)}P^h=T\!\mathfrak{R}_b\,(T\!_uP^h).\label{equivariant}\end{equation}
Let $X$ be a vector field on $T\cM$. One can prove that  there is a
unique vector field $\tilde X$ on $TP$ such that $T\!\pi(\tilde X)=X$ and
$\tilde X(\tilde m)$ is horizontal for every $\tilde m$. It is the {\it
horizontal lift} of $X$.

\bigskip

The equivariance of horizontal subspaces, (\ref{equivariant}), implies
(\ref{lift2}) as we wanted to show. One can also show that any horizontal
vector field on $\tilde\cM$ satisfying the invariance condition
(\ref{lift2}) is the lift of a vector field on $\cM$.

One can prove that if $\tilde X$ and $\tilde Y$ are horizontal lifts of
$X$ and $Y$ respectively, then $[\tilde X,\tilde Y]$ is the horizontal
lift of $[X,Y]$. So if $\tilde \cD$ is a flat connection (as
$\tilde\nabla$) then the induced connection $\cD$ on $\cH$ is also a flat
connection.

\begin{example}Horizontal lift in $\tilde\cM$.\label{examhorlift}\end{example}

As an example that we will use in the following, we are going to compute
the horizontal lift of a holomorphic vector $X$ on $T\cM$ to $T\tilde
\cM$ for the Hermitian connection.

In the coordinates (\ref{etayz}) we have
%
%As before, we take $z^i$, $i=1,\dots n$ to be complex coordinates on
%$U\stackrel{\mathrm{open}}\subset \cM$. Let $s:U\rightarrow \pi^{-1}(U)$
%be a local section on $\cM$, giving the local trivialization
%$\pi^{-1}(U)\thickapprox U\times \C^\times$, so
%$$\tilde m=(m,y s(m)),\qquad m=\pi(\tilde m), \qquad  y\in
%\C^\times.$$ $(z^i, y)$ are coordinates on
%$\pi^{-1}(U)\stackrel{\mathrm{open}}\subset \tilde\cM$ such that
\begin{equation}
  \mathfrak{R}_b(m,y)=(m,by), \qquad \rmd\mathfrak{R}_b=\rmd z^i\otimes
\partial_i + b\rmd y\otimes\partial_y.
 \label{componentRb}
\end{equation}

The connection 1-form and its pull back are
$$\omega=\omega_y \rmd y+\kappa_i\rmd z^i ,\qquad
s^*\omega=\kappa_i\rmd z^i.$$ $\kappa_i$ is determined by the pull back,
which from (\ref{metricconnection}) and (\ref{theta}) is
$$\kappa_i =\cK^{-1}\partial_i \cK=\partial_i\log|\cK|.
$$

The other component, $\omega_y$, is determined by conditions
\ref{condomega1} and \ref{condomega2} in Definition \ref{defcpb}. Since $
A=y\partial_y$ and $\hat A=1$, (i) implies $\omega _y=y^{-1}$. (ii) is
then satisfied.

%Indeed, for the abelian case it reads $\mathfrak{R}_b{}^*\omega
%=\omega $ and $\mathfrak{R}_b{}^*\omega$ in the point $(m,y)$ is
%$\kappa _i\rmd z^i+b\omega _y\rmd y$ in the point $(m,by)$.
% condition \ref{condomega2} readily implies that
%$\omega_y=c(z)y^{-1}$, while \ref{condomega1}, since $ A=y\partial_y$,
%with $\hat A=1$, then $c(z)=1$.
The connection 1-form is then
\begin{equation}\omega=y^{-1}\rmd y +\kappa_i\rmd z^i=y^{-1}\rmd y +\partial_i\log|\cK|\,\rmd z^i. \label{hermconcoor}
\end{equation}

A vector on $T\tilde\cM$, $v=v^i\partial_i+v^y\partial_y$ is horizontal
if and only if
$$y^{-1}v^y+\kappa_i v^i=0.$$ If $v$ is arbitrary, then
$v=v^h+v^v$ with
\begin{equation}v^h=v^i\partial_i-y\kappa_i v^i\partial_y,\qquad
v^v=(v^y+y\kappa_i v^i)\partial_y.\label{verhor}\end{equation} $v^h$ is
the {\it horizontal projection\index{horizontal projection}} of $v$.

A vector  $\tilde X= \tilde X^y\partial_y+\tilde X^i\partial_i$ is  the
horizontal lift of $X=X^i\partial_i$ if
\begin{eqnarray*}T\!\pi(\tilde
X)=X\quad&
\mbox{i.e.\ }&\quad \tilde X^i=X^i,\nonumber\\
\omega(\tilde X)=0\quad & \mbox{i.e.\ }&\quad \tilde X^y=-y\kappa_i
\tilde X^i,
\end{eqnarray*} so
\begin{equation}\tilde X=X^i\partial_i-y\kappa_i
X^i\partial_y.\label{horlift}\end{equation} \hfill$\square$

\hfill$\blacksquare$

\paragraph{Projecting down to the tangent bundle.}
Let us consider the subbundle  of $\cH$ formed by equivalence classes
$[(\tilde m, v_{\tilde m})]$ such that $v_{\tilde m}$ is a horizontal
vector. Notice that, due to (\ref{equivariant}) $b^{-1}\rmd
\mathfrak{R}_b v_{\tilde m}$ is horizontal if so is $v_{\tilde m}$. We
will denote this bundle by $\rhor(\cH)$. We have the following lemma:

\begin{lemma}
 $\rhor(\cH)\approx L\otimes T\cM$.
\end{lemma}

{\it Proof.} Let $[(\tilde m, v^h_{\tilde m})]$ be an element of
$\rhor(\cH)$. We can map it to $T_m\cM$ with the projection $v_m=\rmd\pi
v^h_{\tilde m}\in T_m\cM$. If we choose another representative of the
same equivalence class, $(\mathfrak{R}_b(\tilde
m),v'^h_{\mathfrak{R}_b(\tilde m)})$, with $v'^h_{\mathfrak{R}_b(\tilde
m)}= b^{-1}\rmd \mathfrak{R}_b v^h_{\tilde m}$ we obtain another vector
on $T_m\cM$, $v'_m=b^{-1}v_m$. The natural projection applied to
$\rhor(\cH)$ defines then a section of $L\otimes T\cM$.

 In the other direction, let $X_m\in T_m\cM$ and  $\sigma\in L$. We consider the horizontal lift of $\sigma\otimes
 X_m$ to $L\otimes T\tilde\cM$ for some choice of $\tilde m\in \pi^{-1}(m)$ and we denote it by $\sigma\otimes X^h_{\tilde m}$.
 Then we consider the equivalence class
$[(\tilde m, \sigma\otimes X^h_{\tilde m})]\in L\otimes\rhor(\cH)$. Let
$\tilde m'=\mathfrak{R}_b(\tilde m)$ another choice and
$\sigma'=b\sigma$. Then we have the equivalence class $[(\tilde m',
\sigma'\otimes X^h_{\tilde m'})]$, with $X^h_{\tilde m'}=\rmd
\mathfrak{R}_bX^h_{\tilde m}$. Since $(\tilde m,  X^h_{\tilde m})\sim
(\tilde m', b^{-1} X^h_{\tilde m'})$, then $(\tilde m', \sigma'\otimes
X^h_{\tilde m'})\sim(\tilde m, \sigma\otimes X^h_{\tilde m})$, as we
wanted to show. \hfill$\square$

\bigskip

Let $\cD$ be a connection on $\cH$ and let $p_h:\cH\rightarrow\rhor(\cH)$
be the natural projection. We can define a connection on $\rhor(\cH)$  as
$$\hat \cD_X\xi=p_h(\cD_X\xi),\qquad\hbox{ with } X\in T\cM,\hbox{ and } \xi\hbox{ a section in }
\rhor(\cH)\subset\cH.$$

We want to compute $\hat \cD$ in coordinates. As before, let $s$ be a
local section of $\tilde \cM$, so $\tilde m=(m,ys(m))\in \tilde\cM$ and
let $\{z^i\}$ be local coordinates on $\cM$. Then $\{y,z^i\}$ are
coordinates on $\tilde\cM$. We need to compute the horizontal projection
of an arbitrary section $\chi$ of $\cH$, $\chi^h=p_h(\chi)$. The section
$\chi$ has a lift $\tilde \chi$ to $T\tilde \cM$ satisfying
(\ref{localsection}). In coordinates, using the action of
$\mathfrak{R}_b$ as in (\ref{componentRb}), these equations imply the
following $y$-dependence:
$$
\tilde\chi(y, z)=y^{-1}\chi^i(z)\partial_i+\chi^y(z)\partial_y,
 $$
 and according to (\ref{verhor}), the horizontal projection is
$$\tilde\chi^h(y,z)=\chi^i(z)\left(y^{-1}\partial_i-\kappa_i(z)\partial_y\right).$$

Let $\xi$ be a section of $\rhor(\cH)$, so
\begin{equation}
  \tilde \xi=\xi^i(z)\left(y^{-1}\partial_i-\kappa_i(z)\partial_y\right).
 \label{xihor}
\end{equation}
We have to compute the horizontal projection of $\cD_X\xi$. Lifting to
$\tilde \cM$, we know that $(\tilde \cD_{\tilde X}\tilde \xi)^h$ must be
of the same form
$$
\left( \tilde \cD_{\tilde X} \tilde\xi\right)
^h=\chi^i\left(y^{-1}\partial_i-\kappa_i(z)\partial_y\right),
 $$
for some $\chi ^i$. So we can identify
$$\chi^i=\left( \hat\cD_X\xi\right) ^i=X^j(\partial_j\xi^i+\hat\Gamma^i_{jk}\xi^k).$$

For vectors of the form (\ref{xihor}) and (\ref{horlift})
\begin{eqnarray*}
 \tilde \cD_{\tilde X}\tilde \xi^j & = & y^{-1}X^i\partial_i\xi^j+y^{-1}\tilde\Gamma_{ik}^j\xi^kX^i
-\tilde\Gamma_{i0}^jX^i\kappa_k\xi^k\\&&+y^{-1}\kappa_i
X^i\xi^j-\tilde\Gamma_{0k}^j\xi^kX^i\kappa_i+y\kappa_i X^i\tilde \Gamma
_{00}^j\xi ^k\kappa _k\nonumber\\
&=&y^{-1}\hat{D}_X\xi ^j,
\end{eqnarray*}
where we used the coordinates $\{y,z^i\}$ in ${\cal M}$ and used the
index $0$ for the components with respect to $y$. We obtain therefore for
the connection coefficients:
\begin{equation}
\hat\Gamma_{ik}^j=\tilde \Gamma_{ik}^j-y\tilde\Gamma_{i0}^j\kappa_k
-y\tilde\Gamma_{0k}^j\kappa_i +y^2\kappa_i\kappa _k \tilde \Gamma
_{00}^j+\kappa_i \delta _k^j .
 \label{projconn1}
\end{equation}
The last term is just the connection on $L$, while the rest defines a
connection on $T\cM$,
\begin{equation}\Gamma_{ik}^j=\tilde\Gamma_{ik}^j
-y\tilde\Gamma_{i0}^j\kappa_k-y\tilde\Gamma_{0k}^j\kappa_i
+y^2\kappa_i\kappa _k \tilde \Gamma_{00}^j.
 \label{projconn}
\end{equation}
We have then written $\hat \cD$ as a connection on $L\otimes T\cM$.

\bigskip

We can now compute explicitly the Levi-Civita connection of $\tilde g$ in
terms of the Levi-Civita connection of $g$, and check that the formula
(\ref{projconn1}) is satisfied in this case. Due to (\ref{kalpothom}),
${\cal K}$ is $y\bar y$ times a function that depends only on $z$ and
$\bar z$. The relation between $\tilde g_{\alpha \bar \beta }$ and
$g_{i\jb }$, given by (\ref{defgprojK}), is
\begin{eqnarray}
  &&\tilde g_{i\jb }={\cal K}\left( \pm g_{i\jb }+\kappa _i\bar \kappa _{\jb }\right)
  ,\qquad \tilde g_{0\ib }= \frac{{\cal K}}{y}\bar \kappa _{\ib },\qquad
  \tilde g_{0\bar 0}=\frac{{\cal K}}{y\bar y}, \label{tildegg}\\
  &&\tilde g^{\ib  j}=\pm \frac{1}{{\cal K}}g^{\ib  j},\qquad
  \tilde g^{\ib  0}=\mp \frac{y}{{\cal K}}g^{\ib j}\kappa _j,\qquad
  \tilde g^{\bar 00}=\frac{y\bar y}{{\cal K}}\left(1\pm \kappa _i\bar \kappa _{\jb }g^{\jb i}\right),
\nonumber
\end{eqnarray}
where $\kappa _i=\partial _i\log|{\cal K}|$. This leads to the
Levi-Civita connection coefficients $\tilde \Gamma_{\alpha \beta }^\gamma
$:
\begin{eqnarray*}&&\tilde \Gamma_{ik}^j=g^{j\bar
m}\left(\frac1{\cK}\partial_i\partial_{\bar
m}\partial_k\cK-\frac1{\cK^2}\partial_{\bar
m}\cK\partial_{i}\partial_k\cK\right)=\Gamma_{ik}^j(g)+\kappa _{i}\delta
_{k}^j+\kappa _k\delta _i^j,
\\
&&\tilde\Gamma_{i0}^j=y^{-1}g^{j\bar m}\partial_{\bar
m}\partial_i\log|\cK|=y^{-1}\delta_i^j,\\
&&\tilde \Gamma ^0_{ij}= y\Gamma _{ij}^k(g)\kappa _k+2y\kappa _i\kappa
_j\pm
\frac{y}{{\cal K}}\partial _i\partial _j{\cal K},\nonumber\\
 &&\tilde\Gamma_{00}^j=\tilde \Gamma _{0i}^0=\tilde \Gamma _{00}^0=0,
\end{eqnarray*}
where $\Gamma (g)$ is the Levi-Civita connection of the metric on $\cM$
We thus find that indeed $\Gamma_{ik}^j$ as determined in
(\ref{projconn}) are the Christoffel symbols of the Levi-Civita
connection on $\cM$,  as we wanted to show. \hfill$\blacksquare$

\subsection{Examples of K{\"a}hler-Hodge manifolds}

\begin{example}Complex Grassmannian\index{complex Grassmannian} as a K{\"a}hler-Hodge manifold.\label{grassmannian}\end{example}

We consider the Grassmannian manifold of complex $p$-planes in
$\C^{p+q}$, denoted by $G(p,q)$. We take
 $$\tilde \cM=\left\{Z\;\big|\;
Z\hbox{ is a $(p+q)\times p$ matrix of rank $p$}\right\}.$$
 We will write
$$Z=\begin{pmatrix}Z_0\\Z_1\end{pmatrix}$$
with $Z_0$ a $p\times p$ matrix and $Z_1$ a $q\times p$ matrix. Each $Z$
defines a $p$-plane in $\C^{p+q}$ as the span of the column vectors.
Taking linear combinations of these vectors gives the same plane. Then,
there is a right action of $\rGL(p,\C)$ on $\tilde \cM$ which does not
change the $p$-plane. $\tilde \cM\rightarrow G(p,q)$ is a principal
bundle with structure group $\rGL(p,\C)$.

The group $\rSL(p+q)$ acts transitively on $G(p,q)$, but also the action
of $\rSU(p+q)$ is transitive, with little group $\rSU(p)\times
\rSU(q)\times \rU(1)$, so we have that $G(p,q)$ is the Hermitian
symmetric space
$$G(p,q)=\frac{\rSU(p+q)}{\rSU(p)\times \rSU(q)\times \rU(1)}.$$
$G(p,q)$ is a K{\"a}hler manifold and we are going to show that  it is in
fact a K{\"a}hler-Hodge manifold.

\bigskip

An open cover of $G(p,q)$ is given by the open sets with some fixed minor
of order $p$ of $Z$ different from zero. Notice that this property is not
changed by the right action of $\rGL(p,\C)$, so it is well defined on the
equivalence classes For concreteness, let us fix
$$U_0=\left\{Z=\begin{pmatrix}Z_0\\Z_1\end{pmatrix}\in G(p,q)\;|\; \det Z_0\neq
0\right\}.$$ A $p$-plane in $U_0$ can be characterized by a $q\times p$
matrix $T$ such that a vector $(z_1,\dots z_p,z_{p+1},\dots ,z_{p+q})$
satisfies
$$\begin{pmatrix}z_{p+1}\\\vdots \\z_{p+q}\end{pmatrix}=T\begin{pmatrix}z_{1}\\\vdots
\\z_{p}\end{pmatrix}.$$ In fact, a matrix $Z$ with $\det Z_0\neq
0$ is a collection of $p$ column vectors satisfying the above property,
so
$$Z_1=TZ_0\quad  \Leftrightarrow\quad T=Z_1Z_0^{-1}.$$
An arbitrary matrix $T$ defines a $p$-plane in $U_0$, so  we have
$$U_0\approx \cM_{q\times p}(\C)\approx\C^{pq},$$
and the entries of $T$ are local coordinates on $U_0$.

The tautological bundle\index{tautological bundle} $H\rightarrow G(p,q)$,
is the vector bundle with the fiber at each point of $G(p,q)$ the plane
that it represents. It is a rank $p$ subbundle of the trivial bundle
$G(p,q)\times \C^{p+q}$. It is a bundle associated to the principal
bundle $\tilde \cM$.

On the trivial bundle there is a  fiber metric
\begin{equation}\langle \zeta,\zeta'\rangle=\zeta^1\bar
\zeta'^1+\cdots +\zeta^{p+q}
\bar\zeta'^{p+q}\label{fibermetric}\end{equation} for $\zeta,\zeta'$
vectors at a point in $G(p,q)$. It induces a fiber metric on the
tautological bundle. A local section on $H$ is given by  functions
$\zeta^1,\dots \zeta^p$, so that $T$ determines the plane:
\begin{equation}\zeta(T)=\begin{pmatrix}\zeta^1\\\vdots\\\zeta^p\\\zeta^{p+1}\\\vdots
\\\zeta^{p+q}\end{pmatrix}
=\begin{pmatrix}\id\\T\end{pmatrix}\begin{pmatrix}\zeta^1\\\vdots\\\zeta^p\end{pmatrix}.\label{sectionzeta}\end{equation}
The Hermitian inner product on the fiber is
\begin{eqnarray*}
 \langle \zeta(T),\zeta'(T)\rangle & = & (\bar\zeta'^1,\dots, \bar\zeta'^p)
(\id,
T^\dag)\begin{pmatrix}\id\\T\end{pmatrix}\begin{pmatrix}\zeta^1\\\vdots\\\zeta^p\end{pmatrix} \\
   & = & (\bar\zeta'^1,\dots, \bar\zeta'^p) (\id+ T^\dag
T)\begin{pmatrix}\zeta_1\\\vdots\\\zeta_p\end{pmatrix}.
\end{eqnarray*}
If $\alpha,\beta=1,\dots p$ then we have the fiber metric
\begin{equation}\langle \zeta(T),\zeta'(T)\rangle=h_{\alpha\bar \beta}
\zeta^\alpha\bar \zeta'^\beta, \qquad h^\mathrm{t}=\id+ T^\dag
T.\label{hmetric}
\end{equation}
We can write the Hermitian fiber metric as
$$h=h_{\alpha\bar \beta}\rmd z^\alpha\rmd \bar z^\beta.$$
We consider now the line bundle $\Lambda^p(H)$ with fiber at a point
$x\in G(p,q)$ $\Lambda^pH_x\approx\Lambda^p\C^p\approx\C,$ i.e. the
determinant. The structure group is
$$\rGL(p,\C)/\rSL(p,\C)\approx\C^\times .$$ Let $\{s_\alpha(T)\}$ be a local
frame on $H$, with $h_{\alpha\bar \beta}=\langle
s_\alpha,s_\beta\rangle$. (To compare with (\ref{hmetric}) it is enough
to take $s_\alpha(T)=\zeta(T)$ as in (\ref{sectionzeta}) with
$\zeta^\alpha=1$ and the rest 0). Then a local section on $\Lambda^p(H)$
is of the form
$$U(T)=u(T) \, s_1\wedge\cdots \wedge s_p.$$ There is an induced fiber
metric on this bundle given by
$$H=\det (h_{\alpha\bar \beta})\rmd u\rmd \bar u.$$
As in (\ref{ricciform}) we get for the Ricci form associated to the
Hermitian connection on the line bundle
\begin{equation}
\rho _{i\jb }=-\rmi\partial_{\jb }\partial_i\log \det
(h_{\alpha\bar\beta})= -\rmi\partial_{\jb }\partial_i\log \det (\id+
T^\dag T),
 \label{riccigrass}
\end{equation}
where $i,j=1,\dots pq$ run over all the entries of the matrix $T$.
\hfill$\square$

\bigskip

\begin{example}Non compact ``Grassmannian" as a K{\"a}hler-Hodge manifold.\end{example}
In the example above, let us change the fiber metric (\ref{fibermetric})
to a pseudo-Euclidean one  with signature $(p,q)$,
\begin{equation*}\langle \zeta,\zeta'\rangle=\zeta^1\bar
\zeta'^1+\cdots +\zeta^p\bar \zeta'^p-\zeta^{p+1}\bar \zeta'^{p+1}-\cdots
-\zeta^{p+q} \bar\zeta'^{p+q}.\end{equation*} Then, instead of
(\ref{hmetric}) we have
$$h^\mathrm{t}=\id- T^\dag T,$$
so on the points where the matrix $\id- T^\dag T$ is positive definite we
have a positive definite, non degenerate fiber metric. The space of
matrices satisfying this property is a domain in $\C^{pq}$. It is the
Hermitian symmetric space\index{Hermitian symmetric space}
$$D(p,q)=\frac{\rSU(p,q)}{\rSU(p)\times
\rSU(q)\times \rU(1)}.$$ The corresponding expression for the Ricci
curvature is proportional, as before, to the standard K{\"a}hler metric on
this symmetric space. \hfill$\square$ \index{projective K{\"a}hler
manifold|)}
\section{Conformal calculus\label{ss:confcal}}\index{conformal calculus|(}

The ideas described in the previous section originate in physics as a
property of certain sigma models of scalar fields coupled to gravity with
a scaling symmetry. It is in fact a simplification of what occurs in
supergravity (see for example Refs.
\cite{Kallosh:2000ve,VanProeyen:2001wr}), but the essential idea can be
grasped in this simplification. We first consider the version with real
scalars and then we move to K{\"a}hler manifolds.
\subsection{Real manifold.\label{ss:realcase}}
We consider a nonlinear sigma model of $n$ real scalar fields $\phi^I$
with lagrangian
$$
\mathcal{L}_{\mathbb{R},0}=-\ft12\sqrt{g}g^{\mu\nu}G_{IJ}\partial_\mu
\phi^I
\partial_\nu \phi^J .
$$
Here $g_{\mu \nu }$ is the metric of space time, gravitational field,
$g^{\mu\nu}$ its inverse, and $g=|\det g_{\mu\nu}|$. The target space is
a real  Riemannian manifold with coordinates $\phi^I$ and $G_{IJ}(\phi)$
is the Riemannian metric.

We will be interested in the case that the Lagrangian has a dilatation
symmetry\index{dilatation symmetry} given at the infinitesimal level by a
vector $K=K^I\partial_I$. Let $D$ be the dimension of space-time. We
assume that the vector $K$ is a homothetic Killing vector of the metric
$G_{IJ}$,
$$\cL_KG_{IJ}=K^L\partial_LG_{IJ}+\partial_IK^LG_{LJ}+\partial_JK^LG_{IL}=cG_{IJ}.$$
We fix \begin{equation}
 c=D-2.
 \label{cisDmin2}\end{equation} Then, the Lagrangian $\cL_{\R,0}$ is invariant
under the  infinitesimal transformations
\begin{eqnarray}
\delta_\epsilon \phi^I&=&\epsilon K^I,\nonumber \\
\delta_\epsilon g_{\mu\nu}&=&-2\epsilon
g_{\mu\nu},\label{changerealinf}\end{eqnarray} for an infinitesimal
parameter $\epsilon$ independent of de point $x$ in spacetime.

A simple example is when the metric on the target space is such that
\begin{equation}G_{IJ}(\lambda\phi)=\lambda^{D-4}G_{IJ}(\phi),\label{metrichom}\end{equation} which means
that the vector $K=\phi^I\partial_I$ is a homothetic Killing vector
(\ref{homotheticK}) with $c=D-2$. Then the lagrangian is invariant under
the set of transformations
\begin{eqnarray}
\phi^I & \longmapsto & \lambda \phi^I,\nonumber\\
\label{changereal} g_{\mu\nu}&\longmapsto & \lambda^{-2} g_{\mu\nu}
\qquad \lambda\in \R,
\end{eqnarray}
for a constant parameter $\lambda$ (independent of $x$). We obtain the
infinitesimal transformations  (\ref{changerealinf}) by writing $\lambda
=1+\epsilon+{\cal O}(\epsilon ^2)$.

Let us now consider transformations of the form (\ref{changerealinf})  but with a parameter $\epsilon(x)$ depending
on the point.  Then
$$\delta_\epsilon \cL=-\sqrt{g}G_{IJ}(\partial_\mu \epsilon ) K^I
\partial_\nu \phi^J g^{\mu\nu}\,.$$ The lagrangian $\cL_{\R, 0}$ is not
invariant, but this can be remedied if  we add an Einstein-Hilbert like
term for the spacetime metric
\begin{equation}
\cL_{\R}=-\ft12\sqrt{g}G_{IJ}\partial_\mu \phi^I
\partial_\nu \phi^J g^{\mu\nu} -\ft12a(G_{IJ}K^IK^J)\sqrt{g}R(g),
\label{lagrangianreal}
\end{equation}
where $R(g)$ is the spacetime curvature and
\begin{equation}
 a=\frac{1}{(D-1)(D-2)}.
 \label{valueaD}
\end{equation}
In addition, we ought to assume that $K$ is a \emph{closed} homothetic
Killing vector (\ref{closedhomothetic1}). In this case we have that
$$\delta_\epsilon\left(\sqrt{g}R(g)\right)=\epsilon(2-D)\sqrt{g}R(g) + 2(D-1)\partial _\nu  \bigl( \sqrt{g}g^{\mu \nu } \partial _\mu \epsilon\bigr),
$$
and using the identity
$$ G_{IJ} K^I \partial_\nu \phi^J= \frac{1}{D-2}\partial_\nu\left(G_{IJ} K^I
  K^J\right),$$
one can prove that $\cL_{\R}$ transforms into a total derivative.

To have a positive definite energy for the gravitational field we must
have $G_{IJ}K^IK^J< 0$, so  one of the scalars is a ghost.

One can fix this gauge invariance by taking
\begin{equation}G_{IJ}K^IK^J=-\frac{1}{a\kappa ^2},\label{gaugefixing}\end{equation}
where $\kappa ^2$ is the gravitational coupling constant. Then the second
term of (\ref{lagrangianreal}) is just the Einstein-Hilbert action and
the first term is a sigma model defined now on the surface
(\ref{gaugefixing}).

\subsection{K{\"a}hler manifolds}
We can consider the same kind of model for $n+1$ complex scalar fields
$X^\alpha$, which are coordinates in a K{\"a}hler manifold with metric
$G_{\alpha \bar \beta }$. We assume now that this metric has a closed
homothetic Killing vector $K$. As mentioned at the end of section
\ref{ss:AffineHomoth}, this implies the presence of the holomorphic
homothetic Killing vector $H=(1-\rmi J)K$.  The lagrangian density has
the form
\begin{equation}
\mathcal{L}_{\C,0}=-\sqrt{g}g^{\mu \nu }G_{\alpha \bar \beta}\partial_\mu
X^\alpha
\partial_\nu \bar X^{\bar \beta}-\ft14aG_{\alpha\bar \beta}H^\alpha \bar
H^{\bar\beta} \sqrt{g}R(g).\label{lagrangiancomplex}
\end{equation}
The dilatation symmetry is generated by
\begin{equation}
  K= \ft12\left( H^\alpha (z)\partial _\alpha +\bar H^{\bar \alpha }(\bar z)\partial _{\bar \alpha
  }\right),
 \label{Kconform}
\end{equation}
but this model has rigid symmetry generated by $JK$, which was not
present in the real case. This leads to the infinitesimal transformations
\begin{eqnarray}
 \delta _\epsilon X^\alpha & =&\ft12H^\alpha \epsilon,\qquad \delta _\varphi X^\alpha= \ft12\rmi H^\alpha\varphi , \nonumber\\
\delta _\epsilon g_{\mu\nu} & = & -2\epsilon g_{\mu\nu}.
\label{changecomplexinf}
\end{eqnarray}
Assuming (\ref{cisDmin2}), the action is invariant under these
transformations where $\epsilon $ can be local, but $\varphi $ is still a
global transformation parameter.

A relevant example is the finite transformation
\begin{eqnarray}
X^\alpha & \longmapsto & \lambda^{c/2} X^\alpha,\nonumber\\
g_{\mu\nu} & \longmapsto & |\lambda|^{-2} g_{\mu\nu}\qquad
\lambda=|\lambda|\rme^{\rmi\varphi}=1+\epsilon +\rmi\varphi +\ldots \in
\C, \label{changecomplex}
\end{eqnarray}
for which
\begin{equation}
H^\alpha =cX^\alpha,\qquad X^\gamma\partial_\gamma G_{\alpha\bar
\beta}=0.
\label{lieder}
\end{equation}
In that case, the transformations (\ref{changecomplexinf}) can be
integrated to a finite transformation. In order to implement the local
invariance under $\varphi$, we introduce a $\rU(1)$ connection $A_\mu$,
which transforms as
\[A_\mu \longmapsto A_\mu +\partial _\mu \varphi,\]
and we couple it minimally to the scalar fields defining
\begin{eqnarray}
 \mathcal{L}_\C&=&
-\sqrt{g}g^{\mu \nu }G_{\alpha \bar \beta}D_\mu X^\alpha D_\nu \bar
X^{\bar \beta}-\ft14aG_{\alpha\bar \beta}H^\alpha \bar
H^{\bar\beta} \sqrt{g}R(g).\nonumber\\
&& D_\mu X^\alpha =\partial_\mu X^\alpha-\ft12\rmi A_\mu H^\alpha ,
\label{LagrKahler}
\end{eqnarray}
For shortness, we denote
$$N=\frac{1}{c^2}G_{\alpha\bar \beta}H^\alpha \bar H^{\bar\beta}. $$
Using (\ref{holohomoki}) for the Levi-Civita connection, we have also
that
$$\partial _\alpha N= \frac{1}{c}G_{\alpha \bar \beta }\bar H^{\bar \beta },\qquad
\partial _\alpha \partial _{\bar \beta }N=G_{\alpha\bar \beta}.$$
Hence, $N$ is the K{\"a}hler potential of the manifold described by the
$X^\alpha $.

The field equation for the auxiliary field $A_\mu$ is algebraic and it
allows us to solve for $A_\mu $:
$$A_\mu=\frac
{\rmi}{c^2N}G_{\alpha\bar\beta}\left(H^\alpha\partial_\mu\bar
X^{\bar\beta}-
\partial_\mu X^\alpha\bar H^{\bar\beta}\right)=\frac{\rmi}{c\,N}\left(\partial_\mu\bar X^{\bar\beta}
\partial _{\bar\beta } N-\partial_\mu X^\alpha\partial _\alpha N\right).  $$
The first term of (\ref{LagrKahler}), the scalar kinetic term
$\cL_{\mathrm{scalar}}$, is then
\begin{eqnarray}
\frac{\cL_{\mathrm{scalar}}}{\sqrt{g}}&=&-
G_{\alpha\bar\beta}\partial_\mu X^\alpha\partial^\mu\bar X^{\bar\beta}
-\frac {1}{4N}\left[\partial_\mu\bar X^{\bar\beta}
\partial _{\bar\beta }N -\partial_\mu X^\alpha\partial _\alpha  N\right]^2\nonumber\\
&=& -\partial_\mu X^\alpha\partial^\mu\bar X^{\bar\beta}\left[\partial
_\alpha\partial _{\bar\beta}N-\frac{1}{N}(\partial _\alpha N)(\partial
_{\bar \beta }N)\right] \nonumber\\&& -\frac {1}{4N}\left[
\partial _\mu N\,\partial ^\mu N\right]\nonumber\\
&=&-N\partial_\mu X^\alpha\partial^\mu\bar X^{\bar\beta}\partial _\alpha
\bar \partial _{\bar \beta }\log |N|-\frac {1}{4N}\left[
\partial _\mu N\,\partial ^\mu N\right].\label{lagcom}
\end{eqnarray}
Notice that $H^\alpha $ is a zero mode of the quantity in square brackets
in the second line.

We can fix the dilation gauge freedom (\ref{changecomplex}) by taking as
before
\begin{equation}
N=\frac{1}{c^2}G_{\alpha\bar\beta}H^\alpha \bar H^{\bar
\beta}=-\frac{2}{ac^2\kappa ^2}. \label{gaugefixing2}
\end{equation}
The second term of (\ref{LagrKahler}) is then the Einstein-Hilbert
action.

\medskip

As $N$ gets a fixed value, a function of $N$ is not convenient as a
K{\"a}hler potential for the restricted manifold. We will show now how to
construct a K{\"a}hler potential, restricting to the case (\ref{lieder}).

In that case, we rescale the coordinates $X^\alpha $, introducing $\eta
^\alpha $ by
%\begin{equation}
\begin{eqnarray*}\Phi_Y:\tilde \cM&\longrightarrow &\tilde \cM\\
  \eta^\alpha &\longrightarrow &X^\alpha =\eta ^\alpha Y(\eta ,\bar\eta ),\end{eqnarray*}
% \label{Xeta}
%\end{equation}
for an arbitrary function $Y(\eta ,\bar\eta )$. Notice that this map is
not holomorphic with respect to $J$. However, it induces a new complex
structure on $\tilde \cM$, denoted as $J'$, by the commutativity of the
diagram
\[\begin{CD}T\tilde \cM@>T\Phi_Y>>T\tilde \cM\\ @ VJ'VV  @ VVJV\\ T\tilde \cM@>>T\Phi_Y>T\tilde \cM\end{CD}\]
The map $T\Phi_Y$ then sends holomorphic vectors with respect to $J'$ to
holomorphic vectors with respect to $J$. In this sense, it is a
holomorphic map.

Defining
\[
  {\cal K}= \frac{N}{Y\bar Y},
\]
then ${\cal K}$ is a function of $(\eta ,\bar \eta )$.
 The homogeneity properties of $N$ imply
$$\cK(\lambda\eta,\bar\lambda\bar\eta)=\lambda\bar\lambda\cK(\eta,\bar\eta),$$
and therefore also
%\begin{equation}
\[
  \frac{\partial }{\partial X^\alpha }N=\bar Y \frac{\partial }{\partial \eta ^\alpha }{\cal
  K}, \qquad
  \frac{\partial^2 }{\partial X^\alpha \partial \bar X^{\bar \beta}}N=
  \frac{\partial^2 }{\partial \eta ^\alpha \partial \bar \eta ^{\bar \beta}}{\cal
  K}.
\]
Hence, $c\eta ^\alpha\frac{\partial}{\partial \eta ^\alpha}$ is a
holomorphic homothetic Killing vector with respect to $J'$, and ${\cal
K}$ defines the K{\"a}hler potential of a projective K{\"a}hler manifold, see
Definition \ref{psdef}.

 The action reduces to
\begin{equation}
\frac{\mathcal{L}_\C}{\sqrt{g}}=
 -N\partial _\mu \eta ^\alpha \partial ^\mu \bar \eta ^{\bar \beta }\frac{\partial^2 }{\partial \eta ^\alpha \partial \bar \eta ^{\bar \beta}}
 \log|{\cal K}|-\frac {1}{4N}\left[\partial _\mu N\,\partial ^\mu
 N\right]-\frac14 ac^2N R(g).
 \label{LCineta}
\end{equation}

The first term in (\ref{LCineta}) is proportional to $-\tilde
\rho_{\alpha\bar\beta}$ in (\ref{exact}), which is the pull back of a
2-form on the quotient manifold $\tilde\rho=\rmi \pi^*\rho$ as in
(\ref{projexact}). If $z^i$, $i=1,\dots n$ are coordinates on the
quotient, then similarly as in (\ref{defgprojK}), a metric is defined.
The appropriate normalization for the K{\"a}hler potential is
%\begin{equation}
\[
g_{i\jb}=\partial _i\partial _{\jb}\left[   -\frac{2}{ac^2\kappa ^2}\log
-\frac{ac^2}{2}{\cal K}\right] .
 \]
% \label{KcalK}
%\end{equation}
On the quotient $N$ is constant and thus $\partial _\mu  N=0$, so the
action reduces to a sigma model in dimension $n$ coupled to gravity in
the standard way,
$$\cL_\C=-\sqrt{g}g_{i\jb }\partial_\mu
z^i{\partial}^\mu\bar z^{j}+\frac{1}{\kappa ^2}\sqrt{g}R(g).$$

Note that the $D=4$ values of (\ref{cisDmin2}) and (\ref{valueaD}) lead
to $ac^2/2=1/3$. That is also the value that one finds in $N=1$
supergravity. For $N=2$ supergravity, one has two scalar manifolds, the
one of the vector multiplets\index{vector multiplet}, and the one of
hypermultiplets\index{hypermultiplet}. There is another auxiliary field,
whose origin is beyond our discussion here, such that when one eliminates
the hypermultiplets, the effective value of $a$ is $1/2$, i.e.\
$ac^2/2=1$.

We remark that we need here the lower signs in (\ref{defgprojK}) in order
to get the positive kinetic energy for gravity, and the other signatures
should all be $+$ in order to have positive kinetic terms of the sigma
model.

\index{conformal calculus|)}

\section{Projective special K{\"a}hler manifolds \label{ss:psk}}
\index{projective special K{\"a}hler manifold|(}
\subsection{Definition of projective special K{\"a}hler manifolds}\index{projective special K{\"a}hler manifold!definition}

A projective special K{\"a}hler manifold $\cM$ is a K{\"a}hler-Hodge manifold
such that the manifold $\tilde\cM$ is rigid special K{\"a}hler. The flat
connection on $\tilde \cM$ is an extra structure that also projects to
$\cM$. Here we have the precise definition.

\begin{definition}\label{defpsk}
Let $\tilde\cM$ be a rigid special pseudo-K{\"a}hler manifold with complex
structure $\tilde J$, metric $\tilde g$, K{\"a}hler form $\tilde \Phi$ and
flat symplectic connection $\tilde \nabla$. We assume that on $\tilde\cM$
there is a free holomorphic action of $\C^\times$ such that the
fundamental vector field $H$ is a non null, holomorphic homothetic
Killing vector for the flat connection,
\begin{eqnarray}
&&\tilde \nabla_YH=cY\qquad \forall Y\in
T^{1,0}\tilde\cM\label{psk1}\\
&&\hbox{and}\quad  \tilde g(H,\bar H)\neq 0.\label{psk2}\nonumber
\end{eqnarray}
 Then we say that on $\cM=\tilde\cM/\C^\times$ there is a projective special K{\"a}hler structure.
\hfill$\square$
\end{definition}

In fact, we will prove that (\ref{psk1}) implies that $H$ is also an
holomorphic homothetic Killing vector for the Levi-Civita connection,
that is,
\begin{equation}\tilde D_YH=cY\quad\hbox{which is equivalent to}\quad \cL_Y\tilde g=c\tilde g, \qquad \forall Y\in
T^{1,0}\tilde\cM, \label{psk3}
\end{equation}
so it is enough to require it for the flat connection. Let us look at
(\ref{psk1}) in special coordinates. From (\ref{flatderivative}) we have
$$\tilde\nabla H=\frac{\partial H^\beta}{\partial
\eta^\alpha}\rmd \eta^\alpha\otimes \frac{\partial}{\partial \eta^\beta}
-\frac 12 \frac {\partial
\tau_{\beta\gamma}}{\partial\eta^\alpha}H^\gamma\rmd
\eta^\alpha\otimes\frac{\partial}{\partial y_\beta}=c\,\rmd
\eta^\alpha\otimes \frac{\partial}{\partial \eta^\alpha}.$$
 The first term is holomorphic, while in the second there is an
holomorphic and an antiholomorphic part, since $y_\beta$ is real. To
cancel the antiholomorphic part necessarily
\begin{equation}
\frac{\partial\tau_{\beta\gamma}}{\partial\eta^\alpha}H^\gamma=0,
\label{homogeneous1}
\end{equation}
and then
\begin{equation}
\frac{\partial H^\beta}{\partial \eta^\alpha}=c \delta_\alpha^\beta\quad
\Leftrightarrow\quad H^\beta=c\eta ^\beta \hbox{ (up to a constant)}.
\label{homogeneous2}
\end{equation}
We can always shift $\eta^\alpha$ by a constant, it is still a special
coordinate. In particular, (\ref{homogeneous2}) implies that special
coordinates are homogeneous coordinates as defined in (\ref{coorhom}).
(\ref{homogeneous1}) and (\ref{homogeneous2}) imply that
$\tau_{\alpha\beta}$ are homogeneous functions of $\eta$ of degree 0,
$$\tau_{\alpha\beta}(\lambda
\eta)=\tau_{\alpha\beta}(\eta),\qquad \lambda\in \C^\times,$$ so they depend on the
prepotential $\cF$ as in (\ref{lambdataurigid}):
$$\tau_{\alpha\beta}=-8\frac
{\partial^2\cF}{\partial\eta^\alpha\partial\eta^\beta},$$ must be an
homogeneous function of $\eta$ of degree 2,
\begin{equation}
  \cF(\lambda
\eta)=\lambda^2\cF(\eta),\qquad \lambda\in \C^\times.
 \label{homogF}
\end{equation}

For the Levi-Civita connection, we have
$$\DLC H=\frac{\partial
H^\beta}{\partial \eta^\alpha}\rmd \eta^\alpha\otimes
\frac{\partial}{\partial \eta^\beta} +\frac 1{8}\rmi g^{\beta\bar \delta}
\frac {\partial \tau_{\delta\gamma}}{\partial\eta^\alpha}H^\gamma\rmd
\eta^\alpha\otimes\frac{\partial}{\partial \eta_\beta},
 $$
and using (\ref{homogeneous1}) and (\ref{homogeneous2}) we get
$$\DLC H=c\,\rmd
\eta^\alpha\otimes \frac{\partial}{\partial \eta^\alpha},$$ which proves
(\ref{psk3}).

\bigskip

Using (\ref{homogeneous2}) we can compute the integral surfaces
(\ref{integralcurve}) of $H$ in special coordinates,
$$H(\rho_\lambda(x))=\frac {\rmd \rho_\lambda}{\rmd
\lambda}\;\Leftrightarrow\;c\rho^\alpha_\lambda(x)=\frac {\rmd
\rho^\alpha_\lambda}{\rmd
\lambda}\;\Leftrightarrow\;\rho^\alpha_\lambda(x)={\rm
e}^{c\lambda}\eta^\alpha(x),$$ since $\rho_0(x)=x$ and
$\rho^\alpha_0(x)=\eta^\alpha(x)$.

We will denote also by $\mathfrak{R}_b(x)=\rho_\lambda(x)$ with
$b=\rme^{c\lambda}\in \C^\times$.

As we saw in general in (\ref{theta}), $\theta =c^2\cK$.

Since $\nabla$ is flat, it descends to $\cH$ as a flat connection, and
then it defines a connection on $L\otimes T\cM$ as in (\ref{projconn}).
This connection is not necessarily flat.

Next we will see that also the holomorphic cubic form descends to an
appropriate bundle over $\cM$, and we will compute  the curvature tensor
in terms of it.

\paragraph{The holomorphic cubic form\index{holomorphic cubic form}.}
We consider the holomorphic cubic form $\Xi$ defined in section
\ref{cubic}. We want to see how it descends to the manifold $\cM$. If $X$
is a vector field on $\cM$, its horizontal lift is (\ref{horlift})
\begin{equation}
  \tilde
  X=X^i\partial_i-y\kappa_iX^i\partial_y=X^i(\partial_i\eta^\alpha-\kappa_i\eta^\alpha)\partial_\alpha,
 \label{holliftX}
\end{equation}
%
%
%In special coordinates we have
%$$\mathfrak{R}_b^\alpha(\tilde m)=b\eta^\alpha, \qquad \rmd\mathfrak{R}_b(X)=bX=bX^\alpha\frac {\partial}{\partial\eta^\alpha},\quad X\in
%T\tilde \cM.$$ Then, using (\ref{xicom})
%$$\Xi_{\mathfrak{R}_b(\tilde
%m)}(\rmd\mathfrak{R}_bX,\rmd\mathfrak{R}_bY,\rmd\mathfrak{R}_bZ)=-8\frac{\partial^3\cF}{\partial\eta^\alpha\partial\eta^\beta\partial\eta^\gamma}
%(b\eta)\left(b^3X^\alpha Y^\beta Z^\gamma\right).$$ But due to the
%homogeneity property of $\cF$ we have that
%$$-8\frac{\partial^3\cF}{\partial\eta^\alpha\partial\eta^\beta\partial\eta^\gamma}
%(b\eta)=-8b^{-1}\frac{\partial^3\cF}{\partial\eta^\alpha\partial\eta^\beta\partial\eta^\gamma}
%(\eta),$$ so
%$$\Xi_{\mathfrak{R}_b(\tilde
%m)}(\rmd\mathfrak{R}_bX,\rmd\mathfrak{R}_bY,\rmd\mathfrak{R}_bZ)=b^2\Xi_{\tilde
%m}(X,Y,Z).$$ this means that $\Xi$ is cubic form with values in
%$(L^*)^{\otimes 2}$.\hfill$\blacksquare$
so
$$\frac{\partial\cF}{\partial\eta^\alpha}\rmd
\eta^\alpha(\tilde
X)=X^i(\partial_i\cF-2\kappa_i\cF)=X^iD_i^L(\cF)=D^L_X\cF,$$ where we
have used the fact that $\cF$ is homogeneous of degree 2. It is in fact a
section of $(L^*)^{\otimes 2}$, and $D^L$ denotes the covariant
derivative with respect to the Hermitian connection. We can also write
$$D^L_i\cF=\rme^{2\log|\cK|}\partial_i(\rme^{-2\log|\cK|}\cF)=
\cK^2\partial_i(\frac1{\cK^2}\cF).$$

If $Y,Z$ are also vector fields on $\cM$ and $\tilde Y$, $\tilde Z$ are
their horizontal lifts respectively, we have
\[
 \frac{\partial^3\cF}{\partial\eta^\alpha\partial\eta^\beta\partial\eta^\gamma}\rmd
\eta^\alpha\rmd \eta^\beta\rmd \eta^\gamma(\tilde X,\tilde Y,\tilde Z) =
D^L_XD^L_YD^L_Z(\cF).
\]
This shows that the holomorphic cubic form on $\tilde\cM$, $\tilde \Xi$
descends to a section of $(L^*)^{\otimes2}\otimes (T^*\cM)^{\otimes 3}$,
\begin{equation}
  \Xi(X,Y,Z)\equiv\tilde \Xi(\tilde X,\tilde Y,\tilde Z).
 \label{XitildeXi}
\end{equation}

 \hfill$\blacksquare$

This procedure of lifting the vector fields on $\cM$ to $\tilde\cM$ gives
us also another way of computing the metric on $\cM$.

\paragraph{The metric.}
As before, let $\tilde X$ and $\tilde Y$ be the horizontal lifts of $X$
and $Y$, vector fields on $\cM$. Then we have, using (\ref{holliftX}),
(\ref{HonKidentities}) and (\ref{defgprojK}),
\begin{eqnarray}\tilde g(\tilde X,\bar{\tilde Y})&=&\partial_\alpha\partial_{\bar
\beta}\cK(\partial_i\eta^\alpha-\kappa_i\eta^\alpha)(\partial_j\bar\eta^\beta-\bar
\kappa_j\bar\eta^{\bar\beta})X^i\bar
Y^j\nonumber\\
&=&\cK\partial_i\partial_{\jb }(\log|\cK|) X^i\bar Y^j=|\cK| g(X,\bar Y).
\label{smallmetric}
\end{eqnarray}
 \hfill$\blacksquare$

\paragraph{The Riemannian curvature\index{projective special K{\"a}hler manifold!curvature} on $\cM$.} From
(\ref{projconn}) we can compute the curvature tensor of the Levi-Civita
connection on $\cM$. Since $\tilde\cM$ and $\cM$ are K{\"a}hler manifolds we
can use (\ref{curvkm}). We have
$$R^j{}_{ik\bar \ell}=-\partial_{\bar \ell} \Gamma_{ik}^j=\tilde R^j{}_{ik\bar
\ell}+\partial_{\bar\ell}(
\delta_i^j\partial_k\log|\cK|+\delta_k^j\partial_i\log|\cK|).$$
 These are the components of the curvature tensor in the coordinates
$z^i$. To avoid confusion, we will split the coordinates $\eta^\alpha$ as
$(\eta^0,\eta^a)$, $a=1,\dots n$. In this way the indices $i,j,k$ will
always refer to the coordinates $z$.

 We can use (\ref{curvaturecom}) to express $\tilde R^j{}_{ik\bar
\ell}$ in terms of the cubic form ${\tilde\Xi}$. We first have to write
${\tilde\Xi}$ in terms of $(y, z^i)$. We have
\begin{equation}
  \rmd\eta^0=\rmd y,\qquad \rmd\eta^i=z^i\rmd y+y\rmd z^i.
 \label{deta0i}
\end{equation}
Due to the homogeneity condition (\ref{homogF}) and $\tilde \Xi $ being a
third derivative of $\cF$ as in (\ref{xicom}), we have
\begin{equation}
  \eta ^\alpha \tilde \Xi _{\alpha \beta \gamma }=0,
 \label{homogtildeXi}
\end{equation}
and therefore the $\rmd y$ terms in (\ref{deta0i}) do not contribute if
we rewrite
$$\tilde \Xi= \tilde \Xi_{\alpha\beta\gamma}\rmd
 \eta^\alpha\rmd\eta^\beta\rmd \eta^\gamma=
y^3\tilde\Xi_{abc}\delta^a_i\delta^b_j\delta^c_k
 \rmd z^i\rmd z^j\rmd z^k=\Xi_{ijk}\rmd z^i\rmd z^j\rmd z^k,$$
where $\Xi_{ijk}$ has been defined in (\ref{XitildeXi}). This leads to
$$\Xi_{ijk}=2y^2\frac{\partial^3 \cF(1,z^i)}{\partial z^i\partial
 z^j\partial z^k},$$
where $\cF(1,z^i)$ is $\cF(\eta )$ with $\eta ^0$ replaced by 1, and
$\eta ^i$ by $z^i$.

On the other hand, using (\ref{tildegg}), we find
\begin{equation}R^j{}_{ik\bar\ell}=\frac 1{4\cK^2} { g}^{j\jb '}{ g}^{p'\bar
p}{\Xi}_{p'ki}\bar{ \Xi}_{\bar p\jb '\bar\ell}\pm
\delta_i^jg_{k\bar\ell}\pm
\delta_k^jg_{i\bar\ell}.\label{curvedcurvature}
\end{equation}
Notice that in (\ref{curvedcurvature}) all the the dependence in $y, \bar
y$ cancels as ${\cal K}$ is proportional to $y\bar y$.

\hfill$\blacksquare$

\subsection{Examples of projective special K{\"a}hler manifolds}

\begin{example} Projective space and unit ball\index{unit ball} as special K{\"a}hler manifolds.\end{example}
We consider the complex projective space\index{complex projective space}
$\CP^m$ of lines in the complex space $\C^{m+1}$.  It is a special case
of Example \ref{grassmannian}, with $p=1$, $q=m$.

%$\CP^m$ is also seen as the set of equivalence classes of points in
%$\C^{m+1}-\{0\}$ under the equivalence relation
%\begin{equation}(z^0,\cdots, z^{m})\sim (\lambda z^0,\cdots, \lambda
%z^{m}),\qquad \lambda\in \C^\times.\label{equivrel}\end{equation} To
%the equivalence class of a non zero vector, $[(z^0,\cdots, z^{m})]$,
%it corresponds the vector space $\{\lambda(z^0,\cdots,
%z^{m}),\;\lambda\in \C\}$. The equivalence class is obtained from
%the vector space by deleting the point $0$.
%
%
%$\begin{CD}(\C^{m+1}-\{0\})@>\pi>>\CP^m\end{CD}$ is a principal
%bundle over $\CP^m$ with fiber $\C^\times$.
%
%\bigskip

We have a covering  of $\CP^m$ by open sets
$$U_i=\{\hbox{Lines $S$ in $\C^{m+1}$ with } z^i|_S\neq 0\},$$
(these are the lines that do not lie in the hyperplane $z^i=0$). Let us
take a fixed index $i=0$, then we have that
$$z^j|_S=t^{j} z^0|_S,\qquad j\neq 0,$$ so $(t^1,\dots, t^m)$ is a set of coordinates on $U_0$.

The tautological bundle is already a line bundle, $L\subset \CP^m\times
\C^{m+1}$ so there is no need of taking the determinant. On $\CP^m\times
\C^{m+1}$ we have the fiber metric
\begin{equation}\langle\zeta,\zeta'\rangle=\zeta^0\bar\zeta^0+\cdots+
\zeta^m\bar\zeta^m,\label{projfibermetric}
\end{equation}
which we will
restrict to $L$. On $L$ the fiber metric and the Hermitian connection are
$$h(t,\bar t)=1+\sum_{j=1}^mt^j\bar t^j,\qquad\Gamma_i=\left(1+\sum_{j=1}^mt^j\bar t^j\right)^{-1}\bar t^i.$$ The
Ricci form (\ref{riccigrass}) becomes
$$\rho _{i\jb }=-\rmi g_{i\jb }=-\rmi\partial_{\jb }\partial_i\log h=
-\rmi\partial_{\jb }\partial_i\log (1+ t{\bar t})=  \frac {\rmi}{h^2}\bar
t^it^j-\frac {\rmi} h\delta^{ij}.$$ We can define the prepotential as
$$\cF=\frac{1}{4}\rmi (\eta^0 \eta^0+\cdots+\eta^m \eta^m),\qquad \eta^i=
 t^i\eta^0,\quad i=1,\dots m.$$
Then the K{\"a}hler potential in $\tilde\cM=\C^{m+1}$ and in $\CP^m$ is
$$\cK=2\Im\left(\frac{\partial\cF}{\partial\eta}\bar\eta\right)=
 \eta\bar\eta;\qquad \log|\cK|=
 \log(1+t\bar t)+\log \eta^0\bar\eta^0.$$
As the third derivative of the prepotential vanishes, the curvature is
given by the last two terms of (\ref{curvedcurvature}), where we have to
use the $+$ signs.

\bigskip

If we change the fiber metric (\ref{projfibermetric}) to
$$\langle\zeta,\zeta'\rangle=\zeta^0\bar\zeta^0-\cdots-
\zeta^m\bar\zeta^m,$$ we obtain that $h$ is positive on the unit ball
$$h(t,\bar t)=1-t\bar t>0\quad\hbox{for}\quad  t\bar t<1,$$ which is
the symmetric space
$$\frac{\rSU(1,m)}{\rSU(m)\times\rU(1)}.$$ Notice that the metric is
$$g_{i\jb }=\frac 1{h^2}t^i\bar t^j+\frac 1h\delta^{ij}=
 -\partial _i\bar \partial _j\log h.$$
This means that we have to start with the negative K{\"a}hler potential
${\cal K}$. This sign is important in physical applications
(supergravity), as we saw in section \ref{ss:confcal}. In this case, we
have to use the minus signs in the last two terms of
(\ref{curvedcurvature}).

\hfill$\square$

\begin{example}A pseudo-Riemannian special K{\"a}hler manifold\end{example}
We want to describe now the pseudo-Riemannian symmetric space
\begin{equation}\frac{\rSU(1,2)}{\rSU(1,1)\times\rU(1)}.\label{pseudosym}\end{equation}

We start the construction as for the projective space, on which we try to
define a pseudo-Riemannian metric. As we did for the passage to the
unit-ball, we will have to restrict to those points where this metric is
non degenerate.

 As a fiber metric on the trivial bundle $\CP^2\times \C^3$ we take
\begin{equation}\label{metric1}\langle \zeta, \zeta
\rangle=\bar\zeta^1 \zeta^1 - \bar\zeta^2  \zeta^2 + \bar\zeta^3
\zeta^3.
\end{equation}
The space is covered by the three open sets
$$U_i=\left\{ \begin{pmatrix}\zeta^1\\\zeta^2\\\zeta^3\end{pmatrix}\;\big|\;\zeta^i\neq 0\right\},\qquad i=1,2,3$$ as
before.
% for each $U_i$, let $[\zeta]\subset U_i$ denote a
%line through the origin in $U_i$
%$$[\zeta]=\{\zeta\in U_i:\exists \;t_{ji}\in \C\;,\;\zeta_j=t_{ji}\zeta_i\}$$ $S_i$ is the set of these lines $S_i=\{[\zeta]\subset U_i\}$
%such that the pair $\{t_{j_1i}, t_{j_2i}\}$ defines a coordinate system
%for $S_i$
In the patch $U_1$, we have
$$
\begin{pmatrix}
\zeta^2\\
\zeta^3
\end{pmatrix}=
T^1 \zeta^1, \quad\hbox{with} \qquad T^1= \begin{pmatrix} t^{21}\\
t^{31}
\end{pmatrix},
$$
and a local section of the tautological bundle is given by a function
$\zeta^1(T^1)$,
$$\zeta(T_1)=
\begin{pmatrix}
\zeta^1\\
\zeta^2\\
\zeta^3
\end{pmatrix}= \begin{pmatrix}
1\\
T^1\end{pmatrix}\zeta^1(T^1).$$
 The inner product becomes
$$
\langle \zeta(T^1),\zeta'(T^1)\rangle = \bar \zeta^1 (1, T^{1\dagger}) g
\begin{pmatrix} 1\\ T^1 \end{pmatrix} \zeta'^1=\bar\zeta^1
(1-\bar t^{21}t^{21}+\bar t^{31}t^{31})\zeta'^1,
$$
with
$$g=\begin{pmatrix}1&0&0\\0&-1&0\\0&0&1\end{pmatrix}.$$ Doing
the same computation for $U_2, U_3$ we obtain
\begin{eqnarray}&&\hbox{For }U_1 \qquad \langle
\zeta(T^1),\zeta'(T^1)\rangle=\bar\zeta^1 (1-\bar t^{21}t^{21}+\bar
t^{31}t^{31})\zeta'^1\nonumber\\&&\hbox{For }U_2 \qquad \langle
\zeta(T^2),\zeta'(T^2)\rangle=\bar\zeta^2 (\bar t^{12}t^{12}-1+\bar
t^{32}t^{32})\zeta'^2\nonumber\\&&\hbox{For }U_3\qquad \langle
\zeta(T^3),\zeta'(T^3)\rangle=\bar\zeta^3 (\bar t^{13}t^{13}-\bar
t^{23}t^{23}+1)\zeta'^3.\label{metricreduction}
\end{eqnarray}
Equations (\ref{metricreduction}) give us the fiber metric on the
tautological (line) bundle. If  in each $U_i$
\begin{equation}\langle\zeta,\zeta'\rangle=h_i\zeta^i\bar\zeta'^i,  \qquad (\hbox{no sum over }i),
\label{metriccordinates}\end{equation} then
\begin{eqnarray*}h_1(T^1)&=&(1-\bar t^{21}t^{21}+\bar
t^{31}t^{31}),\\
h_2(T^2)&=&(\bar t^{12}t^{12}-1+\bar t^{32}t^{32}),\\
h_3(T^3)&=&(\bar t^{13}t^{13}-\bar t^{23}t^{23}+1).
\end{eqnarray*}
 In the intersections, the change of
coordinates
$$t^{ij}=\frac 1{t^{ji}},\qquad
\zeta^i=t^{ij}\zeta^j$$ leaves (\ref{metriccordinates}) invariant.

We have to restrict ourselves to the space where the fiber metric is
 positive definite. Let
$$\cU_i=\{T^i\in U_i \;|\; h_i(T^i)>0\}.$$ $\cU_1$ and $\cU_3$ are homeomorphic to
$\C^2$, but $\cU_2$  is $U_2$ minus a ball of radius 1 centered at
$T^2=0$. The point $T^2=0$ is the only point in $U_2$ that is not
contained in $U_1$ or $U_3$. So we can safely ignore $\cU_2$, since
$\{\cU_1,\cU_3\}$ form a covering of the space of points where the fiber
metric is definite positive.

Notice that in $U_1$, $\xi^2=t^{21}\xi^1$ serves as a coordinate and the
same in $U_3$, $\xi^2=t^{23}\xi^3$, so $\xi^2$ is a global coordinate and
describes $\C$. The other coordinate, $t^{21}$ or $t^{12}$ respectively
in $U_1$ and $U_3$ describe a sphere $S^2$, so we have that the topology
of (\ref{pseudosym}) is $S^2\times\C$.

Let us compute the Ricci form in $U_1$. For simplicity we will denote
$t^{21}=t^2$, $t^{31}=t^3$.
$$ \rho_{i\jb}=-\rmi\partial_i\partial_\jb\log h=
  \frac {-\rmi}{h^2}\begin{pmatrix}-1-t^3\bar t^3&t^3\bar t^2\\t^2\bar t^3&
 1-t^2\bar t^2\end{pmatrix}.$$ The metric is
 $$g_{i\jb}=\rmi \rho_{i\jb},$$ and it is easy to see that it has one positive and one
negative eigenvalue.

The prepotential is
$$\cF=\rmi(\eta^1\eta^1-\eta^2\eta^2+\eta^3\eta^3),$$ with
$$t^2=\frac{\eta^2}{\eta^1},\qquad t^3=\frac{\eta^3}{\eta^1},$$ and then the
K{\"a}hler potential is
$$\cK=4(\eta^1\bar\eta^1-\eta^2\bar\eta^2+\eta^3\bar\eta^3)=4\eta _1\bar \eta _1 h.$$

We can also see that  $\Xi =0$. Then (\ref{curvedcurvature}) gives
$R_{i\jb }=R^k{}_{ik\jb}=3g_{i\jb }$ and $R=6$.

\index{projective special K{\"a}hler manifold|)}

\hfill$\square$

\newpage

\section{Conclusions}

In this paper we have extended the definition of special K{\"a}hler geometry
to the case of arbitrary signature of the K{\"a}hler metric. For the rigid
case, we have extended the definition given in Ref. \cite{Freed:1997dp},
while for the projective case we have given a definition inspired in the
conformal calculus framework.

We have seen that the non existence of prepotential in some symplectic
coordinates, which was known for projective special geometry
\cite{Ceresole:1995jg} (the special geometry that occurs in
supergravity), is in fact a characteristic of pseudo Riemannian
manifolds, and applies also to the rigid case. This was masked by the
fact that in physical applications of rigid special geometry one is only
interested in the Euclidean signature, which gives positive definite
kinetic energy for the scalar fields.

Projective (or `local', referring to the local supersymmetry invariance
of supergravity) special geometry is obtained from a rigid special
manifold that has a closed homothetic Killing vector $K$. If $K$ is such
vector and $J$ is the complex structure $JK$  is a Killing vector, so
the metric has an extra U(1) symmetry. The result is that the existence
of a closed homothetic Killing vector is equivalent to the existence of
a holomorphic homothetic Killing vector, which we define in
(\ref{holohomoki}). This means essentially that there is an action of
the group $\C^\times$, and the procedure to obtain the projective
special manifold is to take quotient of the rigid `mother' manifold by
this action (and from here, the name of `projective' geometry). The
positive signatures of the kinetic terms of scalars and gravity in
supergravity theories require that the rigid manifold has signature $(2,
2n)$. We extend, however, projective special geometry to arbitrary
signatures. If the signature of the projective manifold is $(s,t)$ ($s$
positive eigenvalues, $t$ negative eigenvalues), then the signature of
the 'mother' rigid manifold is either $(s+1, t)$ or $(s, t+1)$. It is
the later case that occurs in supergravity. The standard formula for the
curvature is generalized to (\ref{curvedcurvature}), the lower choice in
$\pm$ being the standard supergravity case. The other possibility allows
us also to discuss special geometries with a compact isometry group.

In fact, this projectivization can be discussed for general K{\"a}hler
manifolds, not necessarily special. We develop the formalism in this more
general case and, for example, we prove that the projective K{\"a}hler
manifold is automatically K{\"a}hler-Hodge. As this is the method that is
used in conformal calculus, it implies that all the K{\"a}hler manifolds that
are constructed in this way for $N=1$ or $N=2$ supergravity satisfy the
K{\"a}hler-Hodge condition that was introduced in Ref. \cite{Witten:1982hu}.
We also give an interpretation of the Levi-Civita connection in these
projective K{\"a}hler manifolds as induced from the connection of the
`mother' manifold in a particular way, making use of the line bundle and
the Hermitian connection.

\newpage
%%%%%%%%%%%%%%%%%%%%%%%%%%%
\appendix

%%%%%%%%%%%%%%%%%%%%%%%%%%%%%%%%%%%%%%%%%%%%%%%%%%%%%%%%%%%%%%%%%%%%
%%%%%%%%%%%%%%%%%%%%%%Some technical results %%%%%%%%%%%%%%%%%%%%%%%
%%%%%%%%%%%%%%%%%%%%%%%%%%%%%%%%%%%%%%%%%%%%%%%%%%%%%%%%%%%%%%%%%%%%

\section{Some technical results \label{ss:sometec}}

\begin{lemma} (See lemma A1 in  \cite{Craps:1997gp})\label{lemmatoine}
Let $$V=\begin{pmatrix} \alpha\\\beta\end{pmatrix}$$ be an $2n\times n$
matrix of rank $n$ ($\alpha$ and $\beta$ are $n\times n$ matrices). Then,
there is a matrix $S\in \rSp(2n,\R)$ such that the transformed matrix
$$V'=SV=\begin{pmatrix} \alpha'\\\beta'\end{pmatrix}$$
has the property that $\alpha'$ itself has rank $n$.
\end{lemma}

{\it Proof.} We make an outline of the proof.  Let us denote
$$\alpha=\begin{pmatrix}\alpha^1\\\vdots \\\alpha^n\end{pmatrix}, \qquad
\beta=\begin{pmatrix}\beta_1\\\vdots \\\beta_n\end{pmatrix},$$ and let
$r-1\leq n$ be the rank of $\alpha$. If $r-1=n$ we have already the
result, so we will take $r-1<n$. Without loosing generality, we can
 assume that $\alpha^1,\dots \alpha^{r-1}$ are linearly
 independent. Then
\begin{equation}\alpha^r=\sum_{i=1}^{r-1}\lambda_i\alpha^i.\label{linearcombination}\end{equation}

Let $\beta_k$ be such that $\alpha^1,\dots \alpha^{r-1}, \beta_k$ are
also linearly independent. For the particular case $k=r$ the symplectic
matrix
$$S=\begin{pmatrix}\id-E_{r,r}&E_{r,r}\\
-E_{r,r}&\id-E_{r,r}\end{pmatrix}$$ gives an $\alpha'$ with rank $r$.

(We have used the standard notation
$(E_{i,j})_{\phantom{l}k}^l=\delta_i^l\delta_{jk}$.)

In the generic case $k\neq r$, we consider the symplectic matrix
$$S=\begin{pmatrix}\id-E_{r,r}-\frac 1\sigma E_{k,r}&E_{r,k}+\sigma E_{r,r}\\
-\frac 1 \sigma E_{r,r}&\id-E_{r,r}\end{pmatrix},$$ where $\sigma$ is a
parameter $\sigma\neq 0$. It follows that it is always possible to choose
$\sigma$ such that the vectors ${\alpha'}^1,\dots {\alpha'}^r$ are
independent.  In fact, the conditions on $\sigma$ are that $\sigma$ must
be different from some fixed quantities.

By iterating this procedure we see that transforming $V$ with a finite
number of symplectic matrices it is possible to construct a matrix $V'$
such that $\rank(\alpha')=n$. \hfill$\square$

\begin{remark}\label{remarklemmatoine}\end{remark}

When passing from a constant matrix $V$ to a point dependent matrix
$V(z)$, one has first to restrict to a neighbourhood where the same
components of $V$ are independent (not only in number). Otherwise the
theorem could not be applied. So we may have to enlarge the number of
open sets of our covering.

Next, we want to consider a constant symplectic transformation in order
to have flat Darboux coordinates in each open set. The constraints for
$\sigma$ (which must be constant) become now point dependent, namely
$\sigma$ must be different from certain functions of $z$ and $\bar z$.
This is always possible, but perhaps in an open subset of the original
open set. For each point there is a neighbourhood contained in a compact
set where the constraints can be satisfied. One can cover the manifold
with such neighbourhoods and, assuming that the space is locally compact,
one can pick up a subcovering which is locally finite. \hfill$\square$

\begin{lemma}\label{lemmaraja}The subgroup of $\rSp(2n,\R)$ formed by the matrices of the form
$$\left\{\begin{pmatrix}A&0\\C&(A^T)^{-1}\end{pmatrix}\right\}$$
is a maximal subgroup.\end{lemma}

\bigskip

Before going to the proof let us explain a way of  seeing maximality. Let
$\fg$ be a Lie algebra and $\fs$ a Lie subalgebra with respective groups
$G$ and $S$, course $S\subset G$. We are interested in deciding when
$\fs$ is maximal in $\fg$. Note that the adjoint action of $S$ on $\fg$
leaves $\fs$ stable and so $S$ acts on $\fa:=\fg/\fs$.
\bigskip
\begin{theorem}\label{theoremraja}If there is a subgroup $T\subset S$ such that
the action of  $T$ on $\fa$ is irreducible, then $\fs$ is maximal in
$\fg$\end{theorem} (We can operate over the complex numbers as maximality
over the complexes is stronger than maximality over the reals.)

{\it Proof.} Let $\fh$ be a subalgebra such that ${\fs}\subset
{\fh}\subset {\fg}$ with ${\fh}\neq\fs$. We must show that $\fh=\fg$.
Since $\fh$ is invariant under $T$, the image $\fb$ of $\fh$ in $\fa$ is
stable under $T$. Since $\fh$ is strictly larger than $\fs$, the space
$\fb$ is not $0$ and is stable under $T$. By the irreducibility of the
action of $T$ we must have $\fb=\fa$ so that $\fh=\fg$.\hfill$\Box$

\bigskip
Let us go back to the proof of Lemma \ref{lemmaraja}.

\smallskip

{\it Proof.} In our case $S$ is the lower triangular block group and so
we can take $\fa$ to be the space of matrices
$$
\begin{pmatrix} 0&b\\ 0&0\end{pmatrix}\qquad b=b^t.
$$
We take $T$ to be the subgroup
$$
\begin{pmatrix}A&0\\ 0&{A^t}^{-1}\end{pmatrix}\qquad A\in \rGL(n).
$$
Then the action of $T$ on $\fa$ works out to be
$$
A, b\longmapsto AbA^t
$$
which is the representation of ${\rm GL}(n)$ on the symmetric tensors of
the $n$-space, which is known to be irreducible. Applying the Theorem
\ref{theoremraja}, we complete our proof. \hfill$\square$

\section{\label{cpb}Connection on a principal bundle\index{principal bundle!connection} and covariant derivative\index{covariant derivative}}

We will relate now the definition \ref{defcpb} of connection on a
principal bundle to the covariant derivative in associated bundles.

Let $E$ be an associated vector bundle to $P$, with standard fiber $F$,
and let $R:G\rightarrow \mathrm{End}(F)$ be the representation of $G$ on
$F$. For simplicity we will consider $G\thickapprox R(G)$, although this
is not necessary. We want to define the covariant derivative of a section
of $E$ in terms of the connection 1-form. Let $\{e_1, \dots e_k\}$ be a
basis on $F$. A {\it local frame\index{local frame}} of $E$
 is a set of $k=\mathrm{rank}(E)$
independent local sections of $E$. We will denote it by
$\epsilon(m)=\{\epsilon_1(m),\dots \epsilon_k(m)\}$, with $m\in\cM$. It
can be interpreted as an invertible map $\epsilon(m):F\rightarrow E_m$
such that
$$\epsilon(m)(e_a)=\epsilon_a(m),$$ so it provides with an
identification of the fiber $E_m$ with $F$. The set of frames is a
principal bundle with structure group $\rGL_n$. $P$ is a subbundle of the
bundle of frames, so a local section $s:\cM\rightarrow P$ is a local
frame of $E$.

The pull-back  $\Gamma=\sigma^*\omega$ defines a local $\fg$-valued
1-form on $\cM$. The covariant derivative of a local section of $E$,
$\sigma=\sigma^a \epsilon_a$, is given by
\begin{equation}\nabla_i\sigma=\left(\partial_i\sigma^a
+\Gamma_i{}^a{}_ b\sigma^b\right)\epsilon _a,\quad \hbox{so}\quad
\Gamma_i{}^a{}_b=(\nabla_i \epsilon_b)^a.
\label{covder}
\end{equation}
(\ref{covder}) relates the definition of connection as a $\fg$-valued
1-form on $P$ with the notion of covariant derivative that we have been
using through the text. (\ref{covder}) is given in terms of a local
section $\sigma$ on $P$, but changing the section gives the usual gauge
transformation of the local connection 1-form on $\cM$. The coordinate
independent description of the covariant derivative can be found for
example in Ref. \cite{Kobayashi1996}.

%%%%%%%%%%%%%%%%%%%%%%%%%%%%%%%%
\medskip
\section*{Acknowledgments.}

\noindent We are grateful to L. Andrianopoli and S. Ferrara for
interesting and very useful discussions.

This work is supported in part by the European Community's Human
Potential Programme under contract MRTN-CT-2004-005104 `Constituents,
fundamental forces and symmetries of the universe'. The work of A.V.P. is
supported in part by the FWO - Vlaanderen, project G.0235.05 and by the
Federal Office for Scientific, Technical and Cultural Affairs through the
``Interuniversity Attraction Poles Programme -- Belgian Science Policy"
P5/27.

The work of M. A. Ll. and O. M. has been  supported in part by research
grants from the Spanish Ministerio de Educaci{\'o}n y Ciencia (FIS2005-02761
and EU FEDER funds), the Generalitat Valenciana (ACOMP06/187, GV-05/102).

A.V.P. thanks the Universitat de Val{\`e}ncia for hospitality during a visit
that initiated this work. M. A. Ll. and O. M. want to thank the Instituut
voor Theoretische Fysica, Katholieke Universiteit Leuven for its kind
hospitality during part of this work.

%%%%%%%%%%%%%%%%%%%%%%%%%%%%%%%%%%%%%%%%%%%%%%%%%%%%%%%
%\bibliography{refLectParis}
%%Included for WinEdt Gather Purpose (do not remove the comment line below:
%             %input "C:\localtexmf\bibtex\bib\refLectParis.bib"
%\bibliographystyle{toine}
\providecommand{\href}[2]{#2}\begingroup\raggedright\endgroup
\printindex
\end{document}